\documentclass[fleqn,usenatbib]{mnras}

\usepackage[T1]{fontenc}

\DeclareRobustCommand{\VAN}[3]{#2}
\let\VANthebibliography\thebibliography
\def\thebibliography{\DeclareRobustCommand{\VAN}[3]{##3}\VANthebibliography}

\usepackage{graphicx}	
\usepackage{amsmath}	
\usepackage{amssymb}	
\usepackage{multirow}   
\usepackage{newtxtext,newtxmath}

\graphicspath{{figs/}}

\setcitestyle{notesep={}}

\title[$UV$ Luminosity Densities at $z>8$ from the First NIRCam Fields]{UV Luminosity Density Results at $z>8$ from the First {\it JWST}/NIRCam Fields: Limitations of Early Data Sets and the Need for Spectroscopy}
\author[Bouwens et al.]{Rychard Bouwens$^{1}$\thanks{email: bouwens@strw.leidenuniv.nl},
Garth Illingworth$^{2}$, Pascal Oesch$^{3,4}$, Mauro Stefanon$^{5,6}$, Rohan Naidu$^{7,8}$,  
\newauthor
Ivana van Leeuwen$^{1}$, Dan Magee$^{2}$\\
$^{1}$Leiden Observatory, Leiden University, NL-2300 RA Leiden, Netherlands\\
$^{2}$UCO/Lick Observatory, University of California, Santa Cruz, CA 95064\\
$^{3}$Cosmic Dawn Center (DAWN), Niels Bohr Institute, University of Copenhagen, Jagtvej 128, K\o benhavn N, DK-2200, Denmark\\
$^{4}$Department of Astronomy, University of Geneva, Chemin Pegasi 51, 1290 Versoix, Switzerland\\
$^{5}$Departament d’Astronomia i Astrof{\' i}sica, Universitat de Val{\`e}ncia, C. Dr. Moliner 50, E-46100 Burjassot, Val{\` e}ncia, Spain\\
$^{6}$Unidad Asociada CSIC ”Grupo de Astrof{\' i}sica Extragal{\' a}ctica y Cosmolog{\' i}a” (Instituto de F{ \' i}sica de Cantabria - Universitat de Val{\` e}ncia)\\
$^{7}$Center for Astrophysics, Harvard \& Smithsonian, 60 Garden Street, Cambridge, MA 02138, USA\\
$^{8}$MIT Kavli Institute for Astrophysics and Space Research, 77 Massachusetts Ave., Cambridge, MA 02139, USA}

\date{Accepted XXX. Received YYY; in original form ZZZ}

\pubyear{2022}

\begin{document}
\label{firstpage}
\pagerange{\pageref{firstpage}--\pageref{lastpage}}
\maketitle

\begin{abstract}
We have derived luminosity functions, and set constraints on the UV
luminosity and SFR density from $z\sim17$ to $z\sim8$, using the three
most-studied JWST/NIRCam data sets, the SMACS0723, GLASS Parallel, and
CEERS fields.  We first used our own selections on two independent
reductions of these datasets using the latest calibrations. 18
$z\sim8$, 12 $z\sim10$, 5 $z\sim13$, and 1 $z\sim17$ candidate
galaxies are identified over these fields in our primary reductions,
with a similar number of candidates in our secondary reductions. We
then use these two reductions, applying a quantitative discriminator,
to segregate the full set of $z\geq8$ candidates reported over these
fields from the literature, into three different samples, ``robust,''
``solid,'' and ``possible''.  Using all of these samples we then
derive $UV$ LF and luminosity density results at $z\geq8$, finding
substantial differences.  For example, including the full set of
``solid'' and ``possible'' $z\geq12$ candidates from the literature,
we find $UV$ luminosity densities which are $\sim$7$\times$ and $\sim$20$\times$
higher than relying on the ``robust'' candidates alone. These results
indicate the evolution of the $UV$ LF and luminosity densities at
$z\geq8$ is still extremely uncertain, emphasizing the need for 
spectroscopy and deeper NIRCam+optical imaging to obtain reliable
results.  Nonetheless, even with the very conservative ``robust''
approach to selections, both from our own and those of other studies,
we find the luminosity density from luminous ($M_{UV}<-19$)
galaxies to be $\sim$2$\times$ larger than is easily achievable using constant star-formation efficiency models, similar to what other early {\it JWST} results
have suggested.
\end{abstract}



\section{Introduction} \label{sec:intro}

One particularly interesting long-standing question in extragalactic
astronomy has been "how early did massive galaxies
begun to assemble?".  While this question had been the subject of
many investigation prior to the start of {\it James Webb Space
  Telescope} ({\it JWST}) operations \citep{Rigby2022}, with
inferences being made both using very high redshift luminosity
functions
\citep[e.g.,][]{Bouwens2016_LF,Bouwens2019_LF,Bouwens2022_LF,Oesch2018_z10LF,Stefanon2019,Bowler2020_LF,Finkelstein2022,Kauffmann2022,Harikane2022_LF,Donnan2022}
and leveraging constraints on the stellar mass in sources
\citep[e.g.,][]{Duncan2014,Grazian2015,Song2016,Bhatawdekar2019,Kikuchihara2020,Furtak2021,Stefanon2021_SMF,Stefanon2022_IRACz10},
early {\it JWST} investigations are already demonstrating the revolutionary potential of {\it JWST} for addressing the early galaxy assembly timescale
\citep[e.g.][]{Adams2022,Naidu2022_z12,Castellano2022_GLASS,Donnan2022,Atek2022,Morishita2022,Finkelstein2022,Harikane2022_LF,Bouwens2022_XDF,Robertson2022_JADES}.

These early studies have have returned
very luminous and seemingly robust sources at $z\sim10$-13
\citep[e.g.][]{Naidu2022_z12,Castellano2022_GLASS,Finkelstein2022_z12,Bouwens2022_XDF},
and have also identified candidates out to redshifts as high as
$z\sim16$-20
\citep[e.g.][]{Donnan2022,Atek2022,Harikane2022_z9to17,Naidu2022_z17,Zavala2022,Yan2022,
  Finkelstein2022_JWSTLF}.  At the same time, some very massive
sources have been identified at $z\geq 7$ on the basis of what
appear to be substantial Balmer breaks \citep{Labbe2022}, suggesting
substantial early mass assembly in the universe. These 
results are enigmatic, however, potentially exceeding the available
baryons to form stars at $z\geq 13$ (\citealp[][, but see
  also]{BoylanKolchin2022,Naidu2022_z17}
\citealp[][]{Steinhardt2022,Inayoshi2022,Harikane2022_z9to17}).

Despite this flurry of new high-redshift sources, there have been
substantial differences in the $z\geq8$ candidate galaxy samples identified
in different studies over the same fields.  The typical overlap between candidate
lists in the earliest analyses were only $\sim$10-20\% (at least in
the initial versions of these
papers).\footnote{https://twitter.com/stewilkins/status/1554909423759409153}
Broadly such differences can be
indicative either of substantial contamination in $z\geq 8$ selections
or high levels of incompleteness.\footnote{For example, in cases where
source selection is $\gtrsim$80\% complete, different selections
should overlap at $\gtrsim$65\% level.  Conversely, if source overlap
between selections is less than 20\%, then the completeness of
individual selections cannot be generally higher than 50\%.}  In
either case, the inferred luminosity function results could be
substantially mis-estimated (by $\sim$0.3-0.5 dex).  Overall, this entire issue poses a major challenge as we try to understand what is really happening in the first 400-500 Myr at $z\geq10$ where {\it JWST} can provide unique new insights into galaxy buildup. 

The primary purpose of the present paper is to investigate the evolution of of the UV luminosity density and star formation rate density for galaxies at $z\geq8$ from early {\it JWST} data sets, while looking closely at the overall range of constraints allowed based on current observations.  Key to doing this in a quantitative way is to provide an assessment of
the first selections of $z\geq 8$ candidate galaxies over the
first {\it JWST} data sets and any updates to these selections that
have become possible due to improvements e.g. from improved zeropoint
calibrations \citep[e.g][]{Adams2022}, and to identify approaches that lead to the most robust samples.

To this end, we will make use of two independent, recent
reductions of the available {\it JWST} data that use the latest calibrations and experience in dealing with artifacts, over the three most well-studied
fields, the SMACS0723 cluster \citep{Pontop2022}, four NIRCam pointings from the Cosmic Evolution Early Release Science (CEERS) fields \citep{Finkelstein2022_JWSTLF}, and the NIRCam GLASS parallel field \citep{Treu2022_GLASS}.  Not only do select our own set of $z\geq 8$ candidates from these fields, but we also make an assessment of essentially all candidates from previous studies of these same fields, to gauge how well individual selections appear to be working and to characterize potential progress.  In doing so, we present community LF results, showing the impact of including
candidates of various quality on the $UV$ LF and luminosity density
results at $z>8$.  We will also investigate the
extent to which an emergent picture is forming on the basis of the latest results from the collective analyses.

The plan for this paper will be as follows.  In \S2, we summarize the
data sets utilized in this paper and our procedure for performing
photometry of sources in those data sets.  In \S3, we present our
procedure for selecting $z\geq 7$ sources from the data sets we
examine, the $z\sim8$-17 samples we derive, and our $z\sim8$, 10, 13,
and 17 LF results, while performing a detailed assessment of other
candidate $z\sim8$-17 galaxies in the recent literature.  In \S4, we
use those results to derive LF results based on our own selections and
our commmunity samples, discuss the results in \S5, and then provide a
summary in \S6.  For convenience, the {\it HST} F435W, F475W, F606W,
F814W, F098M, F125W, F140W, and F160W filters are written as
$V_{606}$, $I_{814}$, $Y_{098}$, $J_{125}$, $JH_{140}$, and $H_{160}$,
respectively, throughout this work.  Also we quote results in terms of
the approximate characteristic luminosity $L_{z=3}^{*}$ derived at
$z\sim3$ by \citet{Steidel1999}, \citet{Reddy2009}, and many other
studies.  A \citet{Chabrier2003} initial mass function is assumed
throughout.  For ease of comparison to other recent extragalactic
work, we assume a concordance cosmology with $\Omega_{m}=0.3$,
$\Omega_{\Lambda}=0.7$, and $H_0 = 70 \,\textrm{km/s/Mpc}$ throughout.
All magnitude measurements are given using the AB magnitude system
\citep{OkeGunn1983} unless otherwise specified.

\begin{table}
\centering
\caption{Estimated $5\sigma$ depth [in mag] of the three {\it JWST} fields (SMACS0723,
CEERS, and Abell 2744 parallel) we use in searching for galaxies at, nominally, $z\sim8$, $z\sim10$, $z\sim13$, and $z\sim17$.  A 0.35$''$-diameter aperture is adopted for the photometry in computing the depth in each field.}
\label{tab:dataset}
\begin{tabular}{c|c|c|c} \hline
       & \multicolumn{3}{c}{$5\sigma$ Depths (AB mag)$^a$} \\
Filter & SMACS0723 & GLASS-P & CEERS \\
\hline\hline
{\it HST}/F435W        & 26.8 & ---  & 28.4 \\
{\it HST}/F606W        & 27.6 & 28.6 & 28.6 \\
{\it HST}/F775W        & ---  & 28.3 & --- \\
{\it HST}/F814W        & 27.2 & ---  & 28.4 \\
{\it JWST}/F090W & 28.7 & 28.9 & ---\\
{\it HST}/F105W        & 26.9 & ---  & 26.9 \\
{\it JWST}/F115W & 27.6 & 28.9 & 28.7 \\
{\it HST}/F125W        & 26.2 & ---  & 26.8 \\
{\it HST}/F140W        & 26.4 & ---  & 26.2 \\
{\it HST}/F160W        & 26.4 & ---  & 27.2 \\ 
{\it JWST}/F150W & 29.1 & 28.9 & 28.9\\
{\it JWST}/F200W & 29.3 & 29.0 & 29.2 \\
{\it JWST}/F277W & 29.5 & 29.3 & 29.0 \\
{\it JWST}/F356W & 29.4 & 29.2 & 28.9 \\
{\it JWST}/F410M & ---  & ---  & 28.1 \\
{\it JWST}/F444W & 29.1 & 29.5 & 28.4 \\\hline
Area [arcmin$^2$] & 9.5 & 7.0 & 33.8 \\\hline\hline
\end{tabular}
\\\begin{flushleft}
$^a$ These depths include a correction for the flux in point sources lying outside a 0.35$"$-diameter aperture and thus correspond to the total magnitudes of point sources that would be detected at $5\sigma$ for a given band.
\end{flushleft}
\end{table}

\begin{figure*}
\centering
\includegraphics[width=2\columnwidth]{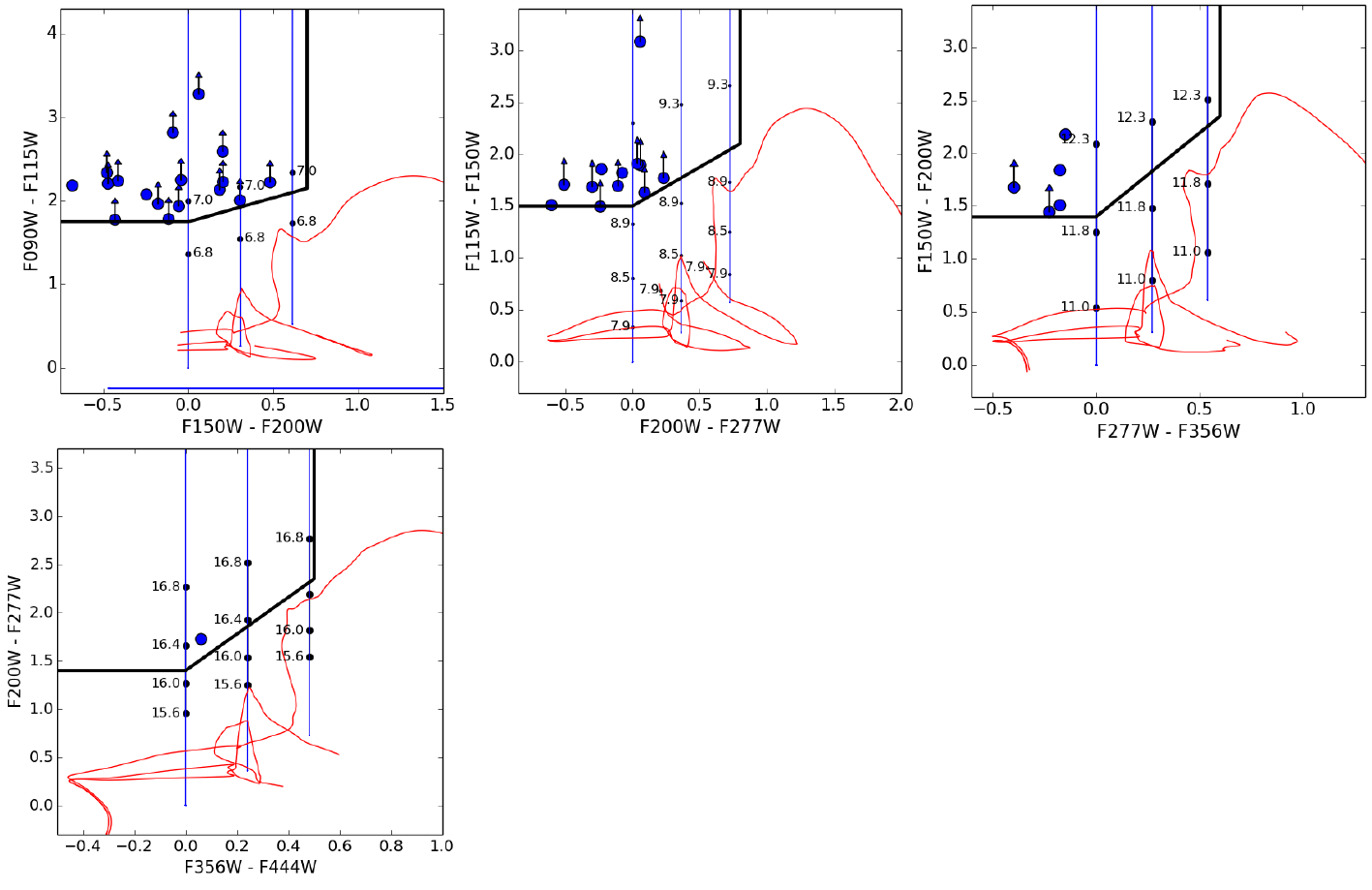}
\caption{Two color Lyman-break selection criteria we use in
  identifying our nominal $z\sim8$, $z\sim10$, $z\sim13$, and $z\sim17$ candidate
  galaxies in three of the {\it JWST}/NIRCam data
  sets that have been widely used for early galaxy selections.  The thick black lines indicate the boundaries of our
  color-color criteria.  The blue lines indicate the expected colors
  of star-forming galaxies with 100 Myr constant SF histories and
  $E(B-V)$ dust extinction of 0, 0.15, and 0.3, with colors at
  specific redshifts indicated by the black dots.  The red lines show
  the expected colors of lower-redshift template galaxies from
  \citet{Coleman1980} out to $z\sim5$.  The solid blue circles show
  the colors of specific sources in our primary selection.  In cases
  where sources are not detected in a band, they are shown with an
  arrow at the $1\sigma$ limit. These nominal selections span the redshift ranges $z\sim7-9$, $z\sim9-11$, $z\sim12-14$ and $z\sim16-19$, respectively (cf. Figure 2). }
    \label{fig:criteria}
\end{figure*}

\section{Data Sets and Photometry}

\subsection{Data Sets\label{sec:data}}

We make use of the three most studied NIRCam data sets in
constructing a selection of $z\geq 7$ galaxies from current {\it JWST}
observations, i.e., the $\sim$12-hour SMACS0723 cluster field featured
in the {\it JWST} early release observations \citep{Pontop2022}, the
4-pointing NIRCam observations taken as part of the CEERS early
release science program
\citep[][]{Finkelstein2022_JWSTLF}, and the
sensitive NIRCam parallel observations as part of the GLASS early
release science program \citep{Treu2022_GLASS}.

These three fields cover a total area of $\sim$51 arcmin$^2$.  The
approximate $5\sigma$ depths of these data sets reach from $\sim$28 to
29.2 mag and are presented in detail in Table~\ref{tab:dataset}.
These depths are derived by measuring the flux variations in
source-free 0.35$''$-diameter apertures across our reduced images of
each field.

We derived PSFs for the fields by taking a PSF from WebbPSF and then
drizzling onto a grid consistent with our NIRCam reductions.  The
FWHMs for the F090W, F115W, F150W, F200W, F277W, F356W, F410M, and
F444W PSFs are 0.06$''$, 0.06$''$, 0.06$''$, 0.07$''$, 0.12$''$,
0.14$''$, 0.15$''$, and 0.16$''$, respectively.  We also extracted
empirical PSFs based on isolated, non-saturated stars located across
the NIRCam images and obtained consistent results.

Our fiducial reductions of each data set are executed using the
\textsc{grizli} software \citep{grizli2022}.  
\textsc{grizli} has procedures in place
both to minimize the $1/f$ noise and to mask ``snowballs'' on
individual NIRCam frames.  \textsc{grizli} combines NIRCam frames
using the \textit{astrodrizzle} software package, after modifying the
headers of the frames to use the required SIP WCS headers.  The
\textsc{grizli} reductions also take advantage of both significantly
improved flat fields (and the jwst\_0942.pmap calibration files) that 
became available in early September and the zeropoint adjustments 
derived by G. Brammer et al.\ (2022, in prep).

To better understand possible systematics and how they impact the
selection of star-forming galaxies at $z\geq 6$, we also make use of
the NIRCam imaging pipeline \textsc{pencil} (Magee et al.\ 2022, in
prep) built for the PRIMER team (PI: Dunlop).  This pipeline leverages
STScI's {\it JWST} Calibration pipeline (v1.6.2), but also includes
additional processing steps which are not part of the standard 
calibration pipeline. This includes the subtraction of $1/f$ noise striping
patterns (both vertical and horizontal) that are not fully removed by
the standard calibration pipeline and the subtraction of “wisps”
artifacts from the short wavelength filters F150W and F200W in the
NRCA3, NRCB3, and NRCB4 detector images.

Additionally, the background sky subtraction is performed by
subtracting the median background over a N$\times$N grid while using a
segmentation map to mask pixels attributed to sources.  Image
alignment is executed in two passes using the calibration pipeline’s
\textsc{TweakReg} step and then using STScI python package
\textsc{TweakWCS}: the first pass uses TweakReg to group overlapping
images for each detector/filter and perform an internal alignment
within the detector/filter group; the second performs alignment
against an external catalog using \textsc{TweakWCS}. The external
catalog is, if possible, generated from an HST ACSWFC image mosaic
which has been registered to the GAIA DR3 catalog.

Finally, the \textsc{Pencil} reductions we utilize take advantage of
calibration files (jwst\_1009.pmap) which have been updated to reflect
new in-flight photometric
zeropoints.\footnote{https://www.stsci.edu/contents/news/jwst/2022/an-improved-nircam-flux-calibration-is-now-available.html}
Before the final NIRCam image mosaics are generated using the
calibration pipeline \textsc{calweb\_image3} stage, we perform an
additional step to identify and mask “snowball” artifacts that are not
identified and masked during the \textsc{calweb\_detector1} stage.

For the {\it HST} Advanced Camera for Surveys (ACS) and Wide Field Camera 3 near-IR (WFC3/IR) observations over SMACS0723, GLASS parallel field, and CEERS Extended Groth Strip (EGS) field, we made use of a reduction of the data generated by \textsc{grizli} for our fiducial set of reductions made with \textsc{grizli}.  For the alternate set of reductions made with \textsc{pencil}, we used reductions made with \textsc{astrodrizzle} for the SMACS023 and GLASS parallel fields, following many of the same procedures used in the product of the XDF data set \citep{Illingworth2013}.  Finally, for the CEERS EGS field, we made use of the ACS and WFC3/IR data products made available by CEERS team prior to the start of science observations by {\it JWST}. 

\subsection{Source Detection and Photometry}

As in previous efforts by our team
\citep[e.g.][]{Bouwens2011_LF,Bouwens2015_LF,Bouwens2019_LF,Bouwens2021_LF,Bouwens2022_LF},
we perform source detection and photometry using SExtractor
\citep{Bertin1996}.  For our F090W, F115W, F150W, and F200W dropout
selections, source detection is performed using the square root of
$\chi^2$ image constructed by coadding PSF-matched F200W, F277W,
F356W, and F444W data over the fields.  PSF matching is done using our
own implementation of the Lucy-Richardson deconvolution algorithm
\citep{Richardson1972,Lucy1974}.

Color measurements for sources are made based on the measured flux in
0.35$''$-diameter apertures, after PSF-correcting the shorter
wavelength data to match the PSF in the F444W band.  These
measurements are then corrected to total using (1) the additional flux
in scalable \citet{Kron1980} apertures with Kron factor of 2.5 and (2)
using the estimated flux outside these scalable apertures based on the
encircled energy distribution in the derived PSFs.

Finally, a foreground dust correction based on extinction maps of
\citet{Schlafly2011} is applied to colors and total magnitude
measurements.

\section{Source Selections}

\subsection{Lyman Break Selections}

We make use of two-color Lyman break selections to identify $z>6$
galaxies from the {\it JWST} data of the three fields identified above.  Lyman-break selections
have been shown to be a very efficient way of identifying star-forming
galaxies in the distant universe
\citep[e.g.,][]{Steidel1999,Bouwens2011_LF,Bouwens2015_LF,Bouwens2021_LF,Schenker2013}
and largely lie at the redshifts targeted by Lyman-break selections, given adequate S/N and bands either side of the break
\citep[e.g.][]{Steidel1999,Steidel2003,Stark2010,Ono2012,Finkelstein2013,Oesch2015,
  Zitrin2015,Oesch2016,Hashimoto2018_z9,Jiang2021}.

In devising color-color criteria for our selection, we follow the
strategy employed in \citet{Bouwens2015_LF,Bouwens2021_LF} and make
use of a two-color selection criterion, the first color probing the
Lyman break and the second color probing the color of the
$UV$-continuum just redward of the break.  In choosing the passbands
to utilize for this second color, we select bands which show no
overlap with either the Lyman or Balmer breaks, ensuring that our
selection would include even sources with prominent Balmer breaks, as
\citet{Labbe2022} find at $z\sim7$-10.

After some experimentation, we made use of the following two color
criteria:
\begin{eqnarray*}
(F090W - F115W > 1.75)\wedge (F150W - F200W < 0.7)\wedge \\ (F090W -
  F115W > 1.75 + 0.57(F150W-F200W))
\end{eqnarray*}
for our nominal $z\sim8$ selection,
\begin{eqnarray*}
(F115W - F150W > 1.5)\wedge (F200W - F277W < 0.8)\wedge \\
(F115W - F150W > 1.5 + 0.75(F150W-F200W))
\end{eqnarray*}
for our nominal $z\sim10$ selection,
\begin{eqnarray*}
(F150W - F200W > 1.4)\wedge (F277W - F356W < 0.8)\wedge \\
(F150W - F200W > 1.4 + 0.75(F277W-F356W))
\end{eqnarray*}
for our nominal $z\sim13$ selection, and
\begin{eqnarray*}
(F200W - F277W > 1.4)\wedge (F356W - F444W < 0.5)\wedge \\
(F200W - F277W > 1.4 + 1.58(F356W-F444W))
\end{eqnarray*}
for our nominal $z\sim17$ selection.  In cases where sources are undetected in
a given band, flux is set to the $1\sigma$ limit when applying the
selection criteria.  As we show below, when we discuss the redshift selection functions, these nominal redshift selections are better characterized as selections with $z\sim7-9$, $z\sim9-11$, $z\sim12-14$ and $z\sim16-19$ (the approximate half-power points of the redshift selection functions).

Given the significant variation in the composition of various $z\geq8$
selections in the literature, we have purposefully required that
sources show especially large Lyman breaks to maximize the
robustness of the sources we select.  The presence of a large spectral
break is perhaps the most model-independent feature of star-forming
galaxy at very high redshifts and will significantly less sensitive to
uncertainties in the NIRCam zeropoints than photometric redshift codes
that rely on fits to the spectral energy distribution.  As in
\citet{Bouwens2015_LF}, sources are excluded from a Lyman break
selection, if they meet the selection criteria of a higher-redshift sample.  

\begin{figure}
\centering
\includegraphics[width=\columnwidth]{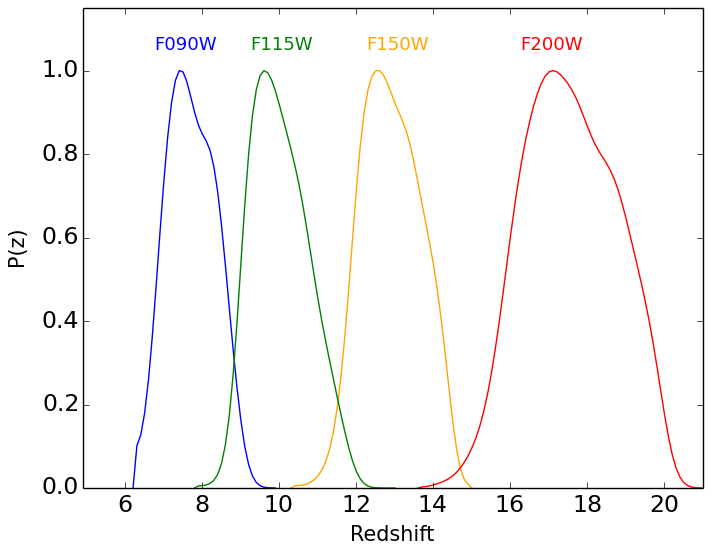}
\caption{Redshift selection functions for our nominal $z\sim8$, $z\sim10$,
  $z\sim13$, and $z\sim17$ selections leveraging a Lyman-break
  selection identifying sources dropping out in the F090W, F115W,
  F150W, F200W, and F277W bands, respectively.  The mean redshift of
  these selections is 7.7, 10.0, 12.9, and 17.6, respectively. These nominal redshift selections can be characterized as ranges with $z\sim7-9$, $z\sim9-11$, $z\sim12-14$ and $z\sim16-19$ (the approximate half-power points of the redshift selection functions) }
    \label{fig:zsel}
\end{figure}

Additionally, we require that sources show no significant flux
blueward of the break.  For this, we co-add the flux blueward of
candidate Lyman breaks using the $\chi^2$ statistic defined
in \citet{Bouwens2011_LF,Bouwens2015_LF} and which is equal to
$\Sigma_{i} \textrm{SGN}(f_{i}) (f_{i}/\sigma_{i})^2$ where $f_{i}$ is
the flux in band $i$ in a consistent aperture, $\sigma_i$ is the
uncertainty in this flux, and SGN($f_{i}$) represents the nominal 
"sign" function from mathematics, being equal to 1 if $f_{i}>0$
and $-1$ if $f_{i}<0$.  Included in this statistic for the following redshift selections are the following bands:
\begin{eqnarray*}
z\sim8&:& B_{435}g_{475}V_{606}I_{814}\\
z\sim10&:& B_{435}g_{475}V_{606}I_{814}z_{090}Y_{105}\\
z\sim13&:& B_{435}g_{475}V_{606}I_{814}z_{090}Y_{105}YJ_{115}J_{125}\\
z\sim17&:& B_{435}g_{475}V_{606}I_{814}z_{090}Y_{105}YJ_{115}J_{125}JH_{140}J_{150}H_{160}
\end{eqnarray*}
Sources are excluded from our selection if
the $\chi^2$ statistics in $0.2''$-diameter, $0.35''$-diameter, or
scalable Kron apertures (using a Kron parameter of 1.2) exceeds 4, 6,
7, 8, and 9, when combining data from 1, 2, 3, 4, and 5 bands,
respectively.

To ensure that sources in our selection corresponded to real objects,
we require that sources be detected at 6$\sigma$ in a stack of all
bands redward of the break in a 0.35$''$-diameter aperture.  We also require sources to be detected at $5\sigma$ in the band just redward of the break to ensure that the break is present at high significance.  Sources are also required to be detected at 3.5$\sigma$ in at least 5, 4, 4, and 3, independent bands redward of the Lyman-break in our $z\sim8$,
$z\sim10$, $z\sim13$, and $z\sim17$ selections.

Following our selection of candidate $z\sim8$-17 sources using the
above criteria, redshift likelihood functions $P(z)$ were computed for
each source using the EAZY photometric redshift code
\citep{Brammer2008}.  In fitting the photometry of individual sources,
use of spectral templates from the EAZY\_v1.0 set and Galaxy
Evolutionary Synthesis Models \citep[GALEV:][]{Kotulla2009} was made.
Nebular continuum and emission lines were added to the templates 
according to the prescription
provided in \citet{Anders2003}, a $0.2 Z_{\odot}$ metallicity, and
scaled to a rest-frame EW for H$\alpha$ of 1300\AA.  Sources are only
retained in our $z>6$ selections, if $>$80\% of the integrated
redshift likelihood is at $z>5.5$, i.e., $P(z>5.5)>0.8$.

Finally, all candidate $z\geq 8$ galaxies are visually examined to
exclude any sources associated with diffraction spikes, on the wings
of early type galaxies, or in regions of the images with elevated
background levels.

The approximate redshift distributions of these selections are
illustrated in Figure~\ref{fig:zsel} and derived using our selection
volume simulations described in \S\ref{sec:ourlfs}.  Using these
selection volume simulations, the mean redshifts inferred for our
F090W, F115W, F150W, and F200W dropout selections are equal to 7.7,
10.0, 12.9, and 17.6, hence our nominal use of $z\sim8$, $z\sim10$,
  $z\sim13$, and $z\sim17$ to identify these samples throughout our paper.

\subsection{$z\geq 8$ Selections\label{sec:oursel}}

Applying our selection criteria to our fiducial reductions using
\textsc{grizli}, we identify 18 $z\sim8$, 12 $z\sim10$, 5 $z\sim13$,
and 1 $z\sim17$ galaxies which satisfy all of our selection criteria.
The apparent magnitude of these sources range from 25.8 to 28.4 mag.
A list of these sources is presented in Table~\ref{tab:cursample} and
will be known as our primary selection.  Figure~\ref{fig:criteria}
shows the colors of our selected $z\sim8$-17 sources relative to the
two color criteria we utilize.

We also indicate in Table~\ref{tab:cursample} which sources from our
selections lie in earlier selections.  Encouragingly, $\sim$67\% of
the $z\geq 8$ candidates from our selection lie in previous $z\geq 8$
selections.  This is a significantly higher level of overlap than was
the case for the first set of $z\geq 8$ selections from the SMACS0723,
CEERS, and Abell 2744 parallel data sets, thanks to the slightly more
complete selections of $z\geq 8$ sources in
\citet{Finkelstein2022_JWSTLF} and \citet[][, v2]{Donnan2022} taking
advantage of a much improved NIRCam zeropoint calibration.

We also pursue an alternate selection of $z\geq 8$ galaxies based on
our NIRCam reductions using the \textsc{pencil} pipeline.  We identify
22 $z\sim8$, 13 $z\sim10$, 3 $z\sim13$, and 1 $z\sim17$ galaxy
candidates in that selection.  Table~\ref{tab:sample2} in Appendix A
provides the coordinates, estimated redshifts, magnitudes, constraints
on the Lyman break amplitudes, and estimated likelihood to
lie at $z>5.5$.

Comparing our two selections, we find $\sim$47\% of the sources in our
primary selection are also present in our alternate selection, while
$\sim$38\% of the sources in our alternate selection also occur in our
primary selection.  The percentage overlap between our selections is
similar to the $\sim$50\% overlap frequently seen between different
$z\sim4$-8 selections executed over the Hubble Ultra Deep Field with
HST data (see \S3.4 of \citealt{Bouwens2015_LF} for a discussion).
The existence of differences between the two selections is not
surprising given our use of two different reductions of the data to
identify sources and measure fluxes.

In an effort to better understand why there is only modest overlap
between the two selections, we compared the photometry of sources in
catalogs where they are selected vs. where they are not selected.
We found that most of the observed differences could be explained 
by variations in the size of the spectral breaks for selected 
sources and the apparent flux blueward of the nominal spectral 
breaks.   

\begin{table*}
\centering
\caption{Selection of nominal $z\sim8$, $z\sim10$, $z\sim13$, and $z\sim17$ sources identified over three most studied imaging fields (SMACS0723,
CEERS, and Abell 2744 parallel) in early {\it JWST} NIRCam observations from our \textsc{grizli} sample.}
\label{tab:cursample}
\begin{tabular}{c|c|c|c|c|c|c|c|c} \hline
     &    &     &            & $m_{UV}$ & Lyman Break$^a$ & $\Delta \chi^2$$^{b}$ &    &  \\
  ID & RA & DEC & $z_{phot}$ & [mag] & [mag] & & $p(z>5.5)$ & Lit$^{1}$ \\\hline\hline
  \multicolumn{9}{c}{$z\sim8$ Selection}\\
  \hline
GLASSP1Z-4049020480 & 00:14:04.908 & $-$30:20:48.04 & 6.7$_{-0.0}^{+0.0}$ & 27.9$\pm$0.2 & $>$2.1 & $-$5.8 & 0.938 & \\
GLASSP2Z-4027418404 & 00:14:02.740 & $-$30:18:40.44 & 6.8$_{-0.1}^{+0.2}$ & 27.8$\pm$0.2 & $>$2.4 & $-$12.1 & 0.996 & B22M\\
GLASSP1Z-4027520346 & 00:14:02.751 & $-$30:20:34.66 & 7.2$_{-0.4}^{+0.3}$ & 28.4$\pm$0.3 & $>$2.2 & $-$3.9 & 0.966 & B22M\\
GLASSP1Z-4030120490 & 00:14:03.018 & $-$30:20:49.03 & 7.2$_{-0.4}^{+0.3}$ & 28.2$\pm$0.2 & $>$2.2 & $-$6.2 & 0.983 & \\
GLASSP1Z-4047621148 & 00:14:04.769 & $-$30:21:14.87 & 7.2$_{-0.3}^{+0.3}$ & 28.3$\pm$0.3 & $>$2.2 & $-$7.7 & 0.993 & \\
GLASSP1Z-4041221456 & 00:14:04.125 & $-$30:21:45.65 & 7.3$_{-0.5}^{+0.4}$ & 28.4$\pm$0.3 & $>$2.0 & $-$3.8 & 0.914 & B22M\\
GLASSP1Z-3589121028 & 00:13:58.913 & $-$30:21:02.82 & 7.4$_{-0.3}^{+0.2}$ & 28.0$\pm$0.2 & $>$2.6 & $-$13.5 & 1.0 & B22M\\
GLASSP1Z-3581321041 & 00:13:58.136 & $-$30:21:04.19 & 7.4$_{-0.4}^{+0.4}$ & 28.4$\pm$0.2 & $>$2.0 & $-$4.8 & 0.962 & B22M\\
GLASSP1Z-4075621492 & 00:14:07.560 & $-$30:21:49.20 & 7.5$_{-0.5}^{+0.5}$ & 28.3$\pm$0.2 & $>$2.0 & $-$4.8 & 0.97 & \\
GLASSP1Z-3577921174 & 00:13:57.792 & $-$30:21:17.44 & 7.6$_{-0.4}^{+0.4}$ & 26.4$\pm$0.1 & 2.3$\pm$0.7 & $-$56.3 & 1.0 & B22M\\
S0723Z-3295626401 & 07:23:29.565 & $-$73:26:40.17 & 7.6$_{-0.6}^{+0.6}$ & 27.7$\pm$0.2 & 2.2$\pm$1.7 & $-$21.4 & 1.0 & \\
GLASSP1Z-4049021453 & 00:14:04.905 & $-$30:21:45.36 & 7.6$_{-0.5}^{+0.5}$ & 27.0$\pm$0.2 & $>$2.3 & $-$2.2 & 0.861 & \\
GLASSP1Z-3575621160 & 00:13:57.567 & $-$30:21:16.02 & 7.8$_{-0.4}^{+0.4}$ & 28.5$\pm$0.2 & $>$1.9 & $-$20.0 & 1.0 & Le22\\
GLASSP1Z-4020821598 & 00:14:02.087 & $-$30:21:59.85 & 7.8$_{-0.5}^{+0.4}$ & 27.2$\pm$0.2 & $>$2.7 & $-$8.0 & 0.992 & B22M\\
S0723Z-3201526042 & 07:23:20.159 & $-$73:26:04.29 & 7.8$_{-0.2}^{+0.3}$ & 26.4$\pm$0.1 & $>$3.2 & $-$41.1 & 1.0 & D22,At22,Ad22,B22M\\
GLASSP2Z-4016419299 & 00:14:01.649 & $-$30:19:29.90 & 8.1$_{-0.4}^{+0.4}$ & 27.8$\pm$0.2 & $>$2.3 & $-$16.7 & 1.0 & \\
S0723Z-3226926062 & 07:23:22.698 & $-$73:26:06.24 & 8.1$_{-0.2}^{+0.3}$ & 25.8$\pm$0.1 & $>$3.6 & $-$48.9 & 1.0 & D22,At22,Ad22,B22M\\
GLASSP1Z-4009521308 & 00:14:00.950 & $-$30:21:30.85 & 8.3$_{-0.3}^{+0.3}$ & 27.3$\pm$0.1 & $>$2.7 & $-$5.0 & 0.937 & \\
\\
\multicolumn{9}{c}{$z\sim10$ Selection}\\
\hline
CEERSYJ-0012159472 & 14:20:01.212 & 52:59:47.29 & 8.9$_{-0.5}^{+0.5}$ & 27.6$\pm$0.3 & $>$1.7 & $-$2.2 & 0.86 & F22,B22M\\
CEERSYJ-9345150450 & 14:19:34.516 & 52:50:45.06 & 9.2$_{-0.6}^{+0.8}$ & 28.2$\pm$0.3 & 1.5$\pm$1.0 & $-$10.5 & 0.997 & \\
CEERSYJ-9586559217 & 14:19:58.654 & 52:59:21.77 & 9.2$_{-0.2}^{+0.1}$ & 27.1$\pm$0.1 & 1.9$\pm$0.7 & $-$15.1 & 0.999 & F22,W22\\
GLASSP1YJ-4003721456 & 00:14:00.378 & $-$30:21:45.60 & 9.2$_{-0.7}^{+0.8}$ & 28.5$\pm$0.2 & $>$1.6 & $-$4.1 & 0.953 & B22M\\
CEERSYJ-9203050435 & 14:19:20.300 & 52:50:43.59 & 9.8$_{-0.7}^{+0.8}$ & 28.5$\pm$0.3 & $>$1.7 & $-$8.9 & 0.989 & \\
CEERSYJ-0012959481 & 14:20:01.290 & 52:59:48.12 & 9.9$_{-0.8}^{+0.8}$ & 27.8$\pm$0.3 & $>$1.8 & $-$4.1 & 0.952 & F22,B22M\\
CEERSYJ-9149352106 & 14:19:14.935 & 52:52:10.63 & 10.0$_{-0.7}^{+0.7}$ & 27.6$\pm$0.2 & 1.8$\pm$2.1 & $-$4.7 & 0.956 & \\
CEERSYJ-9353350378 & 14:19:35.337 & 52:50:37.87 & 10.2$_{-0.7}^{+0.6}$ & 27.9$\pm$0.2 & $>$1.7 & $-$14.2 & 0.999 & D22,F22,W22,B22M\\
GLASSP1YJ-4028622186 & 00:14:02.861 & $-$30:22:18.69 & 10.2$_{-0.4}^{+0.4}$ & 26.7$\pm$0.1 & $>$3.1 & $-$30.4 & 1.0 & N22,C22,D22,H22,B22M\\
GLASSP1YJ-4002721259 & 00:14:00.278 & $-$30:21:25.95 & 10.6$_{-0.7}^{+0.6}$ & 28.0$\pm$0.2 & $>$1.9 & $-$5.6 & 0.975 & D22\\
CEERSYJ-9026550577 & 14:19:02.654 & 52:50:57.74 & 11.2$_{-0.5}^{+0.5}$ & 27.8$\pm$0.2 & $>$1.6 & $-$5.7 & 0.968 & \\
GLASSP1YJ-4069421497 & 00:14:06.945 & $-$30:21:49.73 & 11.2$_{-0.4}^{+0.4}$ & 26.8$\pm$0.1 & $>$1.9 & $-$3.5 & 0.841 & C22,B22M\\
\\
\multicolumn{9}{c}{$z\sim13$ Selection}\\
\hline
CEERSH-9463556328 & 14:19:46.352 & 52:56:32.82 & 11.6$_{-0.5}^{+0.4}$ & 27.7$\pm$0.2 & 1.5$\pm$1.0 & $-$9.9 & 0.995 & D22,F22,H22,B22M\\
GLASSP2H-3597519291 & 00:13:59.756 & $-$30:19:29.14 & 12.1$_{-0.2}^{+0.2}$ & 26.7$\pm$0.1 & 1.8$\pm$0.2 & $-$21.6 & 1.0 & N22,C22,D22,H22,B22M\\
S0723H-2522527555 & 07:22:52.258 & $-$73:27:55.52 & 12.9$_{-0.9}^{+0.9}$ & 28.0$\pm$0.2 & 2.2$\pm$2.4 & $-$4.8 & 0.961 & At22,Y22\\
GLASSP2H-3576218534 & 00:13:57.627 & $-$30:18:53.49 & 13.7$_{-1.0}^{+1.0}$ & 28.2$\pm$0.2 & $>$1.4 & $-$10.2 & 0.996 & \\
S0723H-2394130081 & 07:22:39.416 & $-$73:30:08.17 & 14.9$_{-0.7}^{+0.8}$ & 26.9$\pm$0.1 & $>$1.7 & $-$6.4 & 0.971 & At22\\
\\
\multicolumn{9}{c}{$z\sim17$ Selection}\\
\hline
CEERSK-9394956348 & 14:19:39.491 & 52:56:34.87 & 16.3$_{-0.4}^{+0.3}$ & 26.3$\pm$0.1 & 1.7$\pm$0.1 & $-$8.7 & 0.987 & D22,F22,H22,N22,\\
& & & & & & & & Z22\\
\hline
\end{tabular}
\\\begin{flushleft}
$^1$ Ad22 = \citet{Adams2022}, At22 = \citet{Atek2022}, C22 = \citet{Castellano2022_GLASS}, D22 = \citet{Donnan2022}, F22 = \citet{Finkelstein2022_z12,Finkelstein2022_JWSTLF}, H22 = \citet{Harikane2022_z9to17}, La22 = \citet{Labbe2022}, N22 = \citet{Naidu2022_z12,Naidu2022_z17}, W22 = \citet{Whitler2022_z10}, Y22 = \citet{Yan2022}, B22 = This Work (primary selection), B22M = This Work (alternate selection), Z22 = \citet{Zavala2022}\\
$^a$ Amplitude of Lyman break in mag.  Lower limits are $1\sigma$.\\
$^b$ $\chi^2_{\textrm{best},z>5.5} - \chi^2_{\textrm{best},z<5.5}$\\
\end{flushleft}
\end{table*}

\begin{table*}
\centering
\caption{Sample of $z\sim10$, $z\sim13$, and $z\sim17$ candidate that we deem to be ``robust'' using our own photometry on two separate reductions of the NIRCam Data.}
\label{tab:robust}
\begin{tabular}{c|c|c|c|c|c|c|c|c} \hline
  ID & RA & DEC & $z_{phot}$$^a$ & $M_{UV}$$^*$ & Break [mag]$^{b,c}$ & $\Delta \chi^2$$^{b,d}$ & $p(z>5.5)$$^b$ & Lit\\\hline\hline
  \multicolumn{9}{c}{\citet{Naidu2022_z12}}\\
GL$-$z11 & 00:14:02.857 & $$-$$30:22:18.92 & 10.4/10.2$_{-0.4}^{+0.4}$ & $-$20.7 & $>$3.1,$>$3.3 & $-$30.4,$-$33.7 & 1.000,1.000 &  C22,D22,H22,\\
& & & & & & & & B22,B22M\\
GL$-$z13 & 00:13:59.754 & $$-$$30:19:29.10 & 12.4/12.1$_{-0.2}^{+0.2}$ & $-$21.0  & 1.8$\pm$0.2,2.1$\pm$0.2 & $-$21.6,$-$20.0 & 1.000,1.000 &  C22,D22,H22,\\
& & & & & & & & B22\\\\
\multicolumn{9}{c}{\citet{Adams2022}}\\
6878$^e$ & 07:23:26.238 & $$-$$73:26:56.97 & 8.5/8.9$_{-0.4}^{+0.5}$ & $-$20.4 & $>$2.9,$>$2.8 & $-$26.1,$-$21.3 & 1.000,1.000 &  At22,D22,M22\\
3602 & 07:23:26.705 & $$-$$73:26:10.56 & 9.0/9.5$_{-1.2}^{+1.2}$ & $-$19.5 & $>$1.4,$>$2.0 & $-$18.9,$-$20.4 & 1.000,1.000 &  D22\\
2779 & 07:22:35.053 & $$-$$73:28:32.99 & 9.5/9.2$_{-1.3}^{+1.1}$ & $-$19.5 & $>$2.5,$>$2.6 & $-$6.6,$-$13.2 & 0.991,1.000 &  D22,M22\\\\
\multicolumn{9}{c}{\citet{Atek2022}}\\
SMACS\_z10b & 07:23:22.697 & $$-$$73:26:06.23 & 8.9/8.1$_{-0.2}^{+0.3}$ & $-$20.8 & $>$3.6,$>$3.9 & $-$48.9,$-$49.8 & 1.000,1.000 & D22, M22 \\
SMACS\_z10c & 07:23:20.158 & $$-$$73:26:04.28 & 9.8/7.8$_{-0.2}^{+0.3}$ & $-$20.2 & $>$3.2,3.4$\pm$2.5 & $-$41.1,$-$40.3 & 1.000,1.000 & D22 \\\\
\multicolumn{9}{c}{\citet[][, v2]{Donnan2022}}\\
43031 & 07:23:27.846 & $$-$$73:26:19.78 & 8.6/7.9$_{-0.5}^{+0.5}$ & $-$18.4 & $>$2.5,$>$3.0 & $-$20.7,$-$22.2 & 1.000,1.000 &  \\
22480 & 07:22:45.808 & $$-$$73:27:46.57 & 9.7/8.7$_{-1.3}^{+1.3}$ & $-$18.5 & $>$1.4,$>$1.8 & $-$10.0,$-$11.2 & 0.997,0.998 &  \\
30585 & 14:19:35.334 & 52:50:37.90 & 10.6/10.2$_{-0.7}^{+0.6}$ & $-$19.4 & $>$1.7,2.0$\pm$2.5 & $-$14.2,$-$9.5 & 0.999,0.990 &  W22,F22,B22,B22M\\
32395\_2 & 14:19:46.353 & 52:56:32.81 & 12.3/11.6$_{-0.5}^{+0.4}$ & $-$19.9 & 1.5$\pm$1.0,1.5$\pm$1.2 & $-$9.9,$-$9.1 & 0.995,0.993 &  H22,B22,B22M,F22\\\\
\multicolumn{9}{c}{\citet[][, v1]{Donnan2022}$\dagger$}\\
38681 & 07:23:28.099 & $$-$$73:26:20.10 & 8.6/8.0$_{-0.7}^{+0.6}$ & $-$19.4 & $>$1.8,$>$1.7 & $-$7.8,$-$9.7 & 0.993,0.996\\\\
\multicolumn{9}{c}{\citet{Labbe2022}}\\
35300 & 14:19:19.358 & 52:53:16.01 & 9.3/9.4$_{-1.0}^{+1.0}$ & $-$19.0 & $>$1.4,1.7$\pm$1.5 & $-$7.8,$-$17.2 & 0.993,1.000 &  F22,B22M\\
14924 & 14:19:30.272 & 52:52:51.01 & 9.9/8.8$_{-0.2}^{+0.2}$ & $-$20.1 & 1.1$\pm$0.5,0.9$\pm$0.5 & $-$25.1,$-$22.4 & 1.000,1.000 &  F22\\
16624 & 14:19:22.741 & 52:53:31.60 & 10.0/8.6$_{-0.2}^{+0.3}$ & $-$20.9 & 1.0$\pm$0.2,1.2$\pm$0.2 & $-$17.7,$-$18.4 & 1.000,1.000 &  F22\\
21834 & 14:19:36.533 & 52:56:21.76 & 10.8/9.3$_{-0.9}^{+1.1}$ & $-$19.0 & 1.0$\pm$1.5,0.7$\pm$1.3 & $-$11.4,$-$10.7 & 0.998,0.998 &  F22\\\\
\multicolumn{9}{c}{\citet[][, v1]{Harikane2022_z9to17}$\dagger$}\\
GL$-$z9$-$6 & 00:13:57.110 & $$-$$30:19:31.53 & 8.7/8.1$_{-0.6}^{+0.5}$ & $-$18.8 & $>$1.7,$>$1.9 & $-$9.2,$-$12.8 & 0.993,0.999 &  Le22\\
GL$-$z9$-$12 & 00:14:01.896 & $$-$$30:18:56.89 & 10.2/10.1$_{-0.5}^{+0.6}$ & $-$18.7 & $>$1.7,$>$1.9 & $-$14.6,$-$23.1 & 1.000,1.000\\\\
\multicolumn{9}{c}{\citet{Whitler2022_z10}}\\
EGS$-$39117 & 14:20:02.808 & 52:59:17.91 & 9.0/8.8$_{-0.2}^{+0.1}$ & $-$20.5 & 1.2$\pm$0.1,1.5$\pm$0.2 & $-$23.1,$-$20.9 & 1.000,1.000 &  F22\\\\
\multicolumn{9}{c}{This Work (Primary Selection)}\\
GLASSP2H$-$3576218534 & 00:13:57.629 & $$-$$30:18:53.43 & 13.7/13.7$_{-1.0}^{+1.0}$ & $-$19.6 & $>$1.4,$>$1.8 & $-$10.2,$-$11.3 & 0.996,0.999 &  B22M\\\\
\multicolumn{9}{c}{This Work (Secondary Selection)}\\
GLASSP2Z$-$3553419246 & 00:13:55.345 & $$-$$30:19:24.63 & 8.6/8.3$_{-0.3}^{+0.3}$ & $-$19.8 & 2.1$\pm$1.1,$>$3.0 & $-$15.9,$-$22.1 & 1.000,1.000\\
\hline \hline
\end{tabular}\\
\begin{flushleft}
  $^1$ Ad22 = \citet{Adams2022}, At22 = \citet{Atek2022}, C22 = \citet{Castellano2022_GLASS}, D22 = \citet{Donnan2022}, F22 = \citet{Finkelstein2022_z12,Finkelstein2022_JWSTLF}, H22 = \citet{Harikane2022_z9to17}, La22 = \citet{Labbe2022}, M22 = \citet{Morishita2022}, N22 = \citet{Naidu2022_z12,Naidu2022_z17}, W22 = \citet{Whitler2022_z10}, Y22 = \citet{Yan2022}, B22 = This Work (fiducial selection), B22M = This Work (alternate selection), Z22 = \citet{Zavala2022}\\
$^a$ Presented are the photometric redshift as reported in the earlier work and as estimated here based on the \textsc{grizli} reductions\\
$^b$ The different measurements provided are based on the two different reductions of the NIRCam data utilized in this analysis.\\
$^c$ Amplitude of Lyman break in mag.  Lower limits are $1\sigma$.\\
$^d$ $\chi^2_{\textrm{best},z>5.5} - \chi^2_{\textrm{best},z<5.5}$\\
$^e$ Spectroscopically confirmed to have a redshift $z=8.498$ \citep{Carnall2023}\\
$\dagger$ While these candidates were identified in earlier versions of these manuscripts, they did not make it into the final versions of these manuscripts.  Nonetheless, our analysis suggests they are "robust" $z>5.5$ sources.\\
$^*$ For simplicity, no account is made for lensing magnification for sources over the SMACS0723 and Abell 2744 parallel fields.
\end{flushleft}
\end{table*}

\begin{figure*}
\centering
\includegraphics[width=1.9\columnwidth]{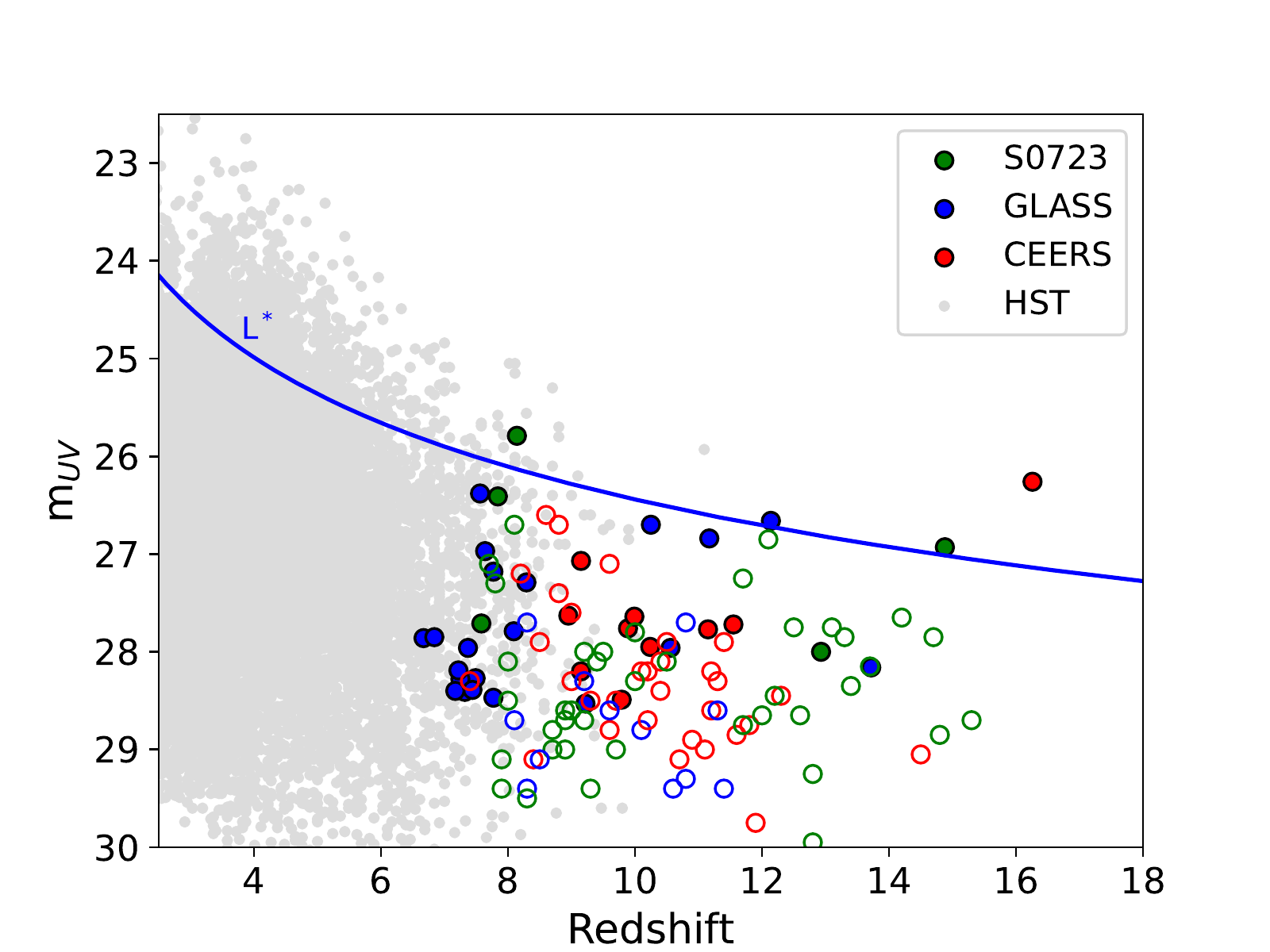}
\caption{Apparent magnitude and estimated redshifts of $z\geq7$
  candidate galaxies identified over the {\it JWST}/NIRCam SMACS0723
  (\textit{solid green circles}), GLASS parallel (\textit{solid blue circles}),
  and CEERS fields (\textit{solid red circles}) from our searches of the \textsc{grizli} data sets.  The open symbols indicate
  other $z\geq8$ galaxy candidates that have been identified in the
  literature and qualify as ``robust'' or ``solid'' $z\geq8$
  candidates according to our own SED fits and photometry.  For
  context, we show the magnitudes and redshifts of sources that have
  derived from a comprehensive set of blank and lensing fields
  observed with HST \citep{Bouwens2015_LF,Bouwens2021_LF,Bouwens2022_LF}.  The blue
  solid line shows the apparent magnitude of sources with an absolute
  magnitude of $-21$, which is an approximate characteristic
  luminosity of galaxies at $z\geq 3$ \citep{Steidel1999,Reddy2009,Bouwens2015_LF,Bouwens2021_LF,Finkelstein2015_LF,Bowler2015}.}
    \label{fig:lum_vs_z}
\end{figure*}

\begin{figure}
\centering
\includegraphics[width=\columnwidth]{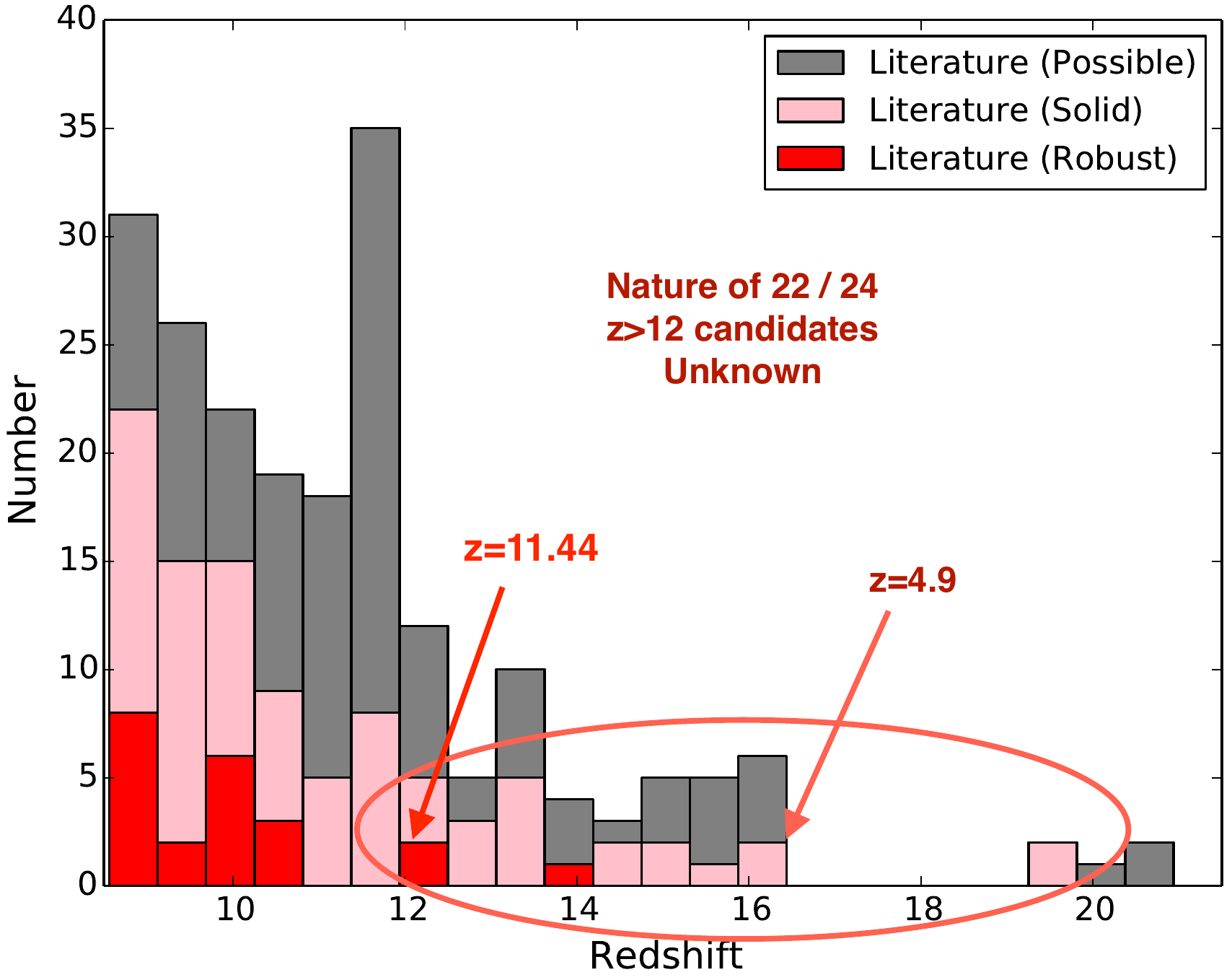}
e\caption{Number of sources in our literature samples of "robust," "solid," and "possible" sources vs. redshift (\textit{red}, \textit{pink}, and \textit{gray} shaded histograms, respectively).  The redshift of any given source is taken to be equal to the redshift of the study where it is presented in the literature for our "solid" and "possible" samples and the geometric mean redshift for sources in our "robust" sample.  The numbers of sources per bin are shown in a cumulative or stacked sense, such that top of the \textit{gray} histograms indicate the total number of sources in the "robust," "solid," and "possible" literature samples.  There are $\sim$8$\times$ more "solid" and $\sim$24$\times$ more "possible" sources reported in the literature at $z>11.5$ than there are "robust" sources, illustrating the potentially large uncertainties in the overall number of bona-fide star-forming galaxies at very high redshifts.  Givven the uncertainties, it is clearly imperative to definitively quantify the redshifts of many of these candidates to determine how rapidly galaxies assemble in the early universe.\label{fig:zdist}}
\end{figure}

\begin{table*}
\centering
\caption{Assessment of the purity and completeness of various $z\sim10$, $z\sim13$, and $z\sim17$ selections on the basis of independent selections made over the same fields and leveraging our own independent photometry.$^{\dagger,\ddagger}$}
\label{tab:assess}
\begin{tabular}{c|c|c|c|c|c} \hline
  & \# of & & & & Fraction of  \\
  Sample & Sources$^a$ & Purity (Robust)$^b$ &  Purity (Robust+Solid)$^c$ & Completeness$^d$ & Total Candidates$^e$\\ \hline\hline
This Work (Fiducial) &         16 &      0.250 &      0.875 &        0.197 &                0.103 \\
This Work (Alternate) &         21 &      0.238 &      0.762 &        0.225 &                0.135 \\
\citet{Adams2022} &          4 &      0.500 &      1.000 &        0.118 &                0.041 \\
\citet[][, v2]{Atek2022} &         10 &      0.300 &      0.500$^f$ &        0.147 &                0.103 \\
\citet[][, v1]{Atek2022} &         15 &      0.200 &      0.333$^f$ &        0.147 &                0.155 \\
\citet[][, v2]{Castellano2022_GLASS} &          6 &      0.333 &      0.667 &        0.333 &                0.286 \\
\citet[][, v1]{Castellano2022_GLASS} &          7 &      0.286 &      0.571 &        0.333 &                0.333 \\
\citet[][, v2]{Donnan2022} &        31 &      0.258 &      0.903 &        0.350 &                0.172\\
\citet[][, v1]{Donnan2022} &         34 &      0.176 &      0.588 &        0.250 &                0.189 \\
\citet{Finkelstein2022_JWSTLF} &         23 &      0.261 &      0.869 &        0.588 &                0.371 \\
\citet[][, v2]{Harikane2022_z9to17} &          13 &      0.231 &      0.461 &        0.187 &                0.151 \\
\citet[][, v1]{Harikane2022_z9to17} &          13 &      0.385 &      0.538 &        0.219 &                0.151 \\
\citet{Labbe2022} &          5 &      0.800 &      1.000 &        0.147$^g$ &                0.081$^g$ \\
\citet[][, v1]{Morishita2022} &   1 & 1.000 & 1.000$^f$ & 0.029 & 0.010 \\
\citet[][, v2]{Morishita2022} &   8 & 0.375 & 0.750$^f$ &  0.176 & 0.082 \\
\citet{Naidu2022_z12}  &          2 &      1.000 &      1.000 &        0.333$^{h}$ &                0.333$^{h}$ \\
\citet[][, v2]{Whitler2022_z10} &          6 &      0.167 &      0.667 &        0.118 &                0.097 \\
\citet[][, v1]{Whitler2022_z10} &          8 &      0.125 &      0.750 &        0.176 &                0.129 \\
\citet{Yan2022} &         64 &      0.000 &      0.234$^{f}$ &        0.441 &                0.660 \\
\hline
\end{tabular}
\begin{flushleft}
$\dagger$ We emphasize that the results we present are completely reliant on the photometry we derive for the candidates from our two reductions and the SED template sets we have available for our redshift likelihood calculations.  As such, these results are merely indicative, and clearly the ultimate arbiter of the purity and completeness of individual selections will be deep spectroscopy with {\it JWST}.\\
 $\ddagger$ In this table, we only report analyses which consider the selection of sources at  $z\sim10$ and higher.  For this reason, we do not report on analyses like \citet{Leethochawalit2022} or \citet{Endsley2022_JWST} where source selection only extends to redshifts of $z\approx 9$.\\
  $^a$ Number of $z\geq8.5$ candidates identified in the magnitude range well probed in most studies in the literature (i.e., $<$29 mag)\\
  $^b$ Fraction of $z\geq8.5$ candidates from this study that satisfy our criteria for being ``robust'' $z\geq 8.5$ candidates.\\
  $^c$ Fraction of $z\geq8.5$ candidates from this study that satisfy our criteria for being ``robust'' or ``solid'' $z\geq 8.5$ candidates.\\
  $^d$ Fraction of the total set of ``robust'' and ``solid'' $z\geq8.5$ candidates (Tables~\ref{tab:robust}, \ref{tab:solid}, and \ref{tab:solid2}) identified in a given study.  Only the search fields utilized in a study are considered for these completeness estimates.\\
  $^e$ Fraction of the total number of $z\geq8.5$ candidates identified in a given study.  Only the search fields utilized in a study are considered for these completeness estimates.\\
    $^f$ The purity of selections focusing on the SMACS0723 cluster and parallel field are likely lower than the other selections due to the lack of especially sensitive F115W data over the fields.\\
  $^g$ We would not expect the \citet{Labbe2022} selection to be an especially complete representation of star-forming galaxies at $z>6$, given their choice to select only those galaxies with prominent Balmer breaks.\\
  $^h$ Since \citet{Naidu2022_z12} expressly only search for $z>10$ sources which are particularly bright and which show high S/N ($>$10) detections in both F356W and F444W sources, we somewhat arbitrarily evaluate the completeness of their selection to 27 mag.\\

\end{flushleft}
\end{table*} 

\subsection{$z\geq 8$ Selections from the Literature\label{sec:othersel}}

Given the many challenges that exist in making use of the first {\it
  JWST} observations in identifying $z\geq 8$ galaxies, the uncertain
NIRCam zeropoints being perhaps the largest \citep[e.g.][]{Adams2022},
it is useful for us to provide an alternate assessment of the many
$z\geq 8$ candidates which have been identified in the early {\it
  JWST} observations.  This can help provide us with insight into both
the reliability and completeness of earlier {\it JWST}/NIRCam
selections.

There have been selections of $z\geq 8$ galaxies
conducted by at least ten different teams, including
\citet{Naidu2022_z12}, \citet{Castellano2022_GLASS},
\citet{Adams2022}, \citet{Atek2022}, \citet{Yan2022},
\citet{Donnan2022}, \citet{Whitler2022_z10}, \citet{Labbe2022},
\citet{Harikane2022_z9to17}, and \citet{Finkelstein2022_JWSTLF}.  In
general, these studies have focused on identifying sources from one or
more of the same three data sets considered here, allowing for an
extensive set of comparisons between the different selections and
independent assessments of various $z\geq 8$ candidates.

\subsubsection{Evaluation and Segregation into Different Subsamples}

In this subsection, we focus on the $z\geq8$ candidates identified
over the three most studied {\it JWST} NIRCam fields (SMACS0723,
CEERS, and Abell 2744 parallel), and provide an
independent evaluation of their robustness.  To provide this
evaluation, we have performed 0.35$''$-diameter aperture photometry on
all the identified candidates from \citet{Naidu2022_z12},
\citet{Castellano2022_GLASS}, \citet{Adams2022}, \citet{Atek2022},
\citet{Yan2022}, \citet{Donnan2022}, \citet{Whitler2022_z10},
\citet{Finkelstein2022_JWSTLF}, \citet{Labbe2022}, and
\citet{Harikane2022_z9to17}.  For papers where changes have occurred to 
the identified high-redshift sources, we consider the catalogs in each 
version of various papers.  Not only is this useful for gaining 
perspective on progress that has been made, but we have found that some 
sources in earlier versions of papers also appear to be credible $z>8$ 
candidates, and therefore we have included these sources in our analysis to 
be as comprehensive as possible.  We perform photometry on both our 
fiducial and alternate reductions using \textsc{grizli} and 
\textsc{pencil}, respectively.

We then look for the presence of a significant spectral break in
sources and compute redshift likelihood distributions for candidates
based on our derived photometry of the recent \textsc{grizli} and \textsc{pencil} reductions.  We then segregate sources into three
different samples:
\begin{itemize}
\item (1) one where the cumulative probability of candidates lying at
  $z>5.5$, i.e., $P(z>5.5)$, exceeds 99\% using the photometry we have
  performed on both NIRCam reductions utilized here,
\item (2) one where $P(z>5.5)$ is in excess of 80\% and 50\% for our fiducial and secondary reductions, in excess of 70\% for both our fiducial and secondary reductions, or in excess of 50\% and 80\% for our fiducial and secondary reductions, but does not the former selection criteria, and
\item (3) one which does not satisfy either of the former selection criteria.
\end{itemize}
\noindent We refer to (1) the first selection of sources from the literature as the ``robust'' sample, (2) the second selection as the ``solid'' sample, and (3) the third selection of sources as the
``possible'' sample.

\subsubsection{Results}

We present these different samples in Table~\ref{tab:robust} and
Tables~\ref{tab:solid}-\ref{tab:possible3} of Appendix B.  
Interestingly enough, only 18 $z\sim10$ and 3 $z\sim13$ candidates
from these fields satisfy our criteria for being robust.  75 of the
reported candidates in the literature qualify as ``solid'' candidates,
while 108 of these candidates qualify as ``possible.''

Of the candidates we classify as ``robust,'' only 10 have an estimated
redshift $z\geq 9$ using our fiducial photometry.  The most consistent
characteristics of sources in our robust lists is that they show
either very pronounced ($\geq$1.5-mag) spectral breaks in the observed
photometry or show two spectral breaks (Lyman + Balmer, as was a key
aspect of the \citealt{Labbe2022} selection).  Additionally, to the
extent that the present compilation of ``robust'' $z\geq9$ candidates
overlap with the redshift ranges and fields examined by
\citet{Harikane2022_z9to17}, three of the four sources from our
compilation, i.e., GL$-$z11, GL$-$z13, and 32395\_2
\citep{Donnan2022,Finkelstein2022_z12}, receive robust designations in
\citet{Harikane2022_z9to17}, with $\Delta\chi(z_{low})^2 -
\chi(z_{high})^2$ equal to 71.9, 72.3, and 14.5,
respectively, each of which is well above their $\Delta\chi^2 >9$
selection criterion for secure sources.  This is reassuring and gives
us confidence that at least for this subset of $z\geq 9$ candidates,
the inferred redshifts might be reasonably secure.

From these numbers, it is clear that the majority of $z\geq 9$
candidates identified to date only qualify as ``possible'' $z\geq9$
candidates and do not meet the higher quality standards required to be
classified as ``robust'' or ``solid.''  Interestingly enough,
essentially all $z\geq 8$ studies presenting significant samples of
$z\geq 8$ candidates from the first {\it JWST} fields, e.g.,
\citet{Atek2022}, \citet{Donnan2022}, \citet{Harikane2022_z9to17},
\citet{Yan2022}, all contain sources that lie in the "possible" category 
given our photometry.  Interestingly enough, of the candidates we grade as "robust," "solid," and "possible" and where we compute photometric redshifts $z\geq9$, 90\%, 28\%, and 12\%, respectively, are also independently reported as a $z\geq 8$ candidate galaxy in a separate manuscript from the literature.

Figure~\ref{fig:lum_vs_z} shows the distribution of the sources we find from our \textsc{grizli} reductions over the three NIRCam fields (SMACS0723, GLASS parallel, and CEERS) in redshift and $UV$ luminosity vs. the comprehensive earlier selection
of $z\sim2$-11 sources from Hubble constructed by
\citet{Bouwens2015_LF,Bouwens2021_LF,Bouwens2022_LF}.  We also 
show the ``robust'' or ``solid'' $z\geq 8$ galaxy candidates
from the earlier studies with {\it JWST}.

In Figure~\ref{fig:zdist}, we show the number of sources that are contained in our literature subsamples of "robust," "solid," and "possible" candidates as a function of redshift.  The number of "solid" and "possible" candidates at $z>11.5$ are $\sim$8$\times$ and $\sim$24$\times$ larger, respectively, than those which we grade as "robust."  Clearly,
it is essential that higher quality JWST data become available for sources in
these samples to determine the fraction that are actually at $z>11.5$.

\subsubsection{Characterization of Literature Subsamples}

To help interpret the quality of the $z\geq 8$ candidates that we
segregated into different categories, we derive median fluxes for
candidates in each category.  Prior to the median stacking, a
renormalization of the fluxes in individual sources is performed such
that the F200W, F277W, and F356W band fluxes for the $z\sim10$,
$z\sim13$, and $z\sim17$ samples, respectively, are 36 nJy.  The
results are presented in Table~\ref{tab:medianstacks} of Appendix C.

The most significant difference between the different stacks is the
flux blueward of the break.  For both the ``robust'' and ``solid''
stacks, no significant flux is present blueward of the putative Lyman
breaks and large Lyman breaks are seen, i.e., $\geq$1.8 mag.  However,
for the ``possible'' stack, not only is the flux in the median stacks
nominally significant 1-2$\sigma$ in individual bands, but the
putative breaks are smaller, i.e., $\sim$1.0-1.5 mag.  Because of such characteristics, the reliability of sources in the "possible" samples is lower, as indicated also by the much greater likelihood these sources show for being at lower redshifts from the individual SED fit results.

To provide some measure of the quality and completeness of earlier
selections of $z\gtrsim 8.5$ galaxies derived from the first {\it JWST}
fields, we have calculated the total number of ``robust''+''solid''
$z\geq 8.5$ candidates that have been identified to $\sim$29 mag over
various {\it JWST} fields by our selection or those in the literature
and quantified the fraction of these candidates that have been
identified in various studies.  We present this fraction as the
completeness of each selection in Table~\ref{tab:assess}.  In
estimating the completeness, we only consider the fields included in
a given selection.  As one example, since \citet{Harikane2022_z9to17}
only search for $z\sim8$-11 F090W-dropout galaxies over the GLASS parallel field as part of their $z\sim9$ selection, we do not treat compelling $z\sim8$-11 galaxies
found over other fields like SMACS0723 or CEERS as contributing to our assessment of
completeness in their study.

\subsubsection{Evaluation of Earlier $z\geq 8$ Selections}

In Table~\ref{tab:assess}, we also present an approximate ``purity''
for each selection by dividing the number of ``robust'' candidates in
each selection by the total number of reported candidates in a study
as well as the number of ``robust'' and ``solid'' candidates in a
given study.  In grading individual candidates 
from various studies, we only include sources which $UV$ magnitudes brightward of 29 mag to limit our analysis to those sources with the highest S/N and to increase the probability that sources will be 
selected as part of multiple studies.

We caution that the results we obtain here are completely reliant on the photometry we derive for the candidates from our two reductions and the SED template sets we utilize in our analysis.  As such, these results (and the remarks in the paragraphs which follow) should be taken as merely indicative, and \textit{clearly the ultimate arbiter} of the purity and completeness of individual selections will be deep spectroscopy with {\it JWST} \citep[e.g.][]{RobertsBorsani2022_SpecJWST,CurtisLake2022_JADES,Tang2023,Bunker2023}.  For the purposes of this calculation, we treat sources with a ``possible'' designation as corresponding to lower-redshift interlopers, but clearly there is some uncertainty in this
designation and many candidates we grade in this category might well prove to be at $z\geq 8$. 

\begin{table*}
\centering
\caption{Fraction of sources in common between various $z\geq8$ selections relative to the total possible}
\label{tab:assess2}
\begin{tabular}{c|c|c|c|c|c|c|c|c|c|c|c|c|c} \hline
     & H22   & D22   & B22   & B22M  & F22   & C22   & At22  & La22  & W22   & Y22 & M22 & Ad22\\\hline\hline
 H22 &  13/13 & ----- & ----- & ----- & ----- & ----- & ----- & ----- & ----- & ----- & ----- & -----\\
 D22 &  5/16   & 31/31 & ----- & ----- & ----- & ----- & ----- & ----- & ----- & ----- & ----- & -----\\
     &  (3/16)$^{b,c}$ \\
 B22 &  6/15 &  7/43 & 21/21 & ----- & ----- & ----- & ----- & ----- & ----- & ----- & ----- & -----\\
     &  (5/16)  & (5/50) \\
 B22M &  4/14$^b$ &  5/40 & 10/27 & 16/16 & ----- & ----- & ----- & ----- & ----- & ----- & ----- & -----\\
     &       & (4/46) \\
 F22 &  2/ 6 &  9/30 &  7/27 &  6/28 & 23/23 & ----- & ----- & ----- & ----- & ----- & ----- & -----\\
       & (2/5)  & (8/35) & (7/28) & (6/29) \\
 C22 &  3/11$^b$ &  2/ 7$^b$ &  3/10$^b$ &  1/12$^b$ & -----$^a$ &  6/ 6 & ----- & ----- & ----- & ----- & ----- & -----\\
     &       & (2/6)$^b$ & (2/11)$^b$ \\
 At22 &  0/ 4 &  3/18 &  2/10 &  0/10 & -----$^a$ & -----$^a$ & 10/10 & ----- & ----- & ----- & -----\\
     &      & (1/29) & (2/15) & (0/15) & ----- \\
 La22 & -----$^a$ &  0/22 &  1/16 &  1/16 &  4/23 & -----$^a$ & -----$^a$ &  5/ 5 & ----- & ----- & ----- & -----\\
     & & (0/25) & & & (4/24) \\
 W22 & -----$^a$ &  1/22 &  3/15 &  2/16 &  3/25 & -----$^a$ & -----$^a$ &  0/11 &  6/ 6 & ----- & ----- & ----- \\
     & & (2/26) & (2/18) & (2/18) & (3/28)  \\
 Y22 &  0/46 &  0/75 &  1/65 &  0/64 & -----$^a$ & -----$^a$ &  1/73 & -----$^a$ & -----$^a$ & 64/64 & ----- & ----- \\
     &      & (0/80) & (1/66) & (0/65) & & & (2/78) \\     
 M22 & 0/3 &  2/17 &  0/ 10 & 0/8 & -----$^a$ & -----$^a$ &  1/17 & -----$^a$ & -----$^a$ &  1/71 & 8/8 & ----- \\
     &     & (1/15) & (0/3) & (0/1) & & & (0/16) & & & (0/66) \\
 Ad22 & -----$^a$ &  2/13 &  0/ 6 & 0/4 & -----$^a$ & -----$^a$ &  0/14 & -----$^a$ & -----$^a$ &  0/68 & 2/10 & 4/4 \\
      &       & (2/17) &      &     &       &       &  (0/19) &     &       &       & (1/4) & 

 \\\hline\hline
\end{tabular}\\
\begin{flushleft}
  $^a$There is no overlap between specific fields and redshift ranges utilized in the two selections being compared.\\
  $^b$Remarkably, approximately half of the overlap between these studies are the two bright sources from \citet{Naidu2022_z12} and \citet{Castellano2022_GLASS}.  If we exclude those two sources from consideration, overlap between the selections is only $\sim$10\%.\\
  $^c$Fraction in parentheses indicates the overlap in the initial versions of the catalogs from these papers.  In the majority of cases, the fraction in the updated versions is higher.
  \end{flushleft}
\end{table*}

There are a few noteworthy results to notice in the results presented
in this table.  First of all, there has been a clear improvement in
both the purity and completeness of most $z\geq8$ samples since NIRCam
data from {\it JWST} became public, as one might expect to
improvements in the NIRCam zeropoint calibrations.  As one example,
the purity of the \citet{Donnan2022} selections -- in terms of sources
graded either "robust" or "solid" -- have improved from 59\% (v1) to
90\% (v2).  Other newer analyses which are able to take advantage of
the improved zeropoint calibrations are the fiducial and secondary
selections from the present analysis as well as those from
\citet{Finkelstein2022_JWSTLF}; these selections feature a purity of
88\%, 76\%, and 87\%, respectively.  Achieving a high purity appears
to have been more difficult for analyses that focus on the SMACS0723
data set (e.g., \citealt{Atek2022,Yan2022} but see however
\citealt{Adams2022}), likely due to the significantly shallower F115W
observations available in the first {\it JWST} data over that field.

Second, selections that focus on the most luminous galaxies at $z\geq8$, i.e., \citet{Naidu2022_z12, Naidu2022_z17,Adams2022}, or selections which focus on sources with multiple spectral breaks \citep[e.g.][]{Labbe2022} show a much higher reliability than those that focus on a broader selection of sources.  Based on the present analyses, we find 100\% purity for all three of these selections in our analysis.  This contrasts with more ambitious selections aiming to select the bulk of the star-forming galaxies at $z\geq8$, e.g., \citet{Donnan2022}, \citet{Harikane2022_z9to17}, \citet{Finkelstein2022_JWSTLF}, and the present selections where $\sim$25\% of the sources in these selections are graded as "robust," 50-60\% of the sources are graded as "solid," and the final $\sim$15\% of the sources in such selections are graded as "possible."

A third striking result is the large differences in the completeness
of selections.  The majority of the analyses only include a fraction
($\lesssim$35\%) of the candidates we grade as "solid" or "robust" in
our analysis.  In many analyses, this appears to have been the result
of a clear choice to include only those sources which appear to be the
most reliable, either because higher amplitude Lyman breaks are
required ({\it this work}) or because the SED fits to $z>8$ solutions
are required to give a much lower values of $\chi^2$ ($\Delta\chi^2 >
9$) than lower redshift fits \citep{Harikane2022_z9to17}.
Nevertheless, the \citet{Finkelstein2022_JWSTLF} selection appears to
perform the best as far as completeness is concerned, showing a
$\sim$2$\times$ higher completeness in their identification of
"solid"+"robust" $z\geq8$ sources than most of the other analyses and
also successfully selecting the $z\geq8$ sources found by
\citet{Labbe2022} with prominent Balmer breaks
(Table~\ref{tab:robust}).  The latter sources mostly miss our own
selections due to their Lyman breaks having a smaller amplitude than
$\gtrsim$1.5 mag required to be included in our own samples.

One consequence of the relatively low estimated completeness for most selections is only a modest ($\sim$20-35\%) overlap between $z\geq 8$ selections.  Table~\ref{tab:assess2} quantifies the number of sources  that are in common for differing selections over the same fields out 
of some total possible.  Nevertheless, it is worthwhile noting that
there has been an improvement in the overlap between samples.
Initially, most of the overlap between studies was confined to a few bright $z\sim10$-12 sources such as have been found by \citet{Naidu2022_z12} and
\citet{Castellano2022_GLASS} and perhaps 10-15\% of the rest, but now
the overlap is approximately $\sim$30\% between selections,
approaching the $\sim$50\% overlap seen in $z\sim4$-8 selections
obtained by HST over the Hubble Ultra Deep Field \citep{Beckwith2006}, e.g., see \S3.4 of
\citet{Bouwens2015_LF} where overlap with other $z\sim7$-8 selections
\citep[e.g.][]{McLure2013,Schenker2013} is discussed.

\section{Luminosity Function Results}

In this section, we make use of the rather small samples of high-likelihood $z\geq8$ galaxy
candidates over the three most well studied {\it JWST} fields to derive $UV$
LF results.  We begin with direct determinations of the LF results
using our own selections and then move onto determinations based on
collective samples of $z\geq 8$ galaxies identified in the present and
previous studies.  We conclude this section with a comparison of these
results with several previous determinations.

\subsection{Results Using Our Own Samples\label{sec:ourlfs}}

We begin by describing LF results derived using our own $z\geq8$
samples constructed from our fiducial reductions of the available {\it JWST} data.

Given the small number of sources in each of our samples, we derive LF
results using the $1/V_{\textrm{max}}$ technique and assuming Poissonian
statistics.  As in our own earlier analyses, we derive LF results by
maximizing the likelihood ${\cal L}$ of producing the observed
distribution of apparent magnitudes given some model LF:
\begin{equation}
{\cal L}=\Pi_{\textrm{field}} \Pi_i p(m_i)
\label{eq:ml}
\end{equation}
where we take the likelihood of LF results derived over the set of
fields we select sources and over a set of apparent magnitude intervals $m_i$.  

Given the lack of F090W observations over the CEERS fields and the limited depth of F115W observations over SMACS0723, we only consider sources in the GLASS parallel field for our $z\sim8$ LF determination and sources over the GLASS parallel and CEERS fields for our $z\sim10$ LF determinations.  For our $z\sim13$ and $z\sim17$ determinations, we consider sources over the GLASS parallel, SMACS0723, and CEERS fields.  For simplicity and because none of our $z\geq12$ candidates lie within $<$60$"$ to the high magnification areas of the Abell 2744 and SMACS0723 clusters, we ignore the impact of lensing magnification on our LF results.

Since we are assuming Poissonian statistics, the probability of finding $n_{\textrm{observed},i}$ sources 
\begin{equation}
p(m_i) = e^{-n_{\textrm{expected},i}} \frac{(n_{\textrm{expected},i})^{n_{\textrm{observed},i}}}{(n_{\textrm{observed},i})!}
\label{eq:mi}
\end{equation}
where $n_{\textrm{observed},i}$ is the number of observed sources in magnitude
interval $i$ while $n_{\textrm{expected},i}$ is the expected number given some
model LF.  We compute the number of expected sources $n_{\textrm{expected},i}$
based on some model LF $\phi_j$ using the equation
\begin{equation}
n_{\textrm{expected},i} = \Sigma _{j} \phi_j V_{i,j}
\label{eq:numcountg}
\end{equation}
where $V_{i,j}$ is the effective volume over which a source in the
magnitude interval $j$ might be both selected and have a measured
magnitude in the interval $i$.

\begin{figure*}
\centering
\includegraphics[width=2\columnwidth]{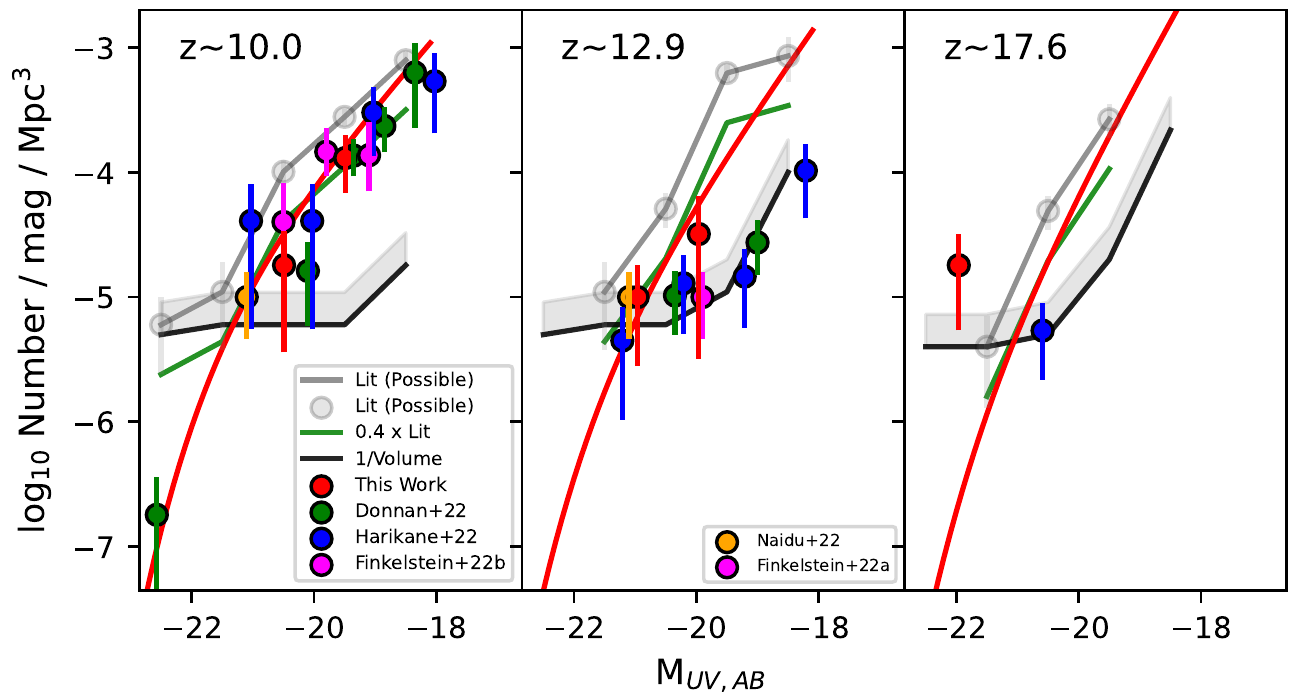}
\caption{Determinations of the $UV$ LF including all of the $z\geq 8$ candidates from the literature (shaded
  grey points) compared against other determinations of the $z\sim 10$
  (\textit{left}), $z\sim13$ (\textit{center}), and $z\sim17$
  (\textit{right}) $UV$ LF using new {\it JWST}/NIRCam data.  
  Schechter and stepwise $UV$ LFs derived using our own fiducial 
  samples are shown with the red lines and circles, respectively,
  while those from
  \citet{Naidu2022_z12}, \citet{Donnan2022},
  \citet{Finkelstein2022_z12, Finkelstein2022_JWSTLF}, and
  \citet{Harikane2022_z9to17} are shown with the gold, green, magenta,
  and blue points, respectively.  The black lines give the LF results one would obtain from the early NIRCam fields if there were just a single source per 1-mag bin, indicating approximately how low in volume density the early fields probe (the equivalent $1\sigma$ upper limits are indicated by the upper edge of the gray shaded region obtained by multiplying these lines by 1.841: \citealt{Gehrels1986}).
  Gray lines are drawn connecting the gray points (to help delineate the LF derived from the full literature sample of sources).  The reason the gray line has a much higher normalization than any LF in the literature is due to there being a much larger number of $z\geq 9$ candidates reported thus far over the SMACS0723, GLASS parallel, and CEERS fields than are present in any one individual analysis.  A better match with the early
  $UV$ LF results from {\it JWST} can be obtained by multiplying the
  shaded gray region by 0.4 (\textit{shown with the green line}).  
  These results suggest that either early {\it JWST} LF
  results in individual analyses are too low (due to incompleteness) 
  or that early selections suffer from substantial ($\geq$50\%) contamination from lower-redshift galaxies.}
    \label{fig:uvlfs}
\end{figure*}

\begin{figure*}
\centering
\includegraphics[width=2\columnwidth]{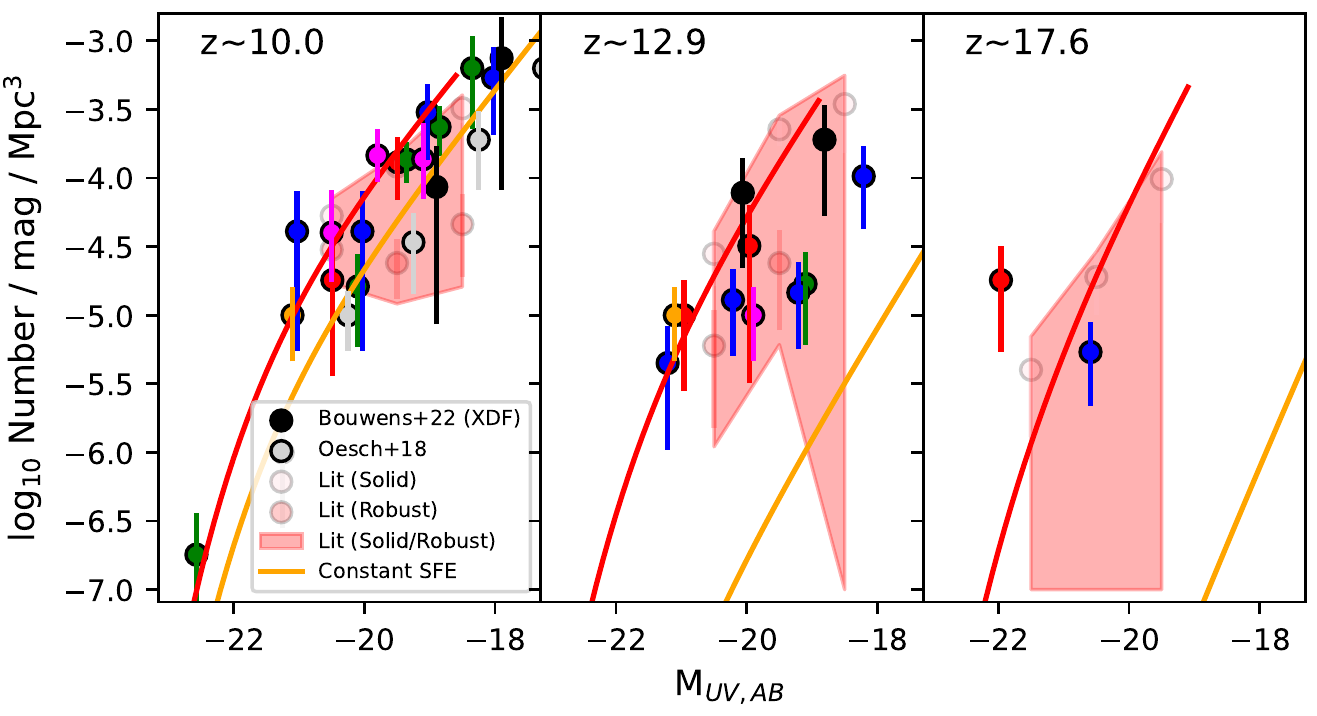}
\caption{Illustration of how much higher new determinations of the
  $UV$ LFs are with {\it JWST} than the expected results assuming
  constant star formation rate efficiency (\textit{yellow line}:
  \citealt{Mason2015}).  The presented LF constraints are similar to 
  those shown on Figure~\ref{fig:uvlfs} but also include the results
  from \citet{Bouwens2022_XDF} (\textit{black solid circles}) over the HUDF/XDF 
  \citep{Beckwith2006,Illingworth2013} 
  and \citet{Oesch2018_z10LF} derived from HST observations alone (\textit{gray solid circles}).
  The partially transparent pink and red points indicate the
  LF results derived alone from the "solid" and "robust" candidates, respectively.  The shaded red region gives $UV$ LF results
  one would obtain including all the literature candidates in our
  ``robust'' list, but no more than in our ``solid'' list of
  candidates.   Given that even this
  low edge of the shaded region exceeds constant star formation rate
  efficiency model predictions at $z\sim10$ suggests that star
  formation in the early universe may be much more efficient than
  suggested in many analyses with HST and ground-based data
\citep{Bouwens2015_LF,Bouwens2021_LF,Harikane2018,Harikane2022_LF,Oesch2018_z10LF,Tacchella2018,Stefanon2021_SMF,Stefanon2022_sSFR}.
  Differences with the constant star formation efficiency models are
  even more substantial at $z\geq 13$.}
    \label{fig:lfscience}
\end{figure*}

\begin{table}
\centering
\caption{Binned LF Results for Galaxies at $z\geq8$}
\label{tab:binnedlfs}
\begin{tabular}{c|c}
\hline
$M_{UV}$ & $\phi^*$ [mag$^{-1}$ Mpc$^{-3}$]\\
\hline\hline
\multicolumn{2}{c}{$z\sim8$ galaxies}\\
$-$21.13 & 0.000046$\pm$0.000036\\
$-$20.13 & 0.000044$\pm$0.000036\\
$-$19.13 & 0.001050$\pm$0.000414\\\\
\multicolumn{2}{c}{$z\sim10$ galaxies}\\
$-$20.49 & 0.000018$\pm$0.000016\\
$-$19.49 & 0.000130$\pm$0.000068\\\\
\multicolumn{2}{c}{$z\sim13$ galaxies}\\
$-$20.96 & 0.000010$\pm$0.000008\\
$-$19.96 & 0.000032$\pm$0.000032\\\\
\multicolumn{2}{c}{$z\sim17$ galaxies}\\
$-$21.96 & 0.000018$\pm$0.000014\\
\hline\hline
\end{tabular}
\end{table}

\begin{table}
\centering
\caption{Best-fit parameters derived for Schecther and double power-law fits to the present $z\geq8$ $UV$ LF results}
\label{tab:lffits}
\begin{tabular}{c|c|c|c|c}
\hline
         & $\phi^*$ & $M^*$ & $\alpha$ & $\beta$ \\
         & [$10^{-5}$ mag$^{-1}$ & & & \\
Redshift &  Mpc$^{-3}$] & [mag] & & \\
\hline
\multicolumn{5}{c}{Schechter}\\
8 & 5.7$_{-2.7}^{+5.2}$ & $-$21.15 (fixed) & $-$2.58$\pm$0.39 & --- \\
10 & 2.6$_{-0.7}^{+1.0}$ & $-$21.15 (fixed) & $-$2.38 (fixed) & --- \\
13 & 1.3$_{-0.6}^{+0.9}$ & $-$21.15 (fixed) & $-$2.71 (fixed) & --- \\
17 & 1.0$_{-0.8}^{+2.3}$ & $-$21.15 (fixed) & $-$3.15 (fixed) & --- \\\\

\multicolumn{5}{c}{Double Power-Law}\\
8 & 102$_{-24}^{+30}$ & $-$19.67 (fixed) & $-$2.17 (fixed) & $-3.75$ (fixed)\\
10 & 28$_{-8}^{+11}$ & $-$19.67 (fixed) & $-$2.35 (fixed) & $-3.75$ (fixed)\\
13 & 22$_{-9}^{+15}$ & $-$19.67 (fixed) & $-$2.62 (fixed) & $-3.75$ (fixed)\\
17 & 26$_{-21}^{+59}$ & $-$19.67 (fixed) & $-$2.98 (fixed) & $-3.75$ (fixed)\\
\hline\hline
\end{tabular}
\end{table}

\begin{table}
\centering
\caption{Binned LF Results derived based on Sources in all Public Analyses$^a$}
\label{tab:lffitresults}
\begin{tabular}{c|c}
\hline
$M_{UV}$ & $\phi^*$ [mag$^{-1}$ Mpc$^{-3}$]\\
\hline\hline
\multicolumn{2}{c}{$z\sim10$ Robust/Solid}\\
$-$20.5 & [0.000017,0.000071] \\
$-$19.5 & [0.000012,0.000148] \\
$-$18.5 & [0.000016,0.000402] \\
\\
\multicolumn{2}{c}{$z\sim13$ Robust/Solid}\\
$-$20.5 & [0.000017,0.000071] \\
$-$19.5 & [0.000012,0.000148] \\
$-$18.5 & [0.000016,0.000402] \\
\\
\multicolumn{2}{c}{$z\sim17$ Robust/Solid}\\
$-$21.5 & [0.000000,0.000007] \\
$-$20.5 & [0.000000,0.000029] \\
$-$19.5 & [0.000000,0.000155] \\
\hline\hline
\end{tabular}
\\\begin{flushleft}
$^a$ Upper and lower bounds correspond to the inferred volume densities
of our ``solid'' and ``robust'' literature selections, respectively, derived
by dividing the number of sources in each of those selections by the available
volume for \textit{detecting} sources down to a given $UV$ luminosity.
\end{flushleft}
\end{table}

We compute the selection volume for our samples by inserting
artificial sources with various redshift and apparent magnitudes at
random positions within the NIRCam images for each of these fields and
then attempting both to detect the sources and select them using our
$z\sim8$, $z\sim10$, $z\sim13$, and $z\sim17$ selection criteria.  We
assume the $UV$-continuum slopes of sources to have a mean value of 
$-2.3$, with a $1\sigma$ scatter of 0.4.  These $UV$-continuum
slopes are in reasonable agreement with determinations available on
the basis of both HST+Spitzer data
\citep[e.g.,][]{Dunlop2013,Wilkins2016,Stefanon2022_sSFR} and now {\it
  JWST} data \citep{Topping2022_blueSlopes,Cullen2022}.  
  
Additionally, we adopt point-source sizes for the artificial sources we inject into various images in our simulation and recovery experiments.  While the present size assumptions are not especially different from that found for galaxies at $z\sim8$-17, both using earlier HST observations and now using {\it JWST} observations \citep{Naidu2022_z12,Naidu2022_z17,Ono2022}, they may lead to a slight overestimate of the total selection volume.  While it is worthwhile keeping this in mind for the discussion which follow, these uncertainties are likely small in comparison to the very large uncertainties in the total number of bona-fide $z\geq8$ galaxies over these fields (amongst the many sources from the literature we have graded as "possible").

We use 0.5-mag bins in deriving our stepwise $UV$ LF results, while for our parametric 
determinations, we adopt both a Schechter and double power-law functional form:
\begin{equation}
\phi(M) = \frac{\phi^*}{10^{0.4(M-M^*)(\alpha+1)} + 10^{0.4(M-M^*)(\beta+1)}}
\end{equation}      
where $\phi^*$ is normalization, $\alpha$ is the faint-end slope, $\beta$ is the bright-end slope, and $M^*$ indicates some characteristic luminosity where there is a transition between the two regimes.   

For the Schechter function results, we fix the $M^*$
to $-21.15$ mag consistent with the $z\geq 7$ $UV$ LF derived by
\citet{Bouwens2021_LF}, while we fix $\alpha$ to $-2.38$, $-2.71$, and
$-3.15$ at $z\sim10$, $z\sim13$, and $z\sim17$ consistent with an
extrapolation of the LF fit results of \citet{Bouwens2021_LF} to the
respective redshifts.  For our double power-law fit results, we fix
$M^*$ to $-19.67$ and $\beta = -3.75$ to match the $UV$ LF fit results
of \citet{Bowler2020_LF} at $z\sim9$.  We fix $\alpha = -2.17$,
$-2.35$, $-2.62$, $-2.98$ at $z\sim8$, $z\sim10$, $z\sim13$, and
$z\sim17$ consistent with the fitting formula \citet{Bowler2020_LF}
provide for evolution of the $UV$ LF using a double power-law
parameterization.

We present our binned $UV$ LF results at $z\sim10$, $z\sim13$, and
$z\sim17$ LF results in both Table~\ref{tab:binnedlfs} and
Figure~\ref{fig:uvlfs}.  The parameterized fit results are presented
in Table~\ref{tab:lffits} and on Figure~\ref{fig:uvlfs} as red lines.  
We also derived $UV$ LF results at $z\sim8$ as a test of
our procedures for deriving $UV$ LFs at $z\geq 10$.  The results are
shown in both Figure~\ref{fig:uvlf8} from Appendix D and
Tables~\ref{tab:binnedlfs}-\ref{tab:lffits}. Encouragingly the
results we obtain are consistent with the earlier determinations we
obtained from HST data in \citet{Bouwens2021_LF}.

The present $UV$ LF results appear to be fairly similar to the $UV$ LF
results of \citet{Donnan2022} and \citet{Harikane2022_z9to17} at
$z\sim9$-11.  At $z\sim13$, we find a $\sim$1.5-2$\times$ higher
volume density of sources than \citet{Donnan2022},
\citet{Harikane2022_z9to17}, and \citet{Finkelstein2022_z12}, and at
$z\sim17$, the volume density we find for sources is $\sim$$3\times$
higher than what \citet{Harikane2022_z9to17} recover.  At the bright
end of the $z\sim10$-13 LFs, our results are very similar to
\citet{Naidu2022_z12}.  In general, there is broad similarity in all
LF results obtained to the present with {\it JWST}, given the limited
statistics available and thus large uncertainties.

\subsection{$UV$ LF Results from Our Literature Samples}

As an alternative to direct determinations of the $UV$ LF from our own
selections of $z\geq 8$ candidates, we also consider the use of the
literature results we analyzed and characterized in (\S\ref{sec:othersel}) to derive $UV$ LF results at
$z\sim10$, $z\sim13$, and $z\sim17$.

As we have already noted, large numbers of $z\geq 9$ candidate
galaxies have been identified in various analyses of the early NIRCam
data, and the purpose of this analysis is to show the implications of
these results for the $z\geq 9$ $UV$ LFs assuming that a significant
fraction of these candidates are at $z\geq 9$.

It is interesting to derive the implied LF results as a function of
the apparent robustness level of the candidates, to demonstrate how
high the volume density of sources is even including only the best
candidates.  We take the $UV$ luminosity of individual candidates to
the values we measure based on our own photometry.  A complete list of
the candidates we utilize and their classification into the groups defined in (\S\ref{sec:othersel}) is provided
in Tables~\ref{tab:robust} and \ref{tab:solid}-\ref{tab:possible3}.

\begin{figure*}
\centering \includegraphics[width=2\columnwidth]{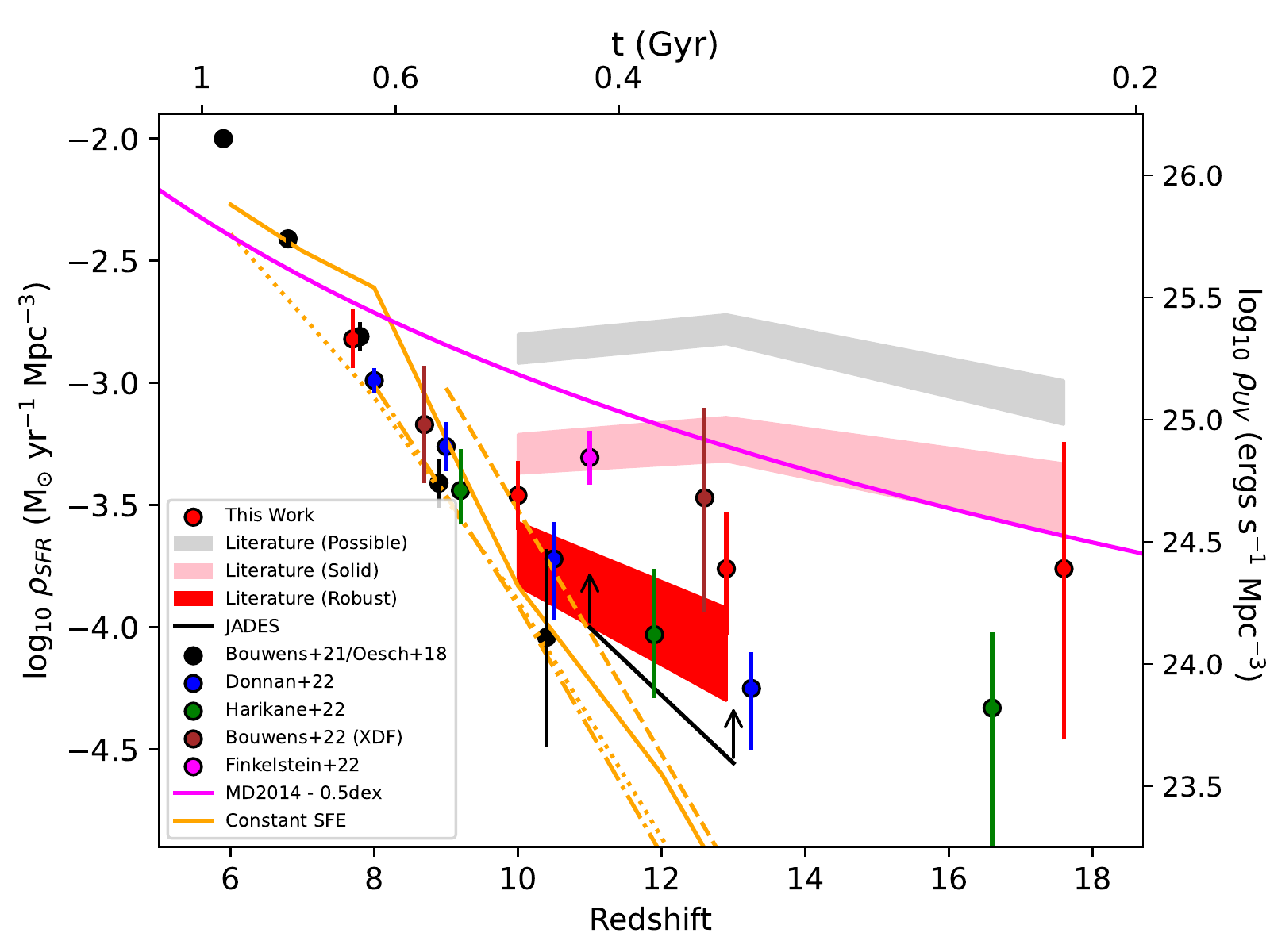}
\caption{$UV$ luminosity and star formation rate density integrated to
  $-19$ mag.  Shown are determinations from the present analysis
  (\textit{red circles}), \citet[][: \textit{blue
      circles}]{Donnan2022}, \citet[][: \textit{green
      circles}]{Harikane2022_z9to17}, \citet[][: \textit{brown
      circles}]{Bouwens2022_XDF}, \citet[][: \textit{magenta
      circles}]{Finkelstein2022_JWSTLF}, and
  \citet{Bouwens2021_LF}/\citet[][: \textit{black
      circles}]{Oesch2018_z10LF}.  The solid black line and lower
  limits are the $UV$ luminosity densities we compute brightward of
  $-19$ mag from the spectroscopically-confirmed sources presented in
  \citet{Robertson2022_JADES,CurtisLake2022_JADES}. The light gray,
  pink, and red shaded regions give the $UV$ luminosity densities at
  $z\sim10$-17 inferred based on literature candidates graded here as
  ``robust,'' ``solid,'' and ``possible,'' respectively.  The reason
  the latter two regions are 3-7$\times$ and 8-20$\times$ higher,
  respectively, in luminosity density than inferred by some analyses
  in the literature, e.g., \citet{Donnan2022} and
  \citet{Harikane2022_z9to17}, is due to many analyses only including
  a fraction of the potentially credible $z\geq 8$ candidates from the
  literature (which are here graded as "solid" or "possible").  The
  magenta line shows the fiducial star formation history derived by
  \citet{Madau2014} extrapolated to $z>8$ and shifted downward by 0.5
  dex to approximately account for the plotted densities being
  integrated down to a 2-mag shallower limit than in the fiducial
  \citet{Madau2014} probe.  The orange lines indicate the expected
  evolution in the $UV$ luminosity density assuming no evolution in
  the star formation efficiency of galaxies across cosmic time using
  the models of \citet[][: \textit{solid}]{Mason2015}, \citet[][:
    \textit{dot-dashed}]{Tacchella2018}, \citet[][: {\it
      dotted}]{Bouwens2021_LF}, and \citet[][: {\it
      dashed}]{Harikane2022_LF}.}
    \label{fig:sfz}
\end{figure*}

Given the diverse selection criteria used to construct these
literature LFs, we take the selection volume to be equal to the
\textit{detection} volume, as we very conservatively assume that all
\textit{detected} sources are \textit{selectable} by one or more of
the diverse selection criteria used in the literature.  By making this
assumption, our derived LFs should be as low as possible given the
available selection volume at high redshift.

To illustrate the implications of including all of the published
$z\geq 8$ candidates to the present in LF determinations, we present
$z\sim10$, $z\sim13$, and $z\sim17$ LF results in
Figure~\ref{fig:uvlfs} using the solid grey circles and error bars.  For clarity, $UV$ LF and luminosity density results derived for this "possible" sample and later for the "solid" literature sample also include the full set of sources from the "solid"+"robust" and "robust" samples, respectively.
 Gray lines are drawn connecting the grey points to help the literature LF results.  Those LF results are some $\sim2$-10$\times$ higher than the LF results reported by
\citet{Donnan2022}, \citet{Harikane2022_z9to17},
\citet{Bouwens2022_XDF}, and \citet{Finkelstein2022_JWSTLF} over the
luminosity range $-20$ to $-18$ mag.  One explanation is that this full sample includes sources we have graded as ``possible'', which we estimate to have a lower probability of being at $z>8$.

Thus, one reason these LFs might be so much in excess of the individual 
LF determinations is because the list of "possible" $z>8$ candidates 
(Table~\ref{tab:possible}-\ref{tab:possible3}) include large numbers of 
lower-redshift interlopers.  This motivates us to also derive LF results 
based on candidates which satisfy much more stringent quality 
requirements, such as those that make up our ``solid'' or ``robust'' 
sample of sources from the literature.

Results for the ``solid'' and ``robust'' samples are shown in
Figures~\ref{fig:lfscience} with the partially transparent pink and red points, respectively.  Even the $z\sim10$ LF results
from the ``solid'' candidates exceed the results from
\citet{Donnan2022} and \citet{Harikane2022_z9to17} by factors of $\sim$1.5 to 2, but agree better with the our own results and those of \citet{Finkelstein2022_JWSTLF}.  At $z\sim13$ and $z\sim17$, the LF results derived from the ``solid'' candidates lies even more clearly in excess of the LF results from \citet{Donnan2022}, \citet{Harikane2022_z9to17}, and our own analysis.

Given current uncertainties over what fraction of current $z\geq
9$ candidate lists are bona-fide, we express the LF results we derive
from the literature in Table~\ref{tab:lffitresults} in terms of a
region spanning the range between our LF results using the ``robust''
candidates and the candidates we classify as ``solid.''  These results
are also shown in Figure~\ref{fig:lfscience}, and we can see it easily
encompasses the range of LF results reported in various studies.

\begin{table}
\centering
\caption{Inferred $UV$ Luminosity Densities$^1$ at $z\geq8$}
\label{tab:uvlumdens}
\begin{tabular}{c|c|c}
\hline
         & $\rho_{UV}$ & $\rho_{SFR}$ \\
Redshift & [ergs $s^{-1}$ Mpc$^{-3}$] & [$M_{\odot}$/yr/Mpc$^{3}$] \\
\hline\hline
\multicolumn{3}{c}{Direct Determinations}\\
8 & 25.21$_{-0.28}^{+0.28}$ & $-$2.94$_{-0.28}^{+0.28}$\\
10.0 & 24.69$_{-0.14}^{+0.14}$ & $-$3.46$_{-0.14}^{+0.14}$\\
12.9 & 24.39$_{-0.27}^{+0.23}$ & $-$3.76$_{-0.27}^{+0.23}$\\
17.6 & 24.39$_{-0.70}^{+0.52}$ & $-$3.76$_{-0.70}^{+0.52}$\\
\\
\multicolumn{3}{c}{``Robust'' Literature Sample}\\
10 & 24.45$\pm$0.13 & $-$3.70$\pm$0.13\\
13 & 24.04$\pm$0.19 & $-$4.36$\pm$0.19\\
\\
\multicolumn{3}{c}{``Solid'' Literature Sample}\\
10 & 24.86$\pm$0.08 & $-$3.29$\pm$0.08\\
13 & 24.92$\pm$0.09 & $-$3.23$\pm$0.09\\
17 & 24.68$\pm$0.14 & $-$3.47$\pm$0.14\\
\\
\multicolumn{3}{c}{``Possible'' Literature Sample}\\
10 & 25.29$\pm$0.06 & $-$2.86$\pm$0.06\\
13 & 25.37$\pm$0.06 & $-$2.80$\pm$0.06\\
17 & 25.07$\pm$0.09 & $-$3.08$\pm$0.09\\
\hline\hline
\end{tabular}
\\\begin{flushleft}
$^1$ Luminosity densities integrated down to $-$19 mag.
\end{flushleft}
\end{table}

\section{Discussion}

\subsection{Evolution of Star-Forming Galaxies from $z\sim17$ to $z\sim8$}

There has been a lot of discussion over the last ten years regarding
how much star formation took place during the earliest epochs of the
universe, when $z>10$.  Some of this discussion had been based on the
evolution of the $UV$ LF at $z>6$ and debate between a slower
evolution in the apparent SFR and $UV$ luminosity density
\citep[e.g.,][]{McLeod2016} and a more rapid evolution
\citep[e.g.,][]{Oesch2014_z910LF,Oesch2018_z10LF,Bouwens2021_LF}.

The relatively small number of apparently robust $z\sim10$ candidates
identified in the wider area data searched by \citet{Oesch2018_z10LF}
seemed to weigh in favor of a faster evolution.  Nevertheless, the
apparent discovery of many luminous galaxy candidates (particularly now with {\it JWST}) in the $z\geq 9$
universe over wide areas 
\citep[e.g.,][]{Bowler2020_LF,RobertsBorsani2022,Harikane2022_LF,Finkelstein2022,Bagley2022,Kauffmann2022,Donnan2022}
and the discovery of apparent Balmer breaks in galaxies at $z\geq 9$
\citep{Zheng2012,Hashimoto2018_z9,Labbe2022} pointed in the other direction,
towards more substantial early star formation activity.

A good baseline for evaluating early star formation activity is
through comparison with the predictions of constant star formation
efficiency models (SFE).  Already, such models have succeeded in providing a
plausible baseline for modeling star formation across cosmic time
\citep[e.g.][]{Mason2015,Bouwens2015_LF,Bouwens2021_LF,Mashian2016,Harikane2018,
Harikane2022_LF,Oesch2018_z10LF,Tacchella2018,Stefanon2021_SMF,Stefanon2022_sSFR}.  While there have been a large number of models using the constant SFE
assumption to model the evolution of the SFR density across cosmic
time, we will test the results against only four: \citet{Mason2015},
\citet{Bouwens2015_LF}, \citet{Tacchella2018}, and
\citet{Harikane2022_LF}.

A comparison of the constant star formation efficiency model results
are shown in Figure~\ref{fig:lfscience}.  As in other recent studies,
the evolution on the $UV$ LF appears to be significantly in excess of
that predicted from constant SFE models at $z\geq 12$ (for galaxies more luminous than $\sim -19$).  Not only does
this clearly appear to be the case for all LF determinations at $z\geq
11$, but it is even true if we only make use of sources from the
literature that we classify as robust.  If one or more of the
candidate $z\sim17$ galaxies is actually at such a high redshift
(\citealt{Donnan2022,Harikane2022_z9to17,Naidu2022_z17}: but see also
\citealt{Zavala2022,Naidu2022_z17}), differences with the constant SFE
models is even larger.

It is unclear whether this indicates the SFE of galaxies is indeed
more efficient or if the IMF of (luminous) galaxies is very different at early
times.  If the stellar masses in $z>8$ galaxies are as high as found
in \citet{Labbe2022}, it would argue in favor of a substantially
higher SFE.  There is clearly a limit to how high the SFE can be based
on the baryon mass in collapsed halos at $z>8$, and interestingly
enough, both \citet{BoylanKolchin2022} and \citet{Naidu2022_z17} find
that some galaxies may be in violation of these limits.  Potential
resolution of this enigma could include an evolution in the stellar
IMF in star-forming galaxies at $z>8$ such that the mass-to-light
ratio in early galaxies is substantially lower than at later times in the history of
the universe
\citep[e.g.][]{Steinhardt2022,Harikane2022_z9to17,Inayoshi2022}.

\subsection{$UV$ Luminosity and SFR Densities of Galaxies at $z\geq 8$}

An alternate way of assessing the star formation activity in the early
universe is by looking at results in terms of the $UV$ luminosity
density and SFR density.  In characterizing the evolution, we only
consider sources and $UV$ LF results brighter than $-$19 mag to avoid
extrapolating the LF faintward of what can be well probed with early
{\it JWST} data, i.e., $\sim$29 mag, as used both in the present
study, \citet{Donnan2022}, and \citet{Finkelstein2022_JWSTLF}.

We have adopted such a limit to avoid substantial extrapolations of
$UV$ LF results to much fainter luminosities where they are less well
constrained.  If we consider extrapolations to $-$17 mag (as
considered in both \citealt{Donnan2022} and
\citealt{Harikane2022_z9to17}), $z\sim17$ SFR density results derived
assuming a faint-end slope $\alpha$ of $-2.1$ (as assumed by
\citealt{Harikane2022_z9to17}) vs. assuming a faint-end slope of $-3$
(as predicted at $z\sim17$ by \citealt{Mason2015}) differ by $\sim$1
dex.  Given that difference would then be driven entirely by the
\textit{assumed} faint-end slope, it is clearly preferable to quote
SFR density results only to luminosity limits which are well probed by
the observations.

In Figure~\ref{fig:sfz}, we present our results for the $UV$
luminosity density evolution both from our direct LF analyses.
Additionally, we include the equivalent SFR density results, assuming
the conversion factor $\cal{K}$$_{FUV}$ is $0.7 \times 10^{-29}
M_{\odot}\,\textrm{year}^{-1} \,\textrm{erg}^{-1}
\,\textrm{s}\,\textrm{Hz}$ from \citet{Madau2014}, which assumes a
\citet{Chabrier2003} IMF, a constant star formation rate, and
metallicity $Z=0.002$ $Z_{\odot}$.  For context, we also include the
results obtained by several other analyses of the {\it JWST}
observations
\citep{Donnan2022,Harikane2022_z9to17,Bouwens2022_XDF,Finkelstein2022_JWSTLF}
and also several constant star formation efficiency (SFE) predictions
for the $UV$ luminosity density evolution
\citep{Mason2015,Tacchella2018,Bouwens2021_LF,Harikane2022_LF}.  For
context, a magenta line is included showing the fiducial star
formation history derived by \citet{Madau2014} extrapolated to $z>8$
but adjusted to be relevant for SFR probes down to $-$19 mag.  We
implement this adjustment as a 0.5 dex offset reflecting the
difference in $UV$ luminosity densities derived by
\citet{Bouwens2021_LF} to $-19$ mag (the limit used here) vs. $-$17
mag (the limit used by \citealt{Madau2014}).  It is interesting to see
how expectations at $z>8$-9 have evolved from a decade earlier and how
uncertain the SFRD still remains in the first $\sim$500 Myr.

In Figure~\ref{fig:sfz}, we also show the $UV$ luminosity density
results derived from our literature samples of the same fields. 
Separate results are presented for candidates categorized as 
``robust,'' ``solid,'' and ``possible'' with the shaded red, pink, and grey regions, respectively.  For additional reference, we include as a solid black line and upward arrows the implied lower limits on the $UV$ luminosity densities at $z>10$ based on the recent {\it JWST} ADvanced Extragalactic Survey (JADES) spectroscopic results over the HUDF/XDF region \citep{Robertson2022_JADES,CurtisLake2022_JADES}.  For those limits, we adopt the $UV$ luminosities measured by \citet{Robertson2022_JADES} and assume a total search area of 2 $\times$ (1.5 arcmin)$^2$ and that sources can selected over the entire volume $z=10$-12 and $z=12$-14.

It is striking how much higher the implied luminosity densities of the "possible" candidates are relative to the results derived from those candidates in the other categories.  Results including all of the candidates are $\sim$3$\times$ and $\sim$8$\times$ higher than those candidates we grade as "solid" and "robust," respectively, at $z\sim10$ and $\sim$7$\times$ and $\sim$20$\times$, respectively, higher at $z\geq 12$.  These same $UV$ luminosity density results are also significantly in excess of our own $UV$ luminosity density results as well as the results of \citet{Donnan2022}, \citet{Harikane2022_z9to17}, and \citet{Finkelstein2022_JWSTLF}. 

Clearly, much of the excess could be due to the presence of potentially substantial numbers of 
lower-redshift contaminants in various $z\geq 8$ selections.  The
detection of possibly significant flux blueward of the breaks in the median stacks of the ``possible'' candidates is indeed suggestive of such a conclusion (cf., \S\ref{sec:othersel}, Appendix C).  There are clearly  large uncertainties in what fraction of these fainter sources are at high redshifts.  As we demonstrate in Appendix E, the assessment of the reliability of specific $z\geq 8$ candidates can vary substantially between the different studies.  It is indicative of the challenges with these early data sets that our independent
evaluation of the candidates from \citet{Donnan2022},
\citet{Harikane2022_z9to17}, \citet{Finkelstein2022_JWSTLF}, and our own selections place a non-negligible fraction of these candidates ($\gtrsim$20\%) in our lowest quality bin (Table~\ref{tab:assess}).

Meanwhile, results using the ``robust'' candidates appear to be in excellent agreement with the collective LF results
of \citet{Donnan2022} and \citet{Harikane2022_z9to17}, while our own results and those of
\citet{Finkelstein2022_JWSTLF} agree better with the results obtained using the ``solid'' candidates.  The
\citet{Finkelstein2022_JWSTLF} LF results appear to be $\approx$2$\times$
higher than the \citet{Donnan2022} results due to the $\sim$2$\times$
higher completeness of the \citet{Finkelstein2022_JWSTLF} selection to
``robust''+''solid'' $z\geq 8$ candidates from the literature
(Table~\ref{tab:assess}).\footnote{We remark in passing that the
\citet{Finkelstein2022_JWSTLF} LF results appears to be more
consistent than with the empirical completeness estimates we derive on
the basis our literature selections (Table~\ref{tab:assess}) than is
the case for either the \citet{Donnan2022} or
\citet{Harikane2022_z9to17} analyses where the completeness of their
selections is $\sim$2$\times$ lower than
assumed in their LF analyses.}

Without spectroscopy, it is difficult to know which of these two
results is more reliable.  A key question is the extent to which
sources in our ``solid'' literature sample are at $z>8$.  Simulation 
results from both \citet{Harikane2022_z9to17} and \citet{Larson2022} 
indicate that $z\geq 8$ selections over CEERS-like data sets might well 
include an appreciable number of lower-redshift
interlopers, even restricting such selections to sources with $>$80\%
of the integrated likelihood at $z>5.5$ (as is required for sources that make up our "solid" literature selections).  Based on the expected contamination in the first {\it JWST} fields (likely due to the limited depth of the data blueward of the break), \citet{Harikane2022_z9to17} require that $z\geq 9$ candidates satisfy an especially demanding
$\Delta(\chi^2 (z_{low})-\chi^2(z_{high})) > 9$ selection criterion to be included in their high-redshift samples.

Another concerning aspect of sources in our "solid" literature
selections is the much less significant overlap between candidates
reported in different studies.  While 90\% of the $z\geq 9$ candidates
in our "robust" literature selections are identified as part of
multiple studies, only 26\% of the $z\geq 9$ candidates in the "solid"
literature selections are found in multiple studies.\footnote{For
reference, the percentage of $z\geq9$ candidates from our "possible"
literature selections that occur in more than one study is just 5\%.
This demonstrates there is really a difference in the quality of the
candidates that make up of our literature subsamples.}  This suggests
that a larger percentage of sources in our "solid" literature sample
may in fact be lower redshift contaminants, but it is a huge open
question what that percentage is.

Even median stacking of the SED results is of little use in ascertaining whether sources in our ``solid'' literature selections are reliable.  
As we show in Appendix C, very similar stack results are obtained using
either the ``robust'' or ``robust''+''solid'' subsamples of literature
candidates.  In both cases, a pronounced $\sim$1.5-mag spectral break is
seen, with no significant flux blueward of the break.  Also both stacks
reveal a blue spectral slope redward of the break.

Fortunately, an increasing amount of spectroscopy is becoming
available for $z>8$ selections, particularly based on the JADES and
CEERS programs
\citep[e.g.,][]{Tang2023,Fujimoto2023,Cameron2023,Saxena2023,Bunker2023},
spectroscopically confirming many sources out to a redshift $z\approx
9.5$ where the strong [OIII]$_{4959,5007}$ doublet can be detected at
high S/N for star-forming sources and in some cases even earlier
\citep[e.g.,][]{Bunker2023}. However, it should be noted that not all
$z\geq 8$ candidates are being confirmed to be at $z\geq 8$.  The
$z\sim8$ candidate 13050 from \citet{Labbe2022} has been found to have
a redshift $z=5.62$ and to be an AGN \citep{Kocevski2023}.  This is a
particularly interesting example since it adds weight to the concern
that our photometric redshift SED templates are not yet as complete as
we would like.

Regardless of what the actual SFR density is at $z\geq 10$, i.e., whether it is closer to the "robust" or "solid" literature results shown in Figure~\ref{fig:sfz}, essentially all of the present results lie in significant excess of the constant SFE models \citep{Mason2015,Tacchella2018,Bouwens2021_LF,Harikane2022_LF} by factors of $\sim$2-6 at $z\sim12$ and by even larger factors at $z>12$.  

It seems likely that at least part of the excess at $z>9$ could be explained due to the impact of noise in driving photometric redshift estimates to somewhat higher values than later found through spectroscopy.  The approximate amplitude of this effect appears to be $\Delta z \sim 1$ at $z>7$ \citep[e.g.,][]{Munoz2008,Bouwens2022_REBELS,Kauffmann2022,Fujimoto2023}.  This appears to be due to typical photometric redshift estimates adopting a flat prior in redshift and thus taking into account the fact that luminous sources are more prevalent at lower redshift than they are at high redshift \citep{Munoz2008}.

There have been a variety of different explanations offered for this deviation from the constant SFE predictions in the literature.  One possibility has been to suppose that the mass-to-light ratios of galaxies in the early universe are much lower than at later points in cosmic time, which could result from a change in the effective IMF of galaxies in the $z>10$ universe to one which is much more top heavy \citep[e.g.][]{Harikane2022_z9to17,Steinhardt2022,Inayoshi2022}.  

Other possibilities have included the hypothesis perhaps AGN contribute much more significant to the light from the earliest generation of galaxies \citep[e.g.][]{Harikane2022_z9to17}, there is much greater scatter in the star formation rates in galaxies in the early universe away from the main star-forming sequence \citep[e.g.][]{Mason2022}, as well a number of other explanations \citep[e.g.][]{Ferrara2022_JWSTmodel,Mirocha2022_modelJWST,Harikane2022_z9to17,Kannan2022,Lovell2023}.  Ascertaining which of these explanations is correct will ultimately require an extensive amount of follow-up observations with ALMA and {\it JWST}, especially involving spectroscopy as e.g. the recent confirmation of a $z=9.76$ source, $z>10$ sources by JADES team demonstrates 
\citep{RobertsBorsani2022_SpecJWST,CurtisLake2022_JADES,Robertson2022_JADES,Bunker2023}, and $z=6$-9 sources in CEERS \citep[e.g.,][]{Tang2023,Fujimoto2023}.

\section{Summary}

We have derived luminosity functions, and set constraints on the UV luminosity and SFR density from $z\sim8$ to $z\sim17$, using the three most well-studied {\it JWST} NIRCam data
sets from the first 5 months of {\it JWST} science operations, namely, the SMACS0723 cluster field \citep{Pontop2022},
the GLASS Abell 2744 parallel field \citep{Treu2022_GLASS}, and four CEERS
\citep{Finkelstein2022_JWSTLF} extragalactic fields.  

We have selected of samples
of $z\sim8$, $z\sim10$, $z\sim13$, and $z\sim17$ galaxies in these fields, and made full use of the very extensive selections done by others to date.
In particular, we have investigated the challenges of the selection of $z\geq 8$ galaxies and derivation of
$UV$ LF results from these early {\it JWST} NIRCam observations.  Even with a very conservative approach to selections, both from our own and similarly sub-selecting those of other studies, we find that luminous galaxies in the first 400-500 Myr are as enigmatic as the first {\it JWST} results suggested.

We first make use of two different reductions of the NIRCam
observations to test the sensitivity of $z\geq8$ selections to the
reduction technique.  The first set of reductions we utilize relies on
the \textsc{grizli} NIRCam pipeline, while the second leverages an
alternate set of reductions made with the \textsc{pencil} NIRCam
pipeline.  Both reductions take advantage of advances made in the
calibrations of the NIRCam zeropoints, as well as including steps to
minimize the impact of $1/f$ noise and ``snowball'' artefacts.

Based on these reductions, we construct substantial samples of nominally
$z\sim8$, $z\sim10$, $z\sim13$, and $z\sim17$ galaxies using two color
Lyman-break selection criteria. Our redshift selection functions indicate that our selections cover redshift ranges of $z\sim7-9$, $z\sim9-11$, $z\sim12-14$, and $z\sim16-19$. Our primary selection using the
\textsc{grizli} reductions includes 18 $z\sim8$, 12 $z\sim10$, 5
$z\sim13$, and 1 $z\sim17$ galaxies, while our alternate selection
using the \textsc{pencil} reductions yields 22 $z\sim8$, 13 $z\sim10$,
3 $z\sim13$, and 1 $z\sim17$ galaxy candidates.  The overlap between these samples, even from two recent reductions, is only $\sim$40-50\%, indicative of the subtleties and challenges of identifying sources at high redshift. 

Using sources from the above selection and using estimates of the
selection volumes in our search fields, we have derived estimates of
the $UV$ LF at $z\sim8$, $z\sim10$, $z\sim13$, and $z\sim17$.  While
the uncertainties are still very large, our $UV$ LF results are
suggestive of factors of 6 and 6 decreases in the normalization of the
$UV$ LF from $z\sim8$ to $z\sim13$ and $z\sim17$, respectively.  Not
surprisingly, the results we obtain are similar to the relatively mild
evolution in the $UV$ luminosity density already reported in
\citet{Naidu2022_z12}, \citet{Donnan2022},
\citet{Harikane2022_z9to17}, \citet{Bouwens2022_XDF}, and
\citet{Finkelstein2022_JWSTLF}.  

We also take these results and set constraints on the UV luminosity and SFR density from $z\sim17$ to $z\sim8$ for galaxies more luminous than $-$19 mag.  Similar to what we found for the $UV$ LF results, the $UV$ luminosity density and SFR density, both our direct determinations and the results based on likely robust $z\sim11$-13 candidates from the literature, lie significantly in excess of the constant star formation efficiency (SFE) models, by factors of $\sim$2-6.  Interpretation of these results is unclear, and it is open question whether the new results indicate the SFE of galaxies is indeed more efficient or if the IMF of (luminous) galaxies is very different at early times. 

As a complement to direct determinations of the $UV$ LF at $z\geq 8$,
we also derive $UV$ LF and luminosity density results, by taking advantage of the full samples of $z\sim10$, $z\sim13$, and $z\sim17$ galaxies that have been identified to date over the three most well studied fields.  We then segregate this sample of candidates into three different samples “robust,” “solid,” and “possible” based on how likely sources are to be at $z>5.5$ based on our photometry of the sources in both our fiducial and secondary reductions of the NIRCam imaging observations.

We first considered the luminosity densities we would derive including all $z\geq 8$ candidates reported over the three most studied fields to the present.  Remarkably, we find $7\times$ and $20\times$ higher luminosity densities at $z\geq 12$ relying on the "solid" and "possible" candidates than relying on the "robust" candidates from the literature alone.  These results demonstrate how uncertain the luminosity densities are at $z>6$ and how much the results depend on the extent to what lower-redshift sources contaminate the $z\geq8$ selections.

Even allowing for a substantial amount of contamination in our selections of "possible" $z\geq 8$ sources from the literature, large ($\sim$0.5-1.0 dex: factors of 3 to 10) differences exist between the luminosity density results derived from sources graded "robust" and those graded "solid."  If the bulk of the $z\geq10$ candidates graded "solid" are instead at lower redshift, the true luminosity density results at $z\geq10$ would be more along the lines of what has been found by \citet{Donnan2022} and \citet{Harikane2022_z9to17}, which are consistent with the recent spectroscopic results of \citet{CurtisLake2022_JADES} and closer to the predictions of the constant SFE models.  Some of the recent simulation results from \citet{Harikane2022_z9to17} and \citet{Larson2022} are suggestive of at least modest levels of contamination in the first {\it JWST} $z\geq 8$ selections with NIRCam.

On the other hand, if the bulk of the $z\geq8$ candidates graded "solid" are bona-fide, then the $UV$ LFs and $UV$ luminosity density at $z\sim10$ and $z\geq 12$ could be up to $\sim$3$\times$ and $\sim$7$\times$ higher and more in the range of the LF results we derive from our own selection of $z\geq8$ sources and also more consistent with the results of \citet{Finkelstein2022_JWSTLF}.  Supportive of these high luminosity density results are the median stack results we obtain for our selection of "solid" candidates from the literature, which appear to have almost identical characteristics to what we obtain from a similar stack of "robust" candidates from the literature.

Whatever the reality is, it is clear that huge open questions remain
regarding the true $UV$ LF and $UV$ luminosity density results at
$z\geq 8$.  To resolve these open questions, deeper imaging
observations and follow-up spectroscopy with {\it JWST} NIRSpec and
the grisms will be required, allowing for a significantly improved
reliability of $z\geq 8$ selections and $UV$ LF determinations going
forwards.  Fortunately, there are already significant on-going efforts
obtaining sensitive imaging over fields like the HUDF
\citep[e.g.][]{Bouwens2022_XDF,Robertson2022_JADES} and sensitive
spectroscopic campaigns by the substantial JADES and other programs
\citep[e.g.][]{CurtisLake2022_JADES,Bunker2023,Cameron2023,Tang2023,Fujimoto2023}
  that provide the needed new data.

\section*{Acknowledgements}

RJB acknowledges support from NWO grants 600.065.140.11N211 (vrijcompetitie) and TOP grant TOP1.16.057. MS acknowledges support from the CIDEGENT/2021/059 grant, from project PID2019-109592GB-I00/AEI/10.13039/501100011033 from the Spanish Ministerio de Ciencia e Innovaci\'on - Agencia Estatal de Investigaci\'on, and from Proyecto ASFAE/2022/025 del Ministerio de Ciencia y Innovación en el marco del Plan de Recuperación, Transformación y Resiliencia del Gobierno de España  RPN acknowledges funding from JWST programs GO-1933 and GO-2279. Support for this work was provided by NASA through the NASA Hubble Fellowship grant HST-HF2-51515.001-A awarded by the Space Telescope Science Institute, which is operated by the Association of Universities for Research in Astronomy, Incorporated, under NASA contract NAS5-26555. PAO acknowledges support from: the Swiss National Science Foundation through project grant 200020 207349.  The Cosmic Dawn Center (DAWN) is funded by the Danish National Research Foundation under grant No. 140.  This work is based on observations made with the NASA/ESA/CSA James Webb Space Telescope. The
data were obtained from the Mikulski Archive for Space Telescopes at the Space Telescope Science Institute, which is operated by the Association of Universities for Research in Astronomy, Inc., under NASA contract NAS
5-03127 for JWST.

\section*{Data Availability}

All data used here are available from the Barbara A. Mikulski Archive for Space Telescopes (MAST: \url{https://mast.stsci.edu}), both in the form of raw and high level science products.

\bibliographystyle{mnras}
\bibliography{main} 
\appendix

\section{$z\geq8$ Candidates Identified in Our Secondary Reductions}

As a test on the sensitivity of $z\geq 8$ selections to the NIRCam
reductions utilized, we perform a second search for $z\geq 8$ galaxies
but using reductions made with \textsc{pencil} (\S\ref{sec:data}).  22
$z\sim8$, 13 $z\sim10$, 3 $z\sim13$, and 1 $z\sim17$ galaxies are
identified in these reductions.

The coordinates, photometric redshifts, apparent magnitudes, and
spectral break amplitudes of the $z\geq 8$ candidates we find are
indicated in Table~\ref{tab:sample2}.  Also presented is the
difference between the minimum $\chi^2$ found for $z>5.5$ and $z<5.5$
fits to the observed SEDs of the sources, the estimated likelihood
that a candidate has a redshift in excess of 5.5, and any earlier
studies who identified a given source as part of their $z\geq8$
searches.

\begin{table*}
\centering
\caption{Selection of $z\sim10$, $z\sim13$, and $z\sim17$ sources identified over three most studied imaging fields in early {\it JWST} NIRCam observations using the alternate set of NIRCam reductions \textsc{pencil} (\S\ref{sec:data}).}
\label{tab:sample2}
\begin{tabular}{c|c|c|c|c|c|c|c|c} \hline
     &    &     &            & $m_{UV}$ & Lyman Break & $\Delta \chi^2$ &    &  \\
  ID & RA & DEC & $z_{phot}$ & [mag] & [mag] & & $p(z>5.5)$ & Lit$^{1}$ \\\hline\hline
  \multicolumn{9}{c}{$z\sim8$ Selection}\\
  \hline
GLASSP2Z-4041319253 & 00:14:04.139 & $-$30:19:25.35 & 6.7$_{-0.1}^{+0.1}$ & 28.5$\pm$0.3 & $>$2.0 & $-$6.3 & 0.95 & \\
GLASSP2Z-4027418404 & 00:14:02.741 & $-$30:18:40.41 & 6.7$_{-0.1}^{+0.1}$ & 27.7$\pm$0.1 & 1.8$\pm$0.5 & $-$15.2 & 0.999 & B22\\
S0723Z-3068426486 & 07:23:06.843 & $-$73:26:48.62 & 6.9$_{-0.1}^{+0.1}$ & 27.0$\pm$0.1 & 1.9$\pm$0.2 & $-$12.7 & 0.997 & \\
S0723Z-3154126292 & 07:23:15.413 & $-$73:26:29.28 & 7.0$_{-0.1}^{+0.1}$ & 25.0$\pm$0.1 & 2.1$\pm$0.1 & $-$6.2 & 0.951 & \\
GLASSP2Z-4029818477 & 00:14:02.988 & $-$30:18:47.77 & 7.2$_{-0.4}^{+0.3}$ & 28.5$\pm$0.2 & $>$2.1 & $-$6.2 & 0.984 & \\
GLASSP1Z-3589121028 & 00:13:58.915 & $-$30:21:02.80 & 7.4$_{-0.2}^{+0.2}$ & 28.6$\pm$0.3 & $>$2.3 & $-$11.2 & 0.998 & B22\\
S0723Z-3053127492 & 07:23:05.311 & $-$73:27:49.22 & 7.4$_{-0.3}^{+0.3}$ & 26.2$\pm$0.1 & 2.8$\pm$2.3 & $-$7.7 & 0.984 & \\
GLASSP1Z-4027520346 & 00:14:02.759 & $-$30:20:34.68 & 7.4$_{-0.2}^{+0.2}$ & 27.9$\pm$0.2 & $>$2.3 & $-$15.6 & 1.0 & B22\\
GLASSP1Z-4002221454 & 00:14:00.229 & $-$30:21:45.46 & 7.4$_{-0.3}^{+0.4}$ & 28.3$\pm$0.2 & $>$2.2 & $-$6.1 & 0.976 & \\
GLASSP1Z-3577921174 & 00:13:57.791 & $-$30:21:17.43 & 7.5$_{-0.3}^{+0.4}$ & 26.8$\pm$0.2 & 2.3$\pm$1.9 & $-$38.9 & 1.0 & B22\\
GLASSP2Z-4018619113 & 00:14:01.860 & $-$30:19:11.38 & 7.5$_{-0.3}^{+0.4}$ & 28.3$\pm$0.2 & $>$2.0 & $-$9.9 & 0.997 & \\
GLASSP2Z-4011518065 & 00:14:01.152 & $-$30:18:06.58 & 7.5$_{-0.4}^{+0.4}$ & 28.3$\pm$0.2 & $>$2.1 & $-$15.2 & 1.0 & \\
GLASSP2Z-4005219093 & 00:14:00.520 & $-$30:19:09.31 & 7.5$_{-0.4}^{+0.4}$ & 28.6$\pm$0.3 & $>$2.0 & $-$4.3 & 0.934 & \\
S0723Z-3295626401 & 07:23:29.566 & $-$73:26:40.18 & 7.5$_{-0.6}^{+0.7}$ & 27.7$\pm$0.2 & 2.1$\pm$1.0 & $-$19.0 & 1.0 & B22\\
S0723Z-3201526042 & 07:23:20.158 & $-$73:26:04.28 & 7.6$_{-0.3}^{+0.3}$ & 26.4$\pm$0.1 & 3.4$\pm$2.5 & $-$40.3 & 1.0 & D22,At22,Ad22,B22\\
GLASSP1Z-4041221457 & 00:14:04.126 & $-$30:21:45.77 & 7.8$_{-0.5}^{+0.4}$ & 27.5$\pm$0.2 & 2.1$\pm$1.6 & $-$7.7 & 0.981 & B22\\
GLASSP1Z-4020822000 & 00:14:02.082 & $-$30:22:00.07 & 7.8$_{-0.5}^{+0.5}$ & 26.8$\pm$0.2 & $>$2.4 & $-$6.6 & 0.977 & B22\\
GLASSP1Z-4041820257 & 00:14:04.187 & $-$30:20:25.79 & 7.8$_{-0.3}^{+0.3}$ & 28.1$\pm$0.2 & $>$2.3 & $-$26.5 & 1.0 & \\
GLASSP2Z-4012218476 & 00:14:01.224 & $-$30:18:47.68 & 8.0$_{-0.4}^{+0.4}$ & 27.3$\pm$0.2 & $>$2.2 & $-$4.1 & 0.923 & \\
GLASSP1Z-4062820397 & 00:14:06.286 & $-$30:20:39.79 & 8.0$_{-0.3}^{+0.4}$ & 28.3$\pm$0.1 & $>$2.5 & $-$6.0 & 0.958 & \\
S0723Z-3226926062 & 07:23:22.697 & $-$73:26:06.23 & 8.1$_{-0.2}^{+0.2}$ & 25.9$\pm$0.1 & $>$3.9 & $-$49.8 & 1.0 & D22,At22,Ad22,B22\\
GLASSP2Z-3553419246 & 00:13:55.345 & $-$30:19:24.63 & 8.6$_{-0.2}^{+0.2}$ & 27.6$\pm$0.1 & $>$3.0 & $-$22.1 & 1.0 & Le22\\\\

\multicolumn{9}{c}{$z\sim10$ Selection}\\
\hline
CEERSYJ-9494458188 & 14:19:49.446 & 52:58:18.89 & 8.8$_{-0.4}^{+0.3}$ & 28.0$\pm$0.2 & 1.6$\pm$2.2 & $-$9.4 & 0.993 & W22 (v1)\\
CEERSYJ-9193553160 & 14:19:19.358 & 52:53:16.01 & 9.0$_{-0.5}^{+0.5}$ & 28.2$\pm$0.2 & 1.7$\pm$1.5 & $-$17.2 & 1.0 & La22,F22\\
CEERSYJ-0012959481 & 14:20:01.290 & 52:59:48.13 & 9.2$_{-0.7}^{+0.7}$ & 27.6$\pm$0.2 & $>$1.9 & $-$1.5 & 0.896 & F22,B22\\
CEERSYJ-9353350378 & 14:19:35.336 & 52:50:37.88 & 9.3$_{-0.6}^{+0.9}$ & 27.8$\pm$0.2 & 1.7$\pm$1.6 & $-$8.4 & 0.982 & D22,F22,W22,B22\\
GLASSP1YJ-4045920571 & 00:14:04.599 & $-$30:20:57.10 & 9.4$_{-0.4}^{+0.5}$ & 28.6$\pm$0.2 & $>$2.3 & $-$3.4 & 0.885 & \\
GLASSP1YJ-4016420474 & 00:14:01.649 & $-$30:20:47.44 & 9.6$_{-0.4}^{+0.4}$ & 29.5$\pm$0.4 & $>$2.2 & $-$6.0 & 0.954 & \\
CEERSYJ-9320850526 & 14:19:32.081 & 52:50:52.63 & 9.6$_{-0.5}^{+0.6}$ & 28.2$\pm$0.2 & 2.0$\pm$1.7 & $-$1.6 & 0.859 & \\
CEERSYJ-9132749268 & 14:19:13.277 & 52:49:26.89 & 9.7$_{-0.6}^{+0.7}$ & 28.2$\pm$0.2 & $>$2.0 & $-$4.9 & 0.962 & \\
GLASSP1YJ-4043920397 & 00:14:04.398 & $-$30:20:39.78 & 9.7$_{-0.5}^{+0.6}$ & 29.2$\pm$0.2 & $>$2.2 & $-$5.8 & 0.969 & H22\\
CEERSYJ-9329250443 & 14:19:32.926 & 52:50:44.37 & 10.1$_{-0.6}^{+0.6}$ & 27.5$\pm$0.2 & $>$2.1 & $-$3.5 & 0.947 & \\
GLASSP1YJ-4028522189 & 00:14:02.857 & $-$30:22:18.92 & 10.3$_{-0.5}^{+0.5}$ & 26.6$\pm$0.1 & $>$3.3 & $-$33.7 & 1.0 & N22,C22,D22,H22,B22\\
CEERSYJ-9006450120 & 14:19:00.643 & 52:50:12.00 & 10.4$_{-0.7}^{+0.7}$ & 28.0$\pm$0.2 & $>$1.8 & $-$7.7 & 0.99 & W22\\
CEERSYJ-9240348289 & 14:19:24.032 & 52:48:28.97 & 10.7$_{-0.8}^{+0.7}$ & 28.0$\pm$0.2 & $>$1.5 & $-$9.2 & 0.992 & D22,F22\\\\
\multicolumn{9}{c}{$z\sim13$ Selection}\\
\hline
CEERSH-9463556328 & 14:19:46.353 & 52:56:32.81 & 11.9$_{-0.5}^{+0.4}$ & 27.7$\pm$0.2 & 1.5$\pm$1.2 & $-$9.1 & 0.993 & D22,F22,H22,B22\\
GLASSP2H-3576218534 & 00:13:57.629 & $-$30:18:53.43 & 13.4$_{-0.8}^{+0.8}$ & 28.1$\pm$0.2 & $>$1.8 & $-$11.3 & 0.999 & B22\\
CEERSH-9220551139 & 14:19:22.054 & 52:51:13.94 & 13.5$_{-0.9}^{+0.8}$ & 27.6$\pm$0.2 & $>$1.9 & $-$3.4 & 0.94 & \\\\
\multicolumn{9}{c}{$z\sim17$ Selection}\\
\hline
CEERSK-9394956349 & 14:19:39.491 & 52:56:34.90 & 16.0$_{-0.2}^{+0.4}$ & 26.3$\pm$0.1 & 1.5$\pm$0.1 & $-$4.2 & 0.885 & D22,F22,H22,N22,B22\\
\hline \hline
\end{tabular}
\\$^1$ Ad22 = \citet{Adams2022}, At22 = \citet{Atek2022}, C22 = \citet{Castellano2022_GLASS}, D22 = \citet{Donnan2022}, H22 = \citet{Harikane2022_z9to17}, La22 = \citet{Labbe2022}, N22 = \citet{Naidu2022_z12,Naidu2022_z17}, W22 = \citet{Whitler2022_z10}, Y22 = \citet{Yan2022}, B22 = This Work (fiducial reductions: see Table~\ref{tab:cursample})\\
\end{table*}

\section{Assessments of $z\geq9$ Candidates in the Literature}

Given the considerable uncertainties regarding both the identity and
total number of high-quality $z\geq 9$ candidate galaxies that have
been identified to present, we have performed independent photometry
on $z\geq 8$ candidates reported in a large number of manuscripts, as
described in \S\ref{sec:othersel}.  Use was made both of a reduction
of the NIRCam imaging data with the \textsc{grizli} and
\textsc{pencil} pipelines.  We then categorized these $z\geq 9$
candidates in three categories ``robust,'' ``solid,'' and
``possible.''

The purpose of this appendix is to present candidates from the
literature which we classify as ``solid'' and ``possible.''  Those classified as ``robust'' have already been tabulated in Table~\ref{tab:robust}.  We place
those candidates in Tables~\ref{tab:solid}-\ref{tab:solid2} and
\ref{tab:possible}-\ref{tab:possible3}, respectively.

\begin{table*}
  \centering
\renewcommand{\arraystretch}{0.87} 
\caption{Sample of $z\sim10$, $z\sim13$, and $z\sim17$ candidate that we deem to be ``solid'' using our own photometry on our fiducial NIRCam reductions$^{*}$}
\vspace{-0.2cm}
\label{tab:solid}
\begin{tabular}{c|c|c|c|c|c|c|c|c} \hline
  ID & RA & DEC & $z_{phot}$ & $M_{UV}$ & Break [mag] & $\Delta \chi^2$ & $p(z>5.5)$ & Lit\\\hline\hline
\multicolumn{9}{c}{\citet{Castellano2022_GLASS}}\\
GHZ3 & 00:14:06.940 & $$-$$30:21:49.72 & 9.3/11.3$_{-0.4}^{+0.4}$ & $-$20.8 & $>$1.8,0.7$\pm$2.4 & $-$4.1,$-$1.3 & 0.862,0.814 &  B22\\
GHZ4 & 00:14:03.291 & $$-$$30:21:05.57 & 9.9/10.8$_{-1.1}^{+0.8}$ & $-$19.8 & $>$1.2,$>$1.7 & $-$5.5,$-$6.4 & 0.950,0.987 &  H22\\\\
\multicolumn{9}{c}{\citet{Adams2022}}\\
1514 & 07:22:27.506 & $$-$$73:28:38.61 & 9.9/10.0$_{-0.7}^{+0.7}$ & $-$19.7 & 1.9$\pm$0.8 & $-$1.0 & 0.811 &  M22\\
10234 & 07:22:39.597 & $$-$$73:30:06.25 & 11.5/11.7$_{-0.2}^{+0.3}$ & $-$20.6 & 1.2$\pm$0.2,1.3$\pm$0.3 & $-$7.9,$-$10.6 & 0.971,0.997\\\\
\multicolumn{9}{c}{\citet{Atek2022}}\\
SMACS\_z12b & 07:22:52.245 & $$-$$73:27:55.47 & 12.3/12.9$_{-0.9}^{+0.9}$ & $-$20.0 & 2.2$\pm$2.4,$>$1.4 & $-$4.8,$-$4.5 & 0.961,0.938 & Y22,B22\\
SMACS\_z16b & 07:22:39.416 & $$-$$73:30:08.15 & 15.3/14.9$_{-0.7}^{+0.8}$ & $-$21.0 & $>$1.7,$>$1.2 & $-$6.4,$-$6.3 & 0.971,0.952 &  B22\\\\
\multicolumn{9}{c}{\citet[][, v2]{Donnan2022}}\\
29274\_4 & 14:19:27.385 & 52:51:46.92 & 8.9/10.7$_{-2.7}^{+3.3}$ & $-$18.4 & $>$0.8,$>$1.3 & $-$0.9,$-$2.1 & 0.857,0.821\\
1434\_2 & 14:19:26.110 & 52:52:52.41 & 9.2/10.2$_{-0.6}^{+0.7}$ & $-$18.8 & $>$1.4,$>$1.4 & $-$3.6,$-$0.5 & 0.922,0.688\\
43866 & 07:23:25.597 & $$-$$73:26:12.41 & 9.5/7.9$_{-0.8}^{+0.7}$ & $-$18.1 & 1.2$\pm$0.7,1.5$\pm$0.4 & $-$3.7,$-$2.2 & 0.958,0.771\\
5071 & 07:22:56.854 & $$-$$73:29:23.48 & 9.5/8.3$_{-1.5}^{+1.7}$ & $-$18.0 & 0.3$\pm$0.8,0.5$\pm$0.6 & $-$2.1,$-$1.7 & 0.887,0.794\\
14391 & 07:22:47.724 & $$-$$73:28:28.31 & 9.5/9.2$_{-2.0}^{+1.4}$ & $-$18.8 & 1.3$\pm$0.3,1.9$\pm$0.7 & $-$0.1,$-$6.3 & 0.576,0.975\\
12682 & 07:22:38.948 & $$-$$73:28:30.37 & 9.6/8.9$_{-1.3}^{+1.3}$ & $-$18.9 & $>$1.6,$>$1.7 & $-$5.6,$-$6.7 & 0.985,0.990 &  \\
15019 & 07:22:58.265 & $$-$$73:28:19.53 & 9.7/8.7$_{-1.4}^{+1.5}$ & $-$18.7 & $>$1.0,$>$1.3 & $-$1.8,$-$1.3 & 0.870,0.784 &  \\
6200 & 07:22:41.502 & $$-$$73:29:10.61 & 9.8/8.9$_{-1.3}^{+1.3}$ & $-$18.5 & $>$1.3,$>$1.4 & $-$3.2,$-$3.3 & 0.932,0.931 &  \\
3763 & 07:22:49.132 & $$-$$73:29:31.17 & 9.9/8.0$_{-0.9}^{+1.0}$ & $-$19.0 & 2.0$\pm$1.5,$>$2.3 & $-$0.2,$-$4.7 & 0.767,0.978 & \\
7606 & 07:22:29.557 & $$-$$73:29:05.66 & 9.9/9.3$_{-1.5}^{+1.4}$ & $-$18.1 & $>$0.7,$>$0.6 & $-$3.8,$-$2.9 & 0.945,0.923\\
3710 & 14:19:24.026 & 52:48:28.99 & 10.4/10.3$_{-1.0}^{+1.0}$ & $-$19.1 & $>$1.2,$>$1.6 & $-$5.2,$-$10.6 & 0.965,0.996 &  F22,B22M\\
73150 & 14:19:26.780 & 52:54:16.61 & 10.6/10.4$_{-0.8}^{+0.8}$ & $-$19.1 & $>$1.4,$>$1.8 & $-$6.4,$-$11.8 & 0.973,0.998 &  F22\\
20757 & 07:23:12.463 & $$-$$73:28:01.71 & 10.7/12.8$_{-1.6}^{+1.6}$ & $-$17.9 & $>$1.3,$>$0.9 & 1.6,$-$2.4 & 0.545,0.899\\
26598 & 07:22:50.553 & $$-$$73:27:37.86 & 10.8/9.7$_{-1.6}^{+1.3}$ & $-$18.5 & $>$1.0,$>$1.3 & $-$3.2,$-$5.2 & 0.920,0.966\\
120880 & 14:20:10.558 & 52:59:39.51 & 10.8/10.4$_{-0.4}^{+0.5}$ & $-$19.4 & $>$1.2,$>$1.0 & $-$8.7,$-$6.8 & 0.983,0.969 &  WA22,F22\\
6415 & 00:14:00.275 & $$-$$30:21:25.90 & 10.8/10.6$_{-0.7}^{+0.6}$ & $-$19.1 & $>$1.9,$>$1.6 & $-$5.6,$-$5.4 & 0.975,0.972 &  H22,B22\\
622\_4 & 14:19:16.536 & 52:47:47.38 & 11.3/11.2$_{-1.1}^{+0.9}$ & $-$18.9 & $>$1.0,$>$1.4 & $-$4.3,$-$3.8 & 0.934,0.910\\
77241 & 14:19:41.462 & 52:54:41.51 & 11.3/11.2$_{-0.8}^{+0.7}$ & $-$19.3 & $>$1.0,1.1$\pm$1.8 & $-$10.5,$-$6.9 & 0.995,0.984 &  F22\\
33593\_2 & 14:19:37.588 & 52:56:43.85 & 11.3/11.4$_{-0.4}^{+0.3}$ & $-$19.6 & $>$1.4,$>$1.1 & $-$4.0,$-$2.3 & 0.753,0.721 &  F22\\
5268\_2 & 14:19:19.678 & 52:53:32.14 & 11.4/11.3$_{-0.6}^{+0.6}$ & $-$19.2 & 0.9$\pm$0.9,1.6$\pm$1.4 & $-$2.8,$-$5.1 & 0.840,0.959\\
8347 & 07:22:56.351 & $$-$$73:29:00.49 & 11.9/11.7$_{-0.5}^{+0.5}$ & $-$19.1 & 1.1$\pm$0.5,$>$1.4 & $-$4.5,$-$8.0 & 0.898,0.987\\
26409\_4 & 14:19:38.478 & 52:51:18.16 & 11.9/14.5$_{-2.3}^{+2.3}$ & $-$18.8 & 0.7$\pm$1.3,$>$0.9 & $-$3.4,$-$3.9 & 0.914,0.941\\
10566 & 07:23:03.469 & $$-$$73:28:46.97 & 12.0/13.7$_{-1.0}^{+0.9}$ & $-$19.7 & $>$1.6,$>$1.7 & $-$8.2,$-$5.9 & 0.991,0.978\\
27535\_4 & 14:19:27.307 & 52:51:29.20 & 12.6/12.3$_{-0.7}^{+1.0}$ & $-$19.4 & $>$1.9,$>$2.3 & $-$6.5,$-$6.9 & 0.970,0.987\\
93316 & 14:19:39.481 & 52:56:34.96 & 16.4/16.3$_{-0.4}^{+0.3}$ & $-$21.3 & 1.7$\pm$0.1,1.6$\pm$0.1 & $-$8.7,$-$7.2 & 0.987,0.962 &  H22,F22,B22,B22M,\\
 & & & & & & & & N22,Z22\\\\
\multicolumn{9}{c}{\citet[][, v1]{Donnan2022}}\\
28093 & 07:23:37.889 & $$-$$73:27:21.87 & 9.2/9.0$_{-1.4}^{+1.2}$ & $-$18.9 & $>$1.2,$>$1.5 & $-$5.6,$-$6.0 & 0.973,0.962\\
26225 & 14:19:11.028 & 52:50:22.44 & 9.2/9.6$_{-0.7}^{+0.8}$ & $-$18.7 & $>$1.2,$>$1.9 & 1.2,$-$1.8 & 0.562,0.816\\
78693 & 14:19:39.375 & 52:54:49.55 & 9.6/11.1$_{-1.5}^{+1.9}$ & $-$18.5 & $>$0.9,$>$1.0 & $-$3.2,$-$3.4 & 0.871,0.873\\
45704 & 14:19:28.719 & 52:51:36.87 & 10.0/10.9$_{-1.8}^{+2.2}$ & $-$18.6 & $>$0.8,$>$1.0 & $-$2.9,$-$3.4 & 0.860,0.891\\
30498 & 14:19:28.710 & 52:50:37.16 & 10.9/10.5$_{-0.9}^{+0.8}$ & $-$19.6 & $>$1.5,1.9$\pm$2.3 & $-$3.6,$-$6.2 & 0.921,0.979 &  F22\\
6486 & 07:22:53.246 & $$-$$73:29:08.77 & 12.6/12.8$_{-1.2}^{+1.3}$ & $-$18.6 & $>$1.2,$>$1.0 & $-$2.8,$-$2.4 & 0.845,0.879\\\\
\multicolumn{9}{c}{\citet{Labbe2022}}\\
2859 & 14:19:21.728 & 52:49:04.57 & 8.9/7.4$_{-0.9}^{+0.9}$ & $-$19.2 & 0.3$\pm$0.4,0.9$\pm$0.3 & $-$7.2,$-$5.3 & 0.996,0.989\\\\
\multicolumn{9}{c}{\citet[][, v2]{Morishita2022}}\\
WDF$-$C$-$1152 & 07:23:03.570 & $$-$$73:27:05.00 & 12.7/13.3$_{-1.7}^{+1.7}$ & $-$20.0 & 1.3$\pm$2.6,$>$1.9 & $-$2.0,$-$5.8 & 0.828,0.963\\
WDF$-$P$-$3504 & 07:22:47.117 & $$-$$73:28:36.68 & 13.4/9.4$_{-2.1}^{+1.6}$ & $-$19.4 & 1.0$\pm$1.4,$>$0.8 & $-$3.3,$-$4.8 & 0.879,0.962\\\\
\multicolumn{9}{c}{\citet[][, v2]{Harikane2022_z9to17}}\\
GL$-$z9$-$10 & 00:14:03.473 & $$-$$30:19:00.89 & 8.7/10.8$_{-1.7}^{+1.5}$ & $-$18.2 & $>$0.7,$>$1.4 & $-$4.6,$-$5.1 & 0.949,0.959 &  \\
GL$-$z9$-$13 & 00:13:57.453 & $$-$$30:17:59.97 & 8.7/8.3$_{-1.4}^{+2.5}$ & $-$18.1 & $>$1.5,$>$1.6 & $-$3.0,$-$4.9 & 0.813,0.942 &  \\
GL$-$z9$-$8 & 00:14:00.833 & $$-$$30:21:29.77 & 9.1/11.4$_{-1.2}^{+1.0}$ & $-$18.1 & $>$1.3,1.1$\pm$1.1 & 0.6,$-$6.5 & 0.541,0.975 &  \\
GL$-$z9$-$9 & 00:14:03.728 & $$-$$30:21:03.49 & 9.3/10.6$_{-2.5}^{+2.8}$ & $-$18.1 & $>$0.9,$>$0.2 & $-$2.8,$-$0.7 & 0.915,0.792 &  \\\\
\multicolumn{9}{c}{\citet[][, v1]{Harikane2022_z9to17}}\\
GL$-$z9$-$11 & 00:14:06.587 & $$-$$30:20:56.08 & 8.7/8.5$_{-1.0}^{+0.9}$ & $-$18.4 & $>$1.1,$>$1.4 & $-$4.5,$-$6.8 & 0.929,0.972\\
CR6$-$z12$-$1 & 14:19:17.510 & 52:49:35.38 & 12.0/11.9$_{-1.3}^{+1.4}$ & $-$18.1 & $>$1.2,$>$1.4 & $-$3.4,$-$6.7 & 0.915,0.981\\
\hline\hline
\end{tabular}
\begin{flushleft}
$^*$ Same remarks apply to this table as Table~\ref{tab:robust}
\end{flushleft}
\end{table*}

\begin{table*}
  \centering
\renewcommand{\arraystretch}{0.87} 
\caption{Sample of $z\sim10$, $z\sim13$, and $z\sim17$ candidate that we deem to be ``solid'' using our own photometry on our fiducial NIRCam reductions$^{*}$}
\vspace{-0.2cm}
\label{tab:solid2}
\begin{tabular}{c|c|c|c|c|c|c|c|c} \hline
  ID & RA & DEC & $z_{phot}$ & $M_{UV}$ & Break [mag] & $\Delta \chi^2$ & $p(z>5.5)$ & Lit\\\hline\hline
\multicolumn{9}{c}{\citet[][, v2]{Whitler2022_z10}}\\
EGS$-$37135 & 14:19:58.650 & 52:59:21.80 & 8.9/9.2$_{-0.2}^{+0.1}$ & $-$20.3 & 1.9$\pm$0.7,1.4$\pm$0.4 & $-$15.1,$-$7.1 & 0.999,0.970 &  F22,B22\\
EGS$-$7860 & 14:19:00.643 & 52:50:12.00 & 10.1/10.3$_{-0.7}^{+0.8}$ & $-$19.6 & $>$1.8,$>$1.8 & $-$7.8,$-$7.7 & 0.986,0.990 &  B22M\\\\
\multicolumn{9}{c}{\citet[][, v1]{Whitler2022_z10}}\\
EGS$-$38601 & 14:19:56.427 & 52:59:25.69 & 8.8/8.2$_{-0.2}^{+0.3}$ & $-$20.3 & 0.6$\pm$0.1,0.8$\pm$0.1 & $-$5.9,$-$7.4 & 0.935,0.980\\
EGS$-$11330 & 14:19:35.255 & 52:50:56.47 & 9.5/10.2$_{-1.0}^{+1.1}$ & $-$19.3 & $>$1.4,$>$1.8 & $-$1.8,$-$2.2 & 0.799,0.867\\\\
\multicolumn{9}{c}{\citet{Finkelstein2022_JWSTLF}}\\
CEERS2\_2274 & 14:19:23.077 & 52:53:38.41 & 8.6/8.4$_{-0.5}^{+0.5}$ & $-$18.4 & 0.7$\pm$1.5,0.8$\pm$3.2 & $-$9.4,$-$4.0 & 0.997,0.900\\
CEERS1\_5534 & 14:19:48.017 & 52:56:57.40 & 8.6/8.5$_{-0.4}^{+0.5}$ & $-$19.6 & 0.9$\pm$0.5,1.0$\pm$0.8 & $-$10.4,$-$3.8 & 0.992,0.958\\
CEERS6\_4012 & 14:19:33.148 & 52:51:32.37 & 8.9/9.0$_{-0.4}^{+0.4}$ & $-$19.9 & 1.1$\pm$0.8,1.2$\pm$0.8 & $-$4.7,$-$1.2 & 0.901,0.651\\
CEERS1\_1875 & 14:19:48.459 & 52:58:18.28 & 8.9/9.6$_{-0.9}^{+1.1}$ & $-$20.4 & $>$1.4,$>$1.4 & $-$3.4,$-$0.3 & 0.904,0.704\\
CEERS1\_3908 & 14:20:01.240 & 52:59:47.71 & 9.0/9.8$_{-0.9}^{+0.9}$ & $-$20.2 & $>$1.6,$>$1.9 & $-$10.0,$-$5.2 & 0.993,0.961 & B22,B22M\\
CEERS6\_7603 & 14:19:36.297 & 52:50:49.20 & 11.3/11.6$_{-1.1}^{+1.1}$ & $-$19.0 & 1.1$\pm$1.6,$>$1.0 & $-$3.9,$-$6.0 & 0.880,0.948 &  D22\\\\
\multicolumn{9}{c}{\citet{Yan2022}}\\
F150DB$-$026 & 07:23:23.735 & $$-$$73:27:40.59 & 11.4/10.5$_{-0.8}^{+0.8}$ & $-$19.4 & $>$1.1,$>$1.8 & $-$7.9,$-$3.7 & 0.989,0.965\\
F150DA$-$060 & 07:22:40.758 & $$-$$73:28:23.75 & 11.4/12.6$_{-0.9}^{+0.9}$ & $-$19.2 & $>$1.7,$>$1.2 & $-$5.3,0.3 & 0.957,0.550\\
F150DB$-$013 & 07:23:05.525 & $$-$$73:27:50.66 & 11.4/13.1$_{-1.0}^{+1.1}$ & $-$20.1 & $>$0.9,$>$1.6 & $-$1.2,$-$3.5 & 0.616,0.904\\
F150DB$-$084 & 07:23:07.540 & $$-$$73:26:23.79 & 11.6/12.2$_{-2.4}^{+2.6}$ & $-$19.4 & 0.8$\pm$2.5,$>$0.8 & $-$0.5,0.2 & 0.764,0.707\\
F150DA$-$038 & 07:23:02.955 & $$-$$73:28:46.16 & 13.4/12.0$_{-3.2}^{+2.8}$ & $-$19.2 & $>$0.9,$>$0.5 & $-$1.1,$-$0.8 & 0.792,0.719\\
F150DA$-$008 & 07:22:52.747 & $$-$$73:29:51.64 & 13.4/12.5$_{-4.8}^{+5.1}$ & $-$20.1 & $>$0.4,$>$0.1 & $-$0.3,$-$0.5 & 0.763,0.749\\
F150DA$-$007 & 07:22:44.875 & $$-$$73:29:53.68 & 13.4/13.4$_{-1.2}^{+1.2}$ & $-$19.5 & $>$1.8,1.0$\pm$1.0 & $-$2.2,$-$0.4 & 0.836,0.612\\
F200DA$-$089 & 07:22:32.427 & $$-$$73:28:06.77 & 14.2/14.8$_{-6.0}^{+4.2}$ & $-$19.0 & $>$0.7,$>$1.0 & $-$1.1,$-$1.4 & 0.751,0.766\\
F150DB$-$007 & 07:23:23.966 & $$-$$73:27:58.78 & 14.6/14.7$_{-1.0}^{+1.0}$ & $-$20.0 & 1.1$\pm$1.9,$>$1.4 & $-$0.4,$-$1.3 & 0.783,0.710\\
F150DB$-$048 & 07:23:01.562 & $$-$$73:27:18.02 & 15.0/12.1$_{-0.4}^{+0.2}$ & $-$21.0 & 1.5$\pm$0.3,1.1$\pm$0.4 & $-$7.9,$-$1.1 & 0.969,0.554\\
F200DB$-$086 & 07:23:06.415 & $$-$$73:27:19.85 & 15.4/14.2$_{-2.7}^{+3.6}$ & $-$20.2 & $>$1.2,$>$2.0 & $-$0.4,$-$2.4 & 0.787,0.915\\
F200DB$-$015 & 07:23:07.672 & $$-$$73:28:01.56 & 16.0/18.4$_{-1.1}^{+1.0}$ & $-$19.4 & $>$1.6,$>$1.5 & $-$10.3,$-$6.2 & 0.997,0.975\\
F200DA$-$034 & 07:23:05.208 & $$-$$73:29:13.37 & 19.8/10.0$_{-1.9}^{+1.4}$ & $-$19.2 & 0.5$\pm$1.0,$>$1.1 & 0.2,$-$0.6 & 0.616,0.817\\
F200DA$-$098 & 07:22:34.793 & $$-$$73:28:00.21 & 19.8/15.3$_{-4.4}^{+3.5}$ & $-$19.4 & $>$1.1,$>$0.4 & 0.6,$-$1.2 & 0.702,0.817\\\\
\multicolumn{9}{c}{This Work (Fiducial Sample)}\\
GLASSP1J$-$4003721456 & 00:14:00.378 & $$-$$30:21:45.54 & 9.2/9.2$_{-0.7}^{+0.8}$ & $-$19.2 & $>$1.6,$>$1.3 & $-$4.1,$-$4.4 & 0.953,0.964\\\\
\multicolumn{9}{c}{This Work (Alternate Sample)}\\
CEERSJ$-$9494458188 & 14:19:49.446 & 52:58:18.89 & 8.8/9.0$_{-0.5}^{+0.9}$ & $-$19.2 & 1.2$\pm$1.8,1.6$\pm$2.2 & $-$5.0,$-$9.4 & 0.913,0.993\\
GLASSP1J$-$4016420474 & 00:14:01.649 & $$-$$30:20:47.44 & 9.6/9.6$_{-0.4}^{+0.4}$ & $-$18.9 & $>$2.2 & $-$6.0 & 0.954\\
CEERSJ$-$9132749268 & 14:19:13.277 & 52:49:26.89 & 9.7/9.7$_{-0.6}^{+0.7}$ & $-$19.0 & $>$2.0 & $-$4.9 & 0.962\\
CEERSJ$-$9329250443 & 14:19:32.926 & 52:50:44.37 & 10.1/10.1$_{-0.6}^{+0.6}$ & $-$19.3 & $>$1.6,$>$2.1 & 0.4,$-$3.5 & 0.527,0.947\\
GLASSP1H$-$4015021230 & 00:14:01.507 & $$-$$30:21:22.85 & 13.3/11.3$_{-2.0}^{+3.0}$ & $-$18.9 & $>$0.9,$>$1.4 & $-$0.2,$-$6.3 & 0.741,0.968\\
\hline\hline
\end{tabular}
\begin{flushleft}
$^*$ Same remarks apply to this table as Table~\ref{tab:robust}
\end{flushleft}
\end{table*}

\begin{table*}
\centering
\caption{Sample of $z\sim10$, $z\sim13$, and $z\sim17$ candidate that we deem to be ``possible'' using our own photometry on two separate reductions of the NIRCam data.$^{*}$}
\label{tab:possible}
\begin{tabular}{c|c|c|c|c|c|c|c|c} \hline
  ID & RA & DEC & $z_{phot}$ & $M_{UV}$ & Break [mag] & $\Delta \chi^2$ & $p(z>5.5)$ & Lit\\\hline\hline
\multicolumn{8}{c}{\citet[][, v2]{Castellano2022_GLASS}}\\
GHZ6 & 00:13:54.970 & $-$30:18:53.68 & 9.1/9.7$_{-2.0}^{+4.5}$ & $-$20.1 & 0.9$\pm$1.0,$>$1.0 & 3.8,1.4 & 0.063,0.382 &  \\
GHZ5 & 00:13:58.661 & $-$30:18:27.39 & 9.2/7.2$_{-0.9}^{+1.9}$ & $-$20.2 & 1.0$\pm$1.0,1.6$\pm$1.0 & 5.4,6.2 & 0.116,0.049 &  \\\\
\multicolumn{8}{c}{\citet[][, v1]{Castellano2022_GLASS}}\\
GHZ6 & 00:13:57.689 & $-$30:19:37.71 & 9.1/11.2$_{-4.5}^{+4.8}$ & $-$19.4 & 0.6$\pm$3.4,1.1$\pm$1.9 & 1.7,18.2 & 0.394,0.001\\\\
\multicolumn{8}{c}{\citet[][, v2]{Atek2022}}\\
SMACS\_z10d & 07:22:46.695 & $-$73:28:40.88 & 9.3/10.8$_{-0.3}^{+0.6}$ & $-$19.8 & $>$1.4,0.9$\pm$1.2 & $-$1.4,3.4 & 0.664,0.129 & \\
SMACS\_z10e & 07:22:45.296 & $-$73:29:30.52 & 10.9/10.3$_{-2.5}^{+1.3}$ & $-$18.9 & $>$0.9,$>$1.4 & 2.0,1.2 & 0.328,0.461 & \\
SMACS\_z11a & 07:22:39.500 & $-$73:29:40.20 & 11.1/10.6$_{-0.5}^{+0.7}$ & $-$18.6 & $>$1.5,1.0$\pm$0.9 & $-$2.0,2.4 & 0.714,0.230 & \\
SMACS\_z12a & 07:22:47.373 & $-$73:30:01.76 & 12.2/12.7$_{-1.2}^{+2.1}$ & $-$19.8 & 1.0$\pm$0.9,1.2$\pm$2.3 & 2.6,0.9 & 0.283,0.556 & \\
SMACS\_z16a & 07:23:26.387 & $-$73:28:04.55 & 15.9/10.6$_{-1.1}^{+0.6}$ & $-$20.6 & $>$1.1,$>$1.1 & 1.7,0.3 & 0.311,0.523 & \\\\
\multicolumn{8}{c}{\citet[][, v1]{Atek2022}}\\
SMACS\_z10f & 07:22:57.297 & $-$73:29:16.49 & 10.5/8.9$_{-1.9}^{+1.8}$ & $-$19.2 & 1.5$\pm$2.1,1.0$\pm$0.4 & 0.8,1.8 & 0.525,0.201\\
SMACS\_z11b & 07:22:53.835 & $-$73:28:23.28 & 11.2/9.7$_{-2.0}^{+1.2}$ & $-$22.4 & 1.7$\pm$1.3 & 1.7 & 0.305\\
SMACS\_z11c & 07:23:01.597 & $-$73:26:54.89 & 11.2/10.9$_{-1.0}^{+1.0}$ & $-$19.6 & $>$0.9,$>$1.8 & $-$0.8,$-$0.1 & 0.530,0.540\\
SMACS\_z11d & 07:22:36.750 & $-$73:28:09.17 & 11.3/10.7$_{-0.3}^{+0.5}$ & $-$21.8 & 1.9$\pm$2.6,0.7$\pm$0.2 & 0.3,5.7 & 0.560,0.029\\
SMACS\_z11e & 07:22:49.255 & $-$73:27:44.56 & 11.5/12.3$_{-1.6}^{+2.3}$ & $-$19.2 & 1.0$\pm$1.7,1.2$\pm$1.4 & $-$1.0,1.8 & 0.675,0.621 &  Y22\\\\
\multicolumn{8}{c}{\citet[][, v2]{Donnan2022}}\\
3398 & 07:22:35.369 & $-$73:29:38.62 & 9.7/8.2$_{-1.3}^{+1.4}$ & $-$18.2 & $>$1.4,$>$1.3 & 1.4,$-$0.5 & 0.608,0.717\\
4063 & 07:22:52.307 & $-$73:29:32.38 & 10.4/8.1$_{-1.4}^{+1.7}$ & $-$18.0 & $>$1.0,$>$1.2 & 3.4,1.4 & 0.411,0.526\\
6647 & 14:19:14.663 & 52:48:49.77 & 10.4/9.8$_{-1.3}^{+1.6}$ & $-$18.9 & $>$1.1,$>$1.4 & 1.8,$-$0.5 & 0.295,0.528 &  \\
21071\_2 & 14:19:36.716 & 52:55:22.66 & 10.7/10.2$_{-1.2}^{+1.0}$ & $-$19.3 & 0.9$\pm$1.2,1.0$\pm$1.7 & 5.8,2.8 & 0.112,0.200\\
61486 & 14:19:23.727 & 52:53:00.98 & 11.2/10.5$_{-2.0}^{+2.8}$ & $-$19.6 & $>$1.7,$>$1.6 & 2.0,$-$0.9 & 0.510,0.683 & \\
1566 & 07:22:39.183 & $-$73:30:00.55 & 12.3/10.0$_{-2.6}^{+2.6}$ & $-$18.8 & 1.0$\pm$1.1,1.1$\pm$2.1 & 2.5,0.8 & 0.352,0.561\\\\
\multicolumn{8}{c}{\citet[][, v1]{Donnan2022}}\\
2873 & 14:19:21.556 & 52:48:20.82 & 9.0/9.6$_{-1.4}^{+1.2}$ & $-$18.8 & $>$1.2,1.9$\pm$1.8 & 1.3,3.5 & 0.468,0.326\\
9544 & 07:22:38.627 & $-$73:28:46.53 & 9.1/9.0$_{-2.2}^{+1.7}$ & $-$19.0 & 1.2$\pm$0.5,1.9$\pm$1.2 & 4.1,1.1 & 0.181,0.536\\
111451 & 14:19:50.598 & 52:58:48.64 & 9.1/9.2$_{-0.8}^{+1.0}$ & $-$18.9 & $>$1.5,0.8$\pm$1.2 & 1.1,1.4 & 0.568,0.433\\
106309 & 14:19:52.004 & 52:58:05.85 & 9.2/9.0$_{-0.6}^{+0.8}$ & $-$18.8 & 1.8$\pm$2.3,$>$1.5 & 0.3,0.2 & 0.718,0.631 &  F22\\
110933 & 14:20:00.345 & 52:58:44.25 & 9.3/8.9$_{-1.3}^{+1.3}$ & $-$18.7 & $>$1.2,1.1$\pm$1.5 & 3.9,6.0 & 0.356,0.204\\
108408 & 14:19:46.666 & 52:58:21.42 & 9.3/9.7$_{-0.7}^{+0.9}$ & $-$19.1 & $>$1.7,$>$1.7 & 1.8,1.7 & 0.355,0.483\\
107364 & 14:19:49.568 & 52:58:13.33 & 10.2/8.6$_{-0.4}^{+0.5}$ & $-$19.1 & 0.8$\pm$0.6,0.8$\pm$0.8 & 7.8,5.3 & 0.027,0.055\\
20311 & 14:18:59.916 & 52:49:56.40 & 10.8/10.7$_{-0.8}^{+0.6}$ & $-$19.3 & 1.8$\pm$1.7,1.2$\pm$0.5 & $-$1.1,7.3 & 0.727,0.030\\
78598 & 14:19:30.185 & 52:54:48.79 & 11.0/12.0$_{-1.2}^{+1.7}$ & $-$19.5 & 1.1$\pm$2.9,$>$1.1 & 2.4,3.1 & 0.418,0.279\\
35470 & 07:23:02.966 & $-$73:26:47.49 & 12.0/12.4$_{-1.2}^{+2.1}$ & $-$19.2 & $>$1.4,$>$1.8 & $-$0.1,0.8 & 0.581,0.567\\
21901 & 07:22:46.756 & $-$73:27:49.43 & 12.2/12.4$_{-0.8}^{+1.0}$ & $-$19.0 & 1.5$\pm$1.5,$>$1.9 & 1.2,0.4 & 0.587,0.697\\
40079 & 07:23:13.894 & $-$73:26:05.07 & 14.3/12.3$_{-1.0}^{+2.6}$ & $-$19.6 & $>$1.1,$>$0.8 & 3.2,$-$0.1 & 0.354,0.595\\\\
\multicolumn{8}{c}{\citet[][, v2]{Morishita2022}}\\
WDF$-$P$-$3004 & 07:22:53.064 & $-$73:28:07.66 & 10.0/8.3$_{-1.5}^{+1.7}$ & $-$18.9 & $>$1.3,$>$1.9 & 0.0,$-$0.3 & 0.649,0.765\\
WDF$-$C$-$1730 & 07:23:22.754 & $-$73:26:25.63 & 13.8/11.4$_{-0.7}^{+0.8}$ & $-$19.8 & $>$1.0,$>$1.7 & 1.9,$-$2.6 & 0.296,0.859 &  Y22\\\\
\multicolumn{8}{c}{\citet[][, v2]{Harikane2022_z9to17}}\\
GL$-$z9$-$5 & 00:14:03.114 & $-$30:22:26.29 & 8.7/8.6$_{-0.8}^{+0.9}$ & $-$18.8 & $>$1.7,$>$1.0 & $-$0.4,0.5 & 0.728,0.471 & \\
GL$-$z9$-$3 & 00:14:00.093 & $-$30:19:06.89 & 8.9/5.7$_{-0.3}^{+0.5}$ & $-$18.8 & 0.9$\pm$2.4,1.2$\pm$0.1 & $-$0.1,$-$0.1 & 0.520,0.478 & \\
GL$-$z9$-$6 & 00:14:04.398 & $-$30:20:39.78 & 9.0/8.8$_{-0.4}^{+1.1}$ & $-$18.9 & $>$2.0,$>$2.2 & 3.1,$-$5.8 & 0.286,0.969 &  B22M\\
GL$-$z9$-$12 & 00:14:06.869 & $-$30:22:01.96 & 9.1/7.3$_{-0.8}^{+1.0}$ & $-$18.2 & 1.6$\pm$2.0,$>$0.2 & 0.7,0.5 & 0.507,0.577 &  \\
GL$-$z9$-$11 & 00:14:02.489 & $-$30:22:00.91 & 9.9/10.3$_{-2.6}^{+3.3}$ & $-$18.6 & 0.6$\pm$2.1,$>$0.5 & 1.9,0.4 & 0.308,0.637\\
GL$-$z9$-$7 & 00:14:02.533 & $-$30:21:57.09 & 10.3/10.7$_{-2.2}^{+4.3}$ & $-$18.2 & 1.2$\pm$1.1,$>$1.5 & 1.4,$-$2.1 & 0.479,0.845\\
CR2$-$z12$-$3 & 14:19:41.606 & 52:55:07.60 & 11.7/12.0$_{-1.4}^{+1.7}$ & $-$19.2 & $>$0.6,0.6$\pm$1.1 & $-$1.1,0.4 & 0.781,0.607\\
CR2$-$z12$-$2 & 14:19:42.567 & 52:54:42.03 & 12.0/10.7$_{-2.2}^{+1.7}$ & $-$19.0 & 1.1$\pm$1.8,$>$1.3 & 5.7,3.1 & 0.241,0.429\\
CR2$-$z12$-$4 & 14:19:24.858 & 52:53:13.93 & 12.1/11.3$_{-3.0}^{+3.2}$ & $-$19.0 & $>$0.3,0.4$\pm$1.0 & $-$0.7,0.6 & 0.733,0.649\\
SM$-$z12$-$1 & 07:22:32.471 & $-$73:28:33.18 & 12.5/10.2$_{-1.4}^{+0.9}$ & $-$18.5 & $>$1.3,$>$1.2 & 1.7,0.0 & 0.408,0.744 &  Y22\\
\hline\hline
\end{tabular}
\begin{flushleft}
$^*$ Same remarks apply to this table as Table~\ref{tab:robust}
\end{flushleft}
\end{table*}

\begin{table*}
\centering
\caption{Sample of $z\sim10$, $z\sim13$, and $z\sim17$ candidate that we deem to be ``possible'' using our own photometry on two separate reductions of the NIRCam data.$^*$}
\label{tab:possible2}
\begin{tabular}{c|c|c|c|c|c|c|c|c} \hline
  ID & RA & DEC & $z_{phot}$ & $M_{UV}$ & Break [mag] & $\Delta \chi^2$ & $p(z>5.5)$ & Lit\\\hline\hline
\multicolumn{8}{c}{\citet[][, v1]{Harikane2022_z9to17}}\\
GL$-$z9$-$5 & 00:13:58.209 & $-$30:21:34.29 & 9.2/9.0$_{-0.2}^{+1.7}$ & $-$18.7 & $>$1.5,1.5$\pm$1.6 & 15.2,19.1 & 0.001,0.000\\
GL$-$z9$-$15 & 00:14:00.487 & $-$30:21:24.49 & 9.8/8.9$_{-2.2}^{+2.5}$ & $-$18.1 & $>$0.8,$>$1.1 & 2.0,0.9 & 0.452,0.562\\
SM$-$z12$-$1 & 07:22:58.264 & $-$73:28:32.99 & 11.5/13.0$_{-1.3}^{+1.5}$ & $-$18.0 & $>$1.1,1.2$\pm$2.7 & $-$1.7,$-$0.6 & 0.793,0.647\\
CR2$-$z12$-$2 & 14:19:24.237 & 52:54:24.81 & 11.6/6.8$_{-0.5}^{+1.0}$ & $-$19.4 & 0.9$\pm$1.8,$>$0.6 & $-$0.7,4.5 & 0.692,0.255\\
CR3$-$z12$-$1 & 14:19:11.106 & 52:49:33.70 & 11.7/10.1$_{-1.8}^{+1.8}$ & $-$18.3 & $>$1.0,$>$1.1 & $-$1.5,0.3 & 0.772,0.522 &  H22\\
CR2$-$z12$-$3 & 14:19:30.405 & 52:52:51.35 & 12.1/12.7$_{-1.4}^{+1.9}$ & $-$19.2 & $>$0.6,$>$1.0 & 1.3,$-$0.4 & 0.524,0.674\\\\
\multicolumn{8}{c}{\citet[][, v2]{Whitler2022_z10}}\\
EGS$-$34362 & 14:20:03.000 & 53:00:04.88 & 9.2/9.1$_{-0.5}^{+0.4}$ & $-$20.2 & 1.4$\pm$0.8 & 5.5 & 0.098\\
EGS$-$14506 & 14:19:13.719 & 52:51:44.49 & 10.7/9.2$_{-0.7}^{+1.1}$ & $-$20.2 & 1.0$\pm$0.6,0.9$\pm$0.3 & 2.3,5.5 & 0.331,0.051\\\\
\multicolumn{8}{c}{\citet[][, v1]{Whitler2022_z10}}\\
EGS$-$36603 & 14:20:09.204 & 52:59:54.73 & 9.3/7.9$_{-1.0}^{+1.0}$ & $-$19.9 & 0.7$\pm$1.5 & $-$1.4 & 0.731\\
EGS$-$42956 & 14:19:48.045 & 52:58:27.52 & 10.4/9.2$_{-0.9}^{+1.5}$ & $-$19.9 & 1.3$\pm$1.6,1.5$\pm$2.3 & 1.2,0.8 & 0.396,0.421\\\\
\multicolumn{8}{c}{\citet{Finkelstein2022_JWSTLF}}\\
CEERS1\_7227 & 14:20:08.996 & 52:59:57.79 & 11.2/10.9$_{-0.6}^{+0.7}$ & $-$19.2 & 0.8$\pm$2.3 & 0.6 & 0.335\\
CEERS1\_1730 & 14:20:02.397 & 53:00:49.09 & 13.4/12.7$_{-1.0}^{+1.2}$ & $-$20.2 & 1.0$\pm$0.9 & 1.6 & 0.572\\\\
\multicolumn{8}{c}{\citet{Yan2022}}\\
F150DB$-$C\_4 & 07:23:25.966 & $-$73:26:39.89 & 10.4/9.8$_{-1.2}^{+2.1}$ & $-$19.9 & 1.8$\pm$0.1,1.8$\pm$0.2 & 3.8,4.4 & 0.215,0.151\\
F150DB$-$040 & 07:23:11.937 & $-$73:27:24.93 & 10.8/7.7$_{-1.5}^{+3.4}$ & $-$20.5 & 0.5$\pm$1.6,0.9$\pm$1.6 & 5.3,6.2 & 0.065,0.069\\
F150DA$-$026 & 07:22:46.013 & $-$73:29:08.12 & 11.0/8.2$_{-1.8}^{+3.0}$ & $-$19.1 & 0.2$\pm$0.8,0.9$\pm$3.2 & 1.1,2.3 & 0.258,0.380\\
F150DA$-$036 & 07:23:00.669 & $-$73:28:48.71 & 11.0/11.2$_{-0.4}^{+0.7}$ & $-$19.2 & 0.9$\pm$0.4,0.8$\pm$1.1 & 3.0,3.6 & 0.219,0.205\\
F150DA$-$020 & 07:22:55.876 & $-$73:29:17.45 & 11.2/7.9$_{-1.7}^{+3.7}$ & $-$19.3 & 0.7$\pm$1.0,$>$0.8 & 1.6,1.6 & 0.179,0.399\\
F150DA$-$005 & 07:22:41.008 & $-$73:29:54.98 & 11.2/9.8$_{-2.7}^{+1.9}$ & $-$19.4 & 0.7$\pm$0.3,0.6$\pm$0.4 & 6.5,4.8 & 0.029,0.041\\
F150DA$-$062 & 07:22:54.222 & $-$73:28:23.56 & 11.4/8.6$_{-1.9}^{+2.2}$ & $-$19.9 & $>$0.9,0.9$\pm$2.8 & 1.4,0.6 & 0.437,0.514\\
F150DA$-$057 & 07:22:58.715 & $-$73:28:28.37 & 11.4/9.9$_{-3.3}^{+1.6}$ & $-$20.9 & 1.0$\pm$1.4,1.0$\pm$1.0 & 8.4,9.2 & 0.019,0.014\\
F150DA$-$066 & 07:22:39.616 & $-$73:28:12.16 & 11.4/10.5$_{-1.1}^{+0.9}$ & $-$19.8 & 1.0$\pm$0.3,0.9$\pm$1.2 & 7.0,6.4 & 0.080,0.135\\
F150DB$-$075 & 07:23:02.229 & $-$73:26:41.51 & 11.4/10.8$_{-0.4}^{+0.8}$ & $-$21.2 & $>$1.5,$>$2.1 & 5.5,4.0 & 0.076,0.110\\
F150DA$-$054 & 07:22:38.904 & $-$73:28:30.84 & 11.4/11.1$_{-0.6}^{+0.8}$ & $-$19.2 & 0.9$\pm$0.4,1.4$\pm$1.8 & 0.2,$-$2.2 & 0.481,0.826\\
F150DB$-$090 & 07:23:26.236 & $-$73:26:13.82 & 11.4/11.3$_{-0.2}^{+0.3}$ & $-$21.6 & 1.2$\pm$0.6,1.3$\pm$0.7 & 5.3,7.0 & 0.026,0.011\\
F150DA$-$052 & 07:22:26.930 & $-$73:28:33.80 & 11.4/12.2$_{-1.2}^{+2.1}$ & $-$19.3 & 1.0$\pm$1.2 & $-$0.4 & 0.591\\
F150DA$-$031 & 07:22:40.648 & $-$73:29:00.50 & 11.4/13.1$_{-1.4}^{+1.5}$ & $-$19.6 & 1.1$\pm$0.7,1.1$\pm$1.0 & 1.5,$-$1.3 & 0.632,0.664\\
F150DB$-$054 & 07:23:12.503 & $-$73:27:10.74 & 11.4/15.0$_{-1.2}^{+2.3}$ & $-$19.3 & 1.0$\pm$1.0,0.7$\pm$0.5 & 3.4,7.4 & 0.265,0.026\\
F150DB$-$076 & 07:23:29.414 & $-$73:26:39.77 & 11.6/6.0$_{-0.1}^{+0.4}$ & $-$19.6 & 0.5$\pm$0.5,0.7$\pm$0.5 & 8.3,8.5 & 0.008,0.006\\
F150DB$-$050 & 07:23:24.576 & $-$73:27:15.06 & 11.6/6.7$_{-0.2}^{+0.1}$ & $-$19.5 & 0.8$\pm$0.7,1.4$\pm$1.4 & 14.5,18.3 & 0.001,0.000\\
F150DB$-$044 & 07:23:39.315 & $-$73:27:22.28 & 11.6/7.0$_{-0.7}^{+0.0}$ & $-$19.6 & 0.8$\pm$0.3,0.4$\pm$0.3 & 20.1,20.2 & 0.000,0.000\\
F150DB$-$095 & 07:23:24.766 & $-$73:26:01.29 & 11.6/9.3$_{-2.4}^{+2.1}$ & $-$19.4 & 1.4$\pm$2.5,1.7$\pm$1.0 & 3.9,2.1 & 0.156,0.138\\
F150DB$-$011 & 07:23:27.382 & $-$73:27:58.03 & 11.6/10.8$_{-0.0}^{+1.0}$ & $-$19.5 & 0.7$\pm$0.3,0.7$\pm$0.5 & 7.4,7.4 & 0.030,0.041\\
F150DA$-$019 & 07:22:39.395 & $-$73:29:20.47 & 11.6/11.1$_{-0.7}^{+0.6}$ & $-$19.4 & $>$0.8,$>$1.2 & 4.4,1.9 & 0.165,0.400\\
F150DB$-$088 & 07:23:14.037 & $-$73:26:17.28 & 11.6/11.5$_{-0.1}^{+0.3}$ & $-$19.7 & 1.1$\pm$0.9,1.1$\pm$0.2 & 10.4,10.5 & 0.003,0.004\\
F150DB$-$031 & 07:23:21.438 & $-$73:27:36.32 & 11.6/13.1$_{-1.5}^{+1.8}$ & $-$19.2 & $>$1.2,$>$1.7 & $-$0.3,1.0 & 0.649,0.443\\
F150DB$-$021 & 07:23:12.643 & $-$73:27:45.24 & 11.8/6.2$_{-0.0}^{+0.5}$ & $-$20.5 & 0.5$\pm$0.3,0.3$\pm$0.2 & 0.6,0.8 & 0.265,0.204\\
F150DA$-$078 & 07:22:49.244 & $-$73:27:49.86 & 11.8/8.1$_{-1.5}^{+2.0}$ & $-$19.8 & 1.1$\pm$1.8,1.7$\pm$2.6 & 0.7,2.1 & 0.447,0.158\\
F150DB$-$069 & 07:23:04.258 & $-$73:26:54.20 & 11.8/11.2$_{-0.7}^{+0.6}$ & $-$19.4 & 1.0$\pm$0.6,$>$1.8 & 1.6,$-$2.2 & 0.468,0.784\\
F150DA$-$015 & 07:22:44.735 & $-$73:29:26.86 & 11.8/11.5$_{-0.5}^{+0.4}$ & $-$19.3 & $>$0.9,0.7$\pm$0.5 & $-$1.0,3.7 & 0.574,0.143\\
F150DA$-$083 & 07:22:42.714 & $-$73:27:32.28 & 11.8/13.2$_{-6.3}^{+3.4}$ & $-$20.6 & 1.0$\pm$0.8,$>$0.9 & 2.6,1.7 & 0.358,0.351\\
F150DA$-$010 & 07:22:40.082 & $-$73:29:46.12 & 12.8/10.8$_{-0.2}^{+1.1}$ & $-$19.4 & 0.7$\pm$0.4,0.3$\pm$0.7 & 4.1,5.3 & 0.107,0.064\\
F150DA$-$058 & 07:22:48.273 & $-$73:28:27.35 & 13.4/8.6$_{-2.1}^{+3.1}$ & $-$20.0 & $>$0.7,$>$0.8 & 2.0,1.7 & 0.266,0.334\\
F150DA$-$075 & 07:22:38.343 & $-$73:27:57.09 & 13.4/9.4$_{-2.0}^{+1.9}$ & $-$19.8 & 0.5$\pm$0.7,$>$1.0 & $-$0.2,2.0 & 0.751,0.535\\
F150DA$-$050 & 07:22:44.994 & $-$73:28:36.88 & 13.4/10.0$_{-1.9}^{+1.1}$ & $-$19.9 & $>$1.5,0.8$\pm$0.4 & $-$0.6,2.1 & 0.500,0.276\\
F150DB$-$079 & 07:23:13.156 & $-$73:26:29.64 & 13.8/12.5$_{-1.2}^{+1.8}$ & $-$19.8 & 1.3$\pm$2.2,$>$1.5 & $-$0.1,0.1 & 0.639,0.595\\
F150DB$-$004 & 07:23:14.299 & $-$73:28:06.75 & 14.0/11.7$_{-0.5}^{+2.5}$ & $-$19.6 & 0.9$\pm$0.9,0.9$\pm$0.9 & 8.6,4.2 & 0.017,0.232\\
F150DB$-$033 & 07:23:30.548 & $-$73:27:33.10 & 14.8/10.5$_{-0.2}^{+0.9}$ & $-$19.9 & $>$1.4,$>$1.5 & 3.5,6.1 & 0.109,0.088\\
F150DB$-$052 & 07:23:28.136 & $-$73:27:13.86 & 15.0/11.5$_{-0.3}^{+0.5}$ & $-$20.5 & 0.9$\pm$0.3,$>$1.5 & 3.4,3.4 & 0.133,0.283\\
F150DB$-$058 & 07:23:24.096 & $-$73:27:09.81 & 15.2/8.8$_{-1.2}^{+1.1}$ & $-$20.6 & $>$0.9,$>$1.6 & 6.7,0.9 & 0.127,0.692\\
F200DA$-$056 & 07:22:37.024 & $-$73:28:41.57 & 15.6/6.2$_{-0.1}^{+5.0}$ & $-$19.3 & 0.9$\pm$0.4,1.0$\pm$0.6 & 4.5,2.5 & 0.028,0.052\\
\hline\hline
\end{tabular}
\begin{flushleft}
$^*$ Same remarks apply to this table as Table~\ref{tab:robust}
\end{flushleft}
\end{table*}

\begin{table*}
\centering
\caption{Sample of $z\sim10$, $z\sim13$, and $z\sim17$ candidate that we deem to be ``possible'' using our own photometry on two separate reductions of the NIRCam data.$^{*}$}
\label{tab:possible3}
\begin{tabular}{c|c|c|c|c|c|c|c|c} \hline
  ID & RA & DEC & $z_{phot}$ & $M_{UV}$ & Break [mag] & $\Delta \chi^2$ & $p(z>
5.5)$ & Lit\\\hline\hline
\multicolumn{8}{c}{\citet{Yan2022}}\\
F200DB$-$109 & 07:23:37.033 & $-$73:27:12.22 & 15.8/7.2$_{-0.9}^{+2.7}$ & $-$19.2 & 0.3$\pm$1.6,0.2$\pm$0.9 & 1.0,2.9 & 0.397,0.188\\
F200DB$-$181 & 07:23:12.615 & $-$73:26:31.71 & 15.8/14.5$_{-1.4}^{+1.2}$ & $-$20.5 & $>$0.9,$>$1.1 & $-$1.9,2.7 & 0.790,0.290\\
F200DA$-$061 & 07:22:31.695 & $-$73:28:38.65 & 15.8/16.6$_{-1.1}^{+1.8}$ & $-$19.6 & 1.4$\pm$1.2,1.1$\pm$0.9 & 2.1,0.5 & 0.431,0.589\\
F150DB$-$041 & 07:23:06.626 & $-$73:27:25.43 & 16.0/11.6$_{-0.3}^{+0.3}$ & $-$20.0 & $>$1.1,$>$1.8 & 0.6,3.2 & 0.546,0.214\\
F200DB$-$159 & 07:23:25.344 & $-$73:26:46.02 & 16.0/13.2$_{-5.6}^{+4.8}$ & $-$19.3 & $>$1.1,$>$1.5 & 2.2,$-$0.7 & 0.512,0.782\\
F200DB$-$175 & 07:23:11.086 & $-$73:26:38.01 & 16.2/12.8$_{-1.1}^{+1.3}$ & $-$19.9 & $>$1.6,$>$1.6 & 2.2,1.8 & 0.387,0.345\\
F200DA$-$040 & 07:23:03.926 & $-$73:29:06.14 & 20.0/9.2$_{-2.1}^{+1.8}$ & $-$19.5 & $>$0.7,0.7$\pm$0.6 & 0.8,1.1 & 0.532,0.691\\
F200DB$-$045 & 07:23:22.766 & $-$73:27:39.69 & 20.4/14.2$_{-0.8}^{+3.7}$ & $-$20.3 & 1.4$\pm$1.1,$>$1.3 & 0.4,0.2 & 0.527,0.518\\
F200DA$-$006 & 07:22:40.349 & $-$73:30:10.33 & 20.6/9.2$_{-2.5}^{+3.0}$ & $-$19.6 & 0.6$\pm$0.3,$>$0.9 & 1.0,1.9 & 0.336,0.419\\\\
\multicolumn{8}{c}{This Work (Fiducial Sample)}\\
CEERSJ$-$9345150450 & 14:19:34.513 & 52:50:45.08 & 9.2/9.2$_{-0.6}^{+0.8}$ & $-$19.5 & 1.5$\pm$1.0,1.0$\pm$0.4 & $-$10.5,2.0 & 0.997,0.424\\
CEERSJ$-$9203050435 & 14:19:20.301 & 52:50:43.60 & 9.8/9.8$_{-0.7}^{+0.8}$ & $-$19.0 & $>$1.7,0.8$\pm$0.4 & $-$8.9,6.9 & 0.989,0.044\\
CEERSJ$-$9149352106 & 14:19:14.931 & 52:52:10.66 & 10.0/10.0$_{-0.7}^{+0.7}$ & $-$19.5 & 1.8$\pm$2.1,1.1$\pm$0.5 & $-$4.7,2.7 & 0.956,0.286\\
CEERSJ$-$9026550577 & 14:19:02.650 & 52:50:57.75 & 11.2/11.3$_{-0.5}^{+0.4}$ & $-$18.9 & $>$1.5,0.9$\pm$0.4 & $-$6.9,13.3 & 0.983,0.002\\\\
\multicolumn{8}{c}{This Work (Alternate Sample)}\\
GLASSP1J$-$4045920571 & 00:14:04.599 & $-$30:20:57.10 & 9.4/8.0$_{-1.2}^{+1.0}$ & $-$19.2 & $>$1.9,$>$2.3 & 1.4,$-$3.4 & 0.420,0.885\\
CEERSJ$-$9320850526 & 14:19:32.076 & 52:50:52.65 & 9.6/8.1$_{-1.0}^{+1.2}$ & $-$19.4 & 1.2$\pm$1.2,$>$2.3 & 1.8,$-$3.4 & 0.387,0.930\\
CEERSH$-$9220551139 & 14:19:22.049 & 52:51:13.94 & 13.5/12.2$_{-1.0}^{+1.5}$ & $-$19.5 & $>$1.5,$>$1.9 & 2.5,$-$4.9 & 0.237,0.971\\
\hline\hline
\end{tabular}
\begin{flushleft}
$^*$ Same remarks apply to this table as Table~\ref{tab:robust}
\end{flushleft}
\end{table*}

\section{Median Stack of $z\geq9$ Candidates with Various Quality Flags}

In evaluating the quality of $z\geq 9$ candidates from the literature,
we place the candidates in three different categories ``robust,''
``solid,'' and ``possible'' depending on the relative likelihood we
estimate for these sources to lie at $z<5.5$ or $z>5.5$ using our own
photometry (see \S\ref{sec:othersel} for how these sets are defined).

In order to interpret each of these designations and determine if the
differences are meaningful, we construct median stacks of flux in
different passbands and for candidates in different categories.  The
results are presented in Table~\ref{tab:medianstacks}.

While the median stack results in the ``robust'' and ``solid''
categories show no significant flux blueward of the nominal spectral
breaks in the $z\geq 8$ candidates, the median stack in the
``possible'' category does show tentative 1-1.5$\sigma$ detections in
each of bands blueward of the break.  Additionally, the median stack
results in the ``robust'' and ``solid'' category shows larger spectral
breaks, i.e., $>$1.8 mag, than the median stack results in the
``possible'' category show, where the break only has an amplitude of
$\sim$1-1.5 mag.

\begin{table}
\centering
\caption{Median stack results for $z\geq 8$ candidates from the literature which we segregate into the ``robust,'' ``solid,'' and ``possible'' categories based on our own photometry and SED fits.}
\label{tab:medianstacks}
\begin{tabular}{c|c|c|c} \hline
  & \multicolumn{3}{c}{Median Flux (nJy)}\\
  Band       & ``Robust'' & ``Solid'' & ``Possible'' \\
  \hline\hline
\multicolumn{4}{c}{z=8.5-11.5}\\
F435W+F475W & $-$1$_{-1}^{+1}$ &  0$_{-2}^{+1}$ &  1$_{-1}^{+3}$\\
F606W+F775W+F814W & 0$_{-1}^{+1}$ &  0$_{-1}^{+1}$ &  2$_{-2}^{+2}$\\
F090W & $-$1$_{-2}^{+1}$ &  0$_{-1}^{+1}$ &  6$_{-2}^{+3}$\\
F115W & 6$_{-4}^{+4}$ &  2$_{-1}^{+2}$ &  5$_{-2}^{+3}$\\
F125W+F140W & 10$_{-5}^{+5}$ &  5$_{-4}^{+5}$ &  12$_{-6}^{+7}$\\
F160W & 22$_{-7}^{+8}$ &  25$_{-8}^{+9}$ &  20$_{-8}^{+9}$\\
F150W & 22$_{-6}^{+7}$ &  28$_{-6}^{+6}$ &  29$_{-5}^{+5}$\\
F200W & 21$_{-6}^{+7}$ &  34$_{-6}^{+7}$ &  32$_{-5}^{+6}$\\
F277W & 23$_{-6}^{+6}$ &  31$_{-6}^{+6}$ &  35$_{-6}^{+5}$\\
F356W & 27$_{-7}^{+7}$ &  32$_{-5}^{+5}$ &  36$_{-6}^{+5}$\\
F444W & 53$_{-14}^{+14}$ &  38$_{-7}^{+7}$ &  33$_{-5}^{+5}$\\
\\
\multicolumn{4}{c}{z=11.5-15.0}\\
F435W+F475W & 11$_{-8}^{+7}$ &  0$_{-14}^{+20}$ &  1$_{-3}^{+5}$\\
F606W+F775W+F814W & 0$_{-2}^{+1}$ &  2$_{-3}^{+5}$ &  2$_{-2}^{+3}$\\
F090W & 0$_{-1}^{+1}$ &  $-$5$_{-5}^{+3}$ &  3$_{-3}^{+3}$\\
F115W & $-$3$_{-5}^{+3}$ &  $-$7$_{-21}^{+7}$ &  6$_{-3}^{+4}$\\
F125W+F140W & $-$19$_{-19}^{+11}$ &  $-$8$_{-31}^{+23}$ &  5$_{-5}^{+8}$\\
F160W & 68$_{-53}^{+49}$ &  53$_{-50}^{+75}$ &  25$_{-15}^{+21}$\\
F150W & 4$_{-5}^{+6}$ &  5$_{-2}^{+3}$ &  13$_{-4}^{+5}$\\
F200W & 42$_{-19}^{+19}$ &  33$_{-12}^{+13}$ &  36$_{-8}^{+8}$\\
F277W & 38$_{-17}^{+18}$ &  36$_{-11}^{+11}$ &  37$_{-7}^{+8}$\\
F356W & 33$_{-15}^{+15}$ &  32$_{-10}^{+10}$ &  33$_{-7}^{+7}$\\
F444W & 37$_{-18}^{+17}$ &  31$_{-10}^{+10}$ &  29$_{-6}^{+7}$\\
\\
\multicolumn{4}{c}{z=15.0-20.0}\\
F435W+F475W & 11$_{-8}^{+7}$ &  0$_{-14}^{+20}$ &  1$_{-3}^{+5}$\\
F606W+F775W+F814W & 0$_{-2}^{+1}$ &  2$_{-3}^{+5}$ &  2$_{-2}^{+3}$\\
F090W & 0$_{-1}^{+1}$ &  $-$5$_{-5}^{+3}$ &  3$_{-3}^{+3}$\\
F115W & $-$3$_{-5}^{+3}$ &  $-$7$_{-21}^{+7}$ &  6$_{-3}^{+4}$\\
F125W+F140W & $-$19$_{-19}^{+11}$ &  $-$8$_{-31}^{+23}$ &  5$_{-5}^{+8}$\\
F160W & 68$_{-53}^{+49}$ &  53$_{-50}^{+75}$ &  25$_{-15}^{+21}$\\
F150W & 4$_{-5}^{+6}$ &  5$_{-2}^{+3}$ &  13$_{-4}^{+5}$\\
F200W & 42$_{-19}^{+19}$ &  33$_{-12}^{+13}$ &  36$_{-8}^{+8}$\\
F277W & 38$_{-17}^{+18}$ &  36$_{-11}^{+11}$ &  37$_{-7}^{+8}$\\
F356W & 33$_{-15}^{+15}$ &  32$_{-10}^{+10}$ &  33$_{-7}^{+7}$\\
F444W & 37$_{-18}^{+17}$ &  31$_{-10}^{+10}$ &  29$_{-6}^{+7}$\\
\hline
\end{tabular}
\end{table}

\section{UV LF at $z\sim8$}

It is useful to test of our procedures for deriving the $UV$ LF at
high redshifts using {\it JWST} data to ensure we can arrive at reliable
results.

\begin{figure}
\centering
\includegraphics[width=\columnwidth]{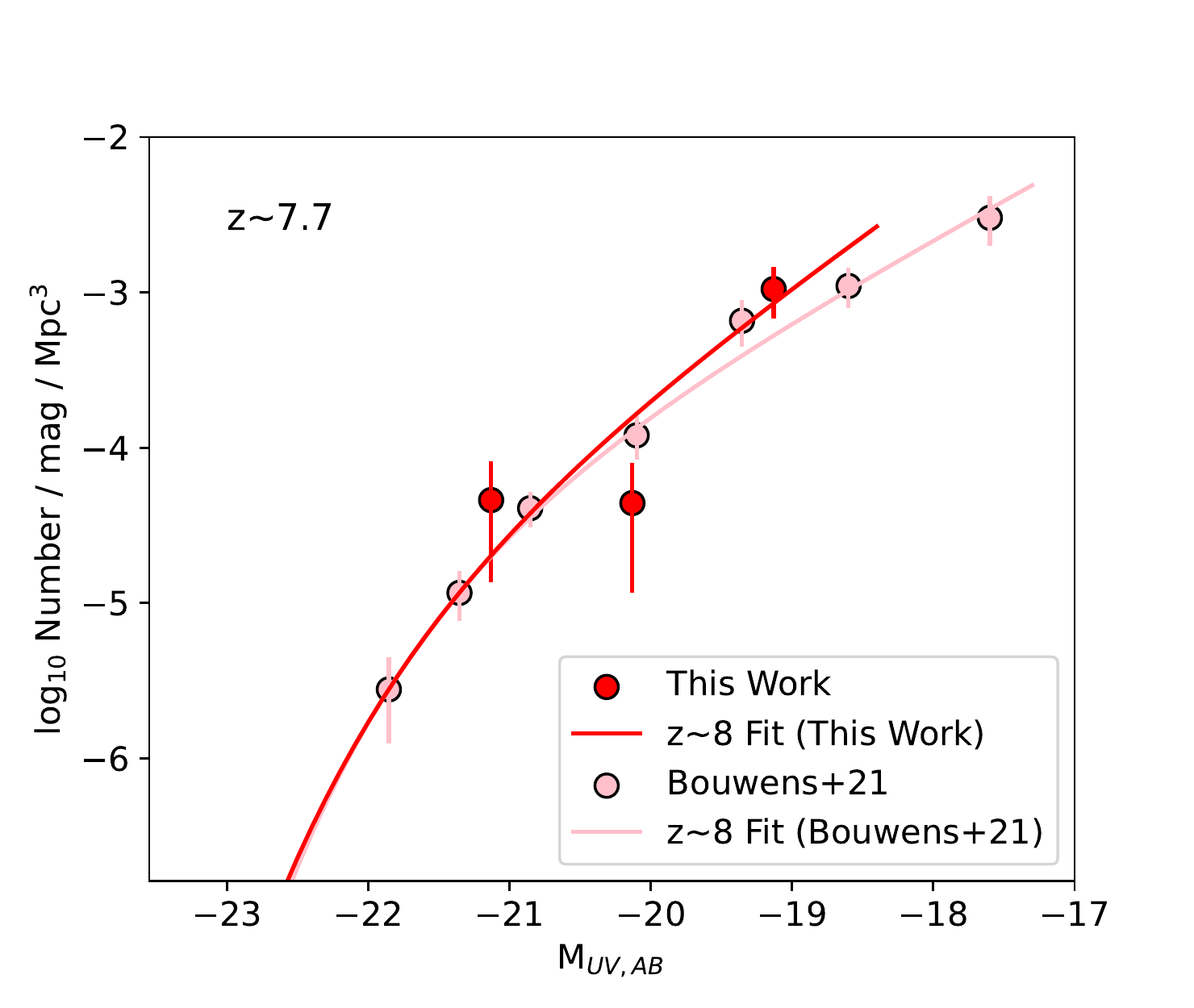}
\caption{Determination of the $UV$ LF at $z\sim8$ (\textit{solid red
    circles}) leveraging the new {\it JWST} observations over the three of
  the most studied early NIRCam observations with {\it JWST}.}
    \label{fig:uvlf8}
\end{figure}

To this end, we made use of our $z\sim8$ F090W-dropout samples and the
same methodology as we use for our $z\geq 8$ analyses to derive the
$UV$ LFs at $z\sim8$.  The results are presented in
Figure~\ref{fig:uvlf8} and in
Tables~\ref{tab:binnedlfs}-\ref{tab:lffits}, and it is clear that the
results are consistent with what we derived earlier in
\citet{Bouwens2021_LF} on the basis of sensitive imaging observations
with HST.

As such, we can conclude that our procedures should produce reliable
$UV$ LF results at $z\geq 8$, assuming we are able to identify
significant samples at $z\geq 8$ which are largely free of
contamination from lower-redshift interlopers.

\section{Study-to-study Scatter in the Assessment of Various $z\geq9$ Candidates}

Ascertaining whether individual $z\geq 8$ candidates are at low or
high-redshift can be challenging to answer in specific cases, due to
uncertainties in both the photometry of individual sources and the optimal SED templates to utilize in performing the fits.

\begin{figure}
\centering
\includegraphics[width=\columnwidth]{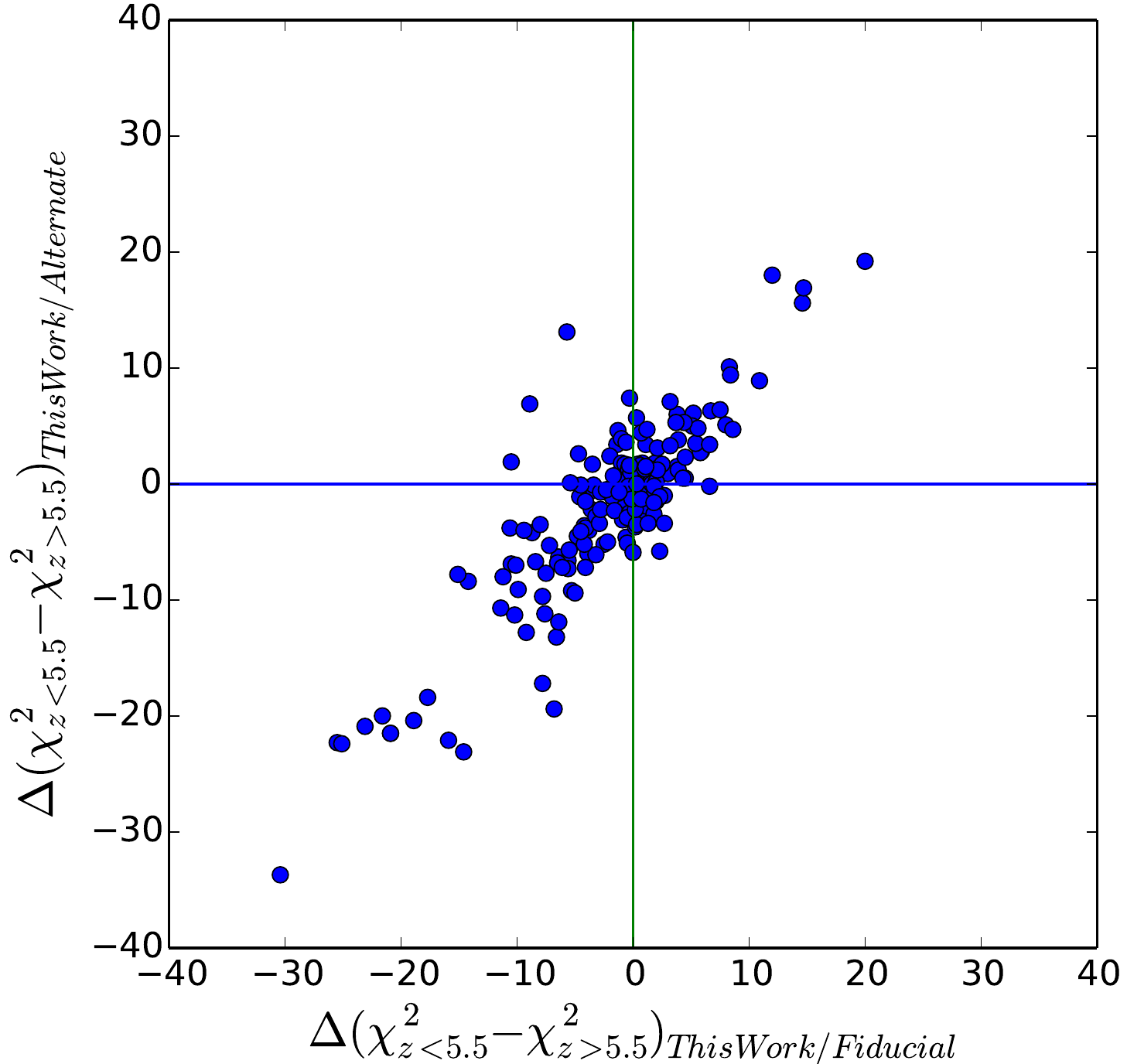}
\caption{Difference between the minimum $\chi^2$ achieved with $z>5.5$
  SED fits to specific sources and that obtained with $z<5.5$ SED
  fits.  $\Delta\chi^2$ values less than $-9$ or $-4$ tend to indicate
  sources are at $z>5.5$ at high confidence.  We note that there is
  nevertheless a substantial dispersion in the derived $\Delta\chi^2$
  values depending on the reductions.  If the uncertainties are
  similar in various literature studies, it could point to there being
  a significant amount of contamination and incompleteness in existing
  $z>5.5$ selections.}
    \label{fig:dchisq}
\end{figure}

\begin{figure}
\centering \includegraphics[width=\columnwidth]{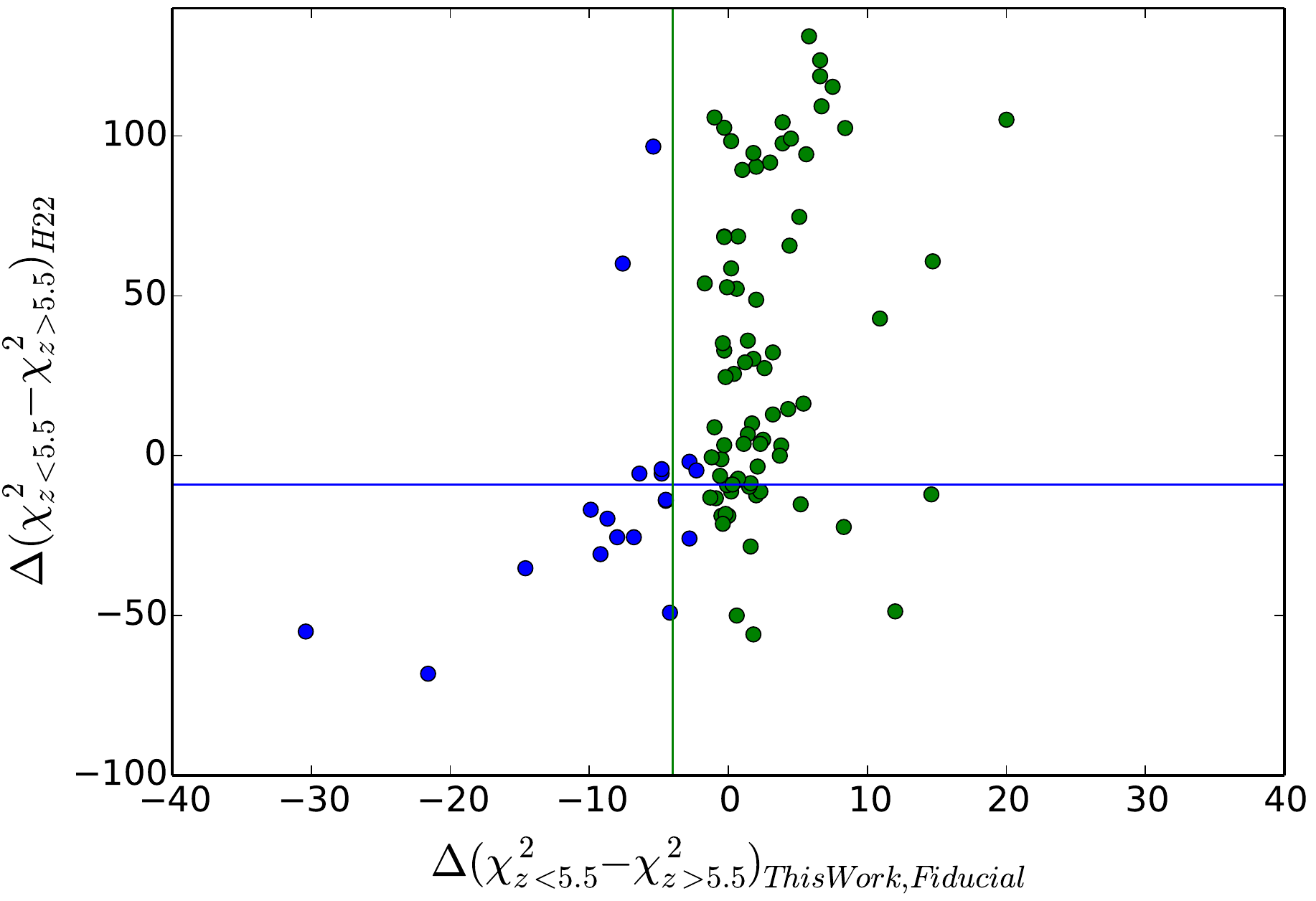}
\caption{Similar to Figure~\ref{fig:dchisq} but comparing the
  source-by-source results of \citet{Harikane2022_z9to17} with those
  we obtain using our fiducial reductions from \textsc{grizli}.
  Sources shown in blue are included in our fiducial sample while
  those shown in green are not.  Note that the substantial dispersion
  in the derived $\Delta\chi^2$ values depending on the analysis.}
    \label{fig:dchisq-h22}
\end{figure}

As an illustration of these uncertainties, Figure~\ref{fig:dchisq}
shows a comparison of the $\Delta(\chi^2_{min,z>5.5} -
\chi_{min,z<5.5}^2)$ results obtained using our photometry on the
\textsc{grizli} reductions and those obtained using the
\textsc{pencil} reductions.  As a second illustration of the
study-to-study differences, Figure~\ref{fig:dchisq-h22} shows a
comparison between the $\Delta(\chi^2_{min,z>5.5} -
\chi_{min,z<5.5}^2)$ results \citet{Harikane2022_z9to17} and the
results we obtain using our \textsc{grizli} reductions.

In both comparisons, there is clearly a significant amount of scatter
in the assessments that are made about specific candidates.

\end{document}